\documentclass[preprint2]{aastex61}
\usepackage{amsmath}
\PassOptionsToPackage{breaklinks=true}{hyperref}
\def\gmrt {\emph{GMRT}}

\def\gax{\gtrsim}

\received{November 01, 2023}
\revised{November 29, 2023}
\accepted{December 04, 2023}
\submitjournal{ApJ}

\shorttitle{Minihalos and large-scale sloshing in A3444 and MS\,1455.0+2232}
\shortauthors{Giacintucci et al.}

\begin{document}

\title{Are radio minihalos confined by cold fronts in galaxy clusters?
Minihalos and large-scale sloshing in A3444 and MS\,1455.0+2232}

\correspondingauthor{Simona Giacintucci}
\email{simona.giacintucci@nrl.navy.mil}

\author{S. Giacintucci}
\affiliation{U.S. Naval Research Laboratory, 
4555 Overlook Avenue SW, Code 7213, 
Washington, DC 20375, USA}

\author{T. Venturi}
\affiliation{INAF - Istituto di Radioastronomia,
via Gobetti 101, I-40129 Bologna, Italy}

\author{M. Markevitch}
\affiliation{NASA/Goddard Space Flight Center,
Greenbelt, MD 20771, USA}

\author{G. Brunetti}
\affiliation{INAF - Istituto di Radioastronomia,
via Gobetti 101, I-40129 Bologna, Italy}

\author{T. E. Clarke}
\affiliation{U.S. Naval Research Laboratory, 
4555 Overlook Avenue SW, Code 7213, 
Washington, DC 20375, USA}

\author{R. Kale}
\affiliation{National Centre for Radio Astrophysics, 
Tata Institute of Fundamental Research, Pune University, 
Pune 411007, India}

\begin{abstract}

  We present radio and X-ray studies of A3444 and MS1455.0+2232, two galaxy clusters with radio minihalos in their cool cores. A3444 is imaged using the Giant Metrewave Radio Telescope (GMRT) at 333, 607 and 1300 MHz and the Very Large Array at 1435 MHz. Most of the minihalo is contained within $r<120$ kpc, but a fainter extension, stretching out to $380$ kpc South-West of the center, is detected at 607 MHz. Using {\em Chandra}, we detect four X-ray sloshing cold fronts: three in the cool core at $r=60$, $120$ and $230$ kpc, and a fourth one at $r=400$ kpc --- in the region of the southwestern radio extension --- suggesting that the intracluster medium (ICM) is sloshing on a cluster-wide scale. The radio emission is contained within the envelope defined by these fronts. We also analyzed archival 383 MHz GMRT and {\em Chandra}\/ observations of MS\,1455.0+2232, which exhibits a known minihalo with its bright part delineated by cold fronts inside the cool core, but with a faint extension beyond the core. Similarly to A3444, we find a cold front at $r\sim 425$ kpc, containing the radio emission. Thus the entire diffuse radio emission seen in these clusters appears to be related to large-scale sloshing of the ICM. The radio spectrum of the A3444 minihalo is a power law with a steep index $\alpha=1.0\pm0.1$. The spectrum steepens with increasing distance from the center, as expected if the minihalo originates from re-acceleration of relativistic particles by the sloshing-induced turbulence in the ICM. 
\end{abstract}


\section{Introduction}\label{sec:intro}

In the cool cores of relaxed clusters of galaxies, the central radio 
galaxy is often surrounded by diffuse, 
steep-spectrum\footnote{Radio spectral index $\alpha \gax 1$, for $S_{\nu} \propto \nu^{-\alpha}$, where $S_{\nu}$ is the flux density at the frequency $\nu$.} 
synchrotron sources, called radio minihalos
\citep[e.g.,][for a review]{2019SSRv..215...16V}.
The number of known minihalos sums up to about 40 
detections including candidates 
\citep[e.g.,][and references therein]{2019ApJ...880...70G, 2020MNRAS.499.2934R};
for recent detections see \cite{2020AJ....160..103P,2020A&A...640A.108G,2021MNRAS.503.4627U,2021PASA...38...53D,2022A&A...657A..56K,2022MNRAS.514.5969K,2023MNRAS.519..767B,2023arXiv230801884L,2023MNRAS.524.6052R}.
They are usually found in high-mass clusters, with a high incidence of $\sim 80\%$ in the most massive systems ($M_{500}>6\times10^{14}$ $M_{\odot}$\footnote{Total mass within $R_{500}$, where
$R_{500}$ is the radius within which the cluster mean total density is 500 times 
the critical density at the cluster redshift.}) among the cool-core cluster 
population \citep[][]{2017ApJ...841...71G}.

Radio minihalos can either arise from electrons that are injected locally by hadronic interactions between cosmic-ray (CR) protons and the cluster thermal protons (secondary models), or from pre-existing, mildly relativistic seed electrons 
that are re-accelerated to ultra-relativistic energies by turbulence in the cluster core
\citep[e.g.,][]{2002A&A...386..456G,2004A&A...413...17P, 2007ApJ...663L..61F, 2010ApJ...722..737K, 2013MNRAS.428..599F, 2014MNRAS.438..124Z, 2013ApJ...762...78Z,2015ApJ...801..146Z, 2017MNRAS.467.1478J, 2020A&A...640A..37I}. 
A review can be found in \cite{2014IJMPD..2330007B}. CR injection by the cluster 
central active galactic nucleus (AGN) may provide the relativistic electrons directly 
as secondary particles \citep[e.g.,][and references therein]{2020A&A...640A..37I}
or as seed electrons that are then re-energized by turbulence \citep[][]{2008A&A...486L..31C,2013ApJ...762...78Z,2021ApJ...914...73Z}. 
Hadronic collisions can be an alternative/additional source of seed electrons for reacceleration \citep[e.g.,][]{2004A&A...413...17P}.

Minihalos are often contained within the cluster central cooling region ($r < 0.2R_{500}$
\footnote{$0.2R_{500}$ is the typical boundary between the cluster core, where 
non-gravitational processes (cooling, AGN and stellar feedback) 
become important, and the outer cluster region where density, 
temperature and pressure profiles of the ICM are self-similar 
\citep[e.g.,][and references therein]{2017ApJ...843...28M}
}
; \cite{2017ApJ...841...71G}), suggesting a close relation between the radio 
emission and the properties of the thermal plasma in the cool cores.  
The existence of positive relations between radio luminosity 
and cluster global and cool-core X-ray luminosity supports a 
connection between the thermal and non-thermal cluster components
\citep[e.g.,][]{2015A&A...579A..92K, 2016MNRAS.455L..41B, 2018A&A...617A..11G, 2020MNRAS.499.2934R, 2019ApJ...880...70G}.

Most of the minihalo emission is often confined by X-ray cold fronts \citep[e.g.,][]{2008ApJ...675L...9M,2013ApJ...777..163H, 2014ApJ...781....9G,2014ApJ...795...73G,2017MNRAS.469.3872G,2018MNRAS.478.2234S,2021A&A...646A..38T,2022MNRAS.512.4210R}, hinting at a link between the origin of the diffuse emission and sloshing motions of the central cool gas  \citep[for reviews of cold fronts see, e.g.,][]{2007PhR...443....1M,2022hxga.book...93Z}. Sloshing can strengthen the magnetic field and inject turbulence in the cool core 
\citep[e.g.,][]{2004ApJ...612L...9F,2010arXiv1011.0729K,2012A&A...544A.103V,2013ApJ...762...78Z},
which in turn can confine and re-accelerate seed electrons, generating diffuse radio 
emission in the region enveloped by the cold fronts \citep{2013ApJ...762...78Z}, as observed
in actual minihalos. 

Recent radio observations have further complicated our understanding of minihalos, 
and of their relation to giant halos in merging clusters \citep[e.g.,][]{2019SSRv..215...16V}, by unveiling diffuse emission outside the sloshing cool 
core in a few minihalo clusters \citep{2017MNRAS.469.3872G, 2018MNRAS.478.2234S,2019A&A...622A..24S,2021MNRAS.508.3995B,2021MNRAS.508.2862P,2022MNRAS.512.4210R}. 
Diffuse emission extending far beyond the inner sloshing region has also 
been found in a few non-cool core clusters \citep[e.g.,][]{2015MNRAS.448.2495S,2017A&A...603A.125V,2022ApJ...934...49G,2023arXiv230807603B,2023arXiv230801884L, 2023MNRAS.524.6052R}. These findings may indicate a co-existence of multiple components of diffuse emission in the same cluster, which may be powered by different processes or mechanisms acting on different spatial and time scales \citep[e.g.,][]{2023arXiv230807603B}. 

Past powerful outbursts of the central AGN may also be responsible for 
very extended, steep-spectrum radio emission located outside 
of the cluster core, as in Ophiuchus \citep{2020ApJ...891....1G} where a $500$ kpc--wide fossil radio lobe was found well beyond the central sloshing region occupied by the minihalo (see also discussion on A\,2319 in \cite{2021MNRAS.504.2800I}).

In this paper, we study the minihalos at the center of A\,3444 and MS\,1455.0+2232, two cool-core clusters at $z=0.254$ and $z=0.258$, respectively.
The minihalo in A\,3444 was first reported by \cite{2007A&A...463..937V} 
using a Giant Metrewave Radio Telescope (GMRT) observation at 610 MHz and by \cite{2009A&A...507.1257G} with the Very Large Array (VLA). It was later 
confirmed by VLA images 1435 MHz presented in \cite{2019ApJ...880...70G},
where it was detected on a scale of $r\sim 120$ kpc. Recently, \cite{2023MNRAS.520.4410T} 
presented new, deeper MeerKAT images at 1283 MHz, in which the diffuse radio emission 
was found to further stretch toward South/South-West into a $\sim 100$ kpc-long, 
faint extension. 

MS\,1455.0+2232 (also known as Z1760) hosts a known minihalo at 607 MHz and 1435 MHz \citep{2008A&A...484..327V,2019ApJ...880...70G}, with its bright part delineated 
by sloshing cold fronts inside the cool core \citep{2001astro.ph..8476M,2008ApJ...675L...9M}.
It was recently imaged at higher sensitivity by \cite{2022MNRAS.512.4210R} 
with MeerKAT at 1283 MHz and the LOw- Frequency ARray (LOFAR) at 145 MHz. 
Similarly to A\,3444, a faint radio extension was found toward 
the South and South-West, extending beyond the sloshing cool core.

Here, we present new GMRT observations of A\,3444 at 607 MHz and 1300 MHz. 
We complement these observations with archival GMRT data at 333 MHz and VLA 1435 
MHz data from \cite{2019ApJ...880...70G} to study the spectral properties of 
the diffuse emission, which can provide information on the origin of the 
radio-emitting electrons. We also present the analysis of 
an archival upgraded GMRT (uGMRT) observation in Band 3 (250--500 MHz) 
containing MS\,1455.0+2232 in its field of view, at $\sim 12^{\prime}$ from the
phase center. For both clusters, we use archival {\em Chandra} X-ray observations to search for cold fronts at large radii and investigate the 
connection of the minihalos (including their extensions past the cool core) with sloshing of the intracluster medium (ICM) on large scale. 

Throughout the paper, we adopt a $\Lambda$CDM cosmology with H$_0$=70 km s$^{-1}$ Mpc$^{-1}$, 
$\Omega_m=0.3$ and $\Omega_{\Lambda}=0.7$. At the redshift of A\,3444 and MS\,1455.0+2232, $1^{\prime\prime}$ corresponds to $\sim 4$ kpc. All errors are quoted at the 68\% confidence level.

\section{Radio observations of A\,3444}\label{sec:obs}

We present new GMRT images at 607 MHz, which allow us to image the minihalo with a higher 
sensitivity than previously achieved in \cite{2007A&A...463..937V}. 
We also present new, high resolution GMRT data at 1300 MHz, and analyze archival 
GMRT data at 333 MHz. We summarize these observations in Table \ref{tab:obs}
along with details on the archival VLA datasets re-analyzed in \cite{2019ApJ...880...70G}. 
For each dataset, we report the largest detectable angular scale ($\theta_{\rm LAS}$) set 
by the minimum $uv$ spacing. All datasets are nominally able to detect extended 
emission on a maximum angular scale of at least $7^{\prime}$ which corresponds to 
a physical scale of $\sim 1.6$ Mpc at the redshift of A\,3444 (note that
the VLA BnC and DnC datasets were combined together in the $uv$ plane to produce the final
images). This scale is much larger than the extent of the minihalo inferred from 
previously-published images, i.e., $\sim 1^{\prime}=240$ kpc in diameter 
\citep{2007A&A...463..937V,2019ApJ...880...70G}.


\begin{table*}
\caption{Radio observations of A\,3444}
\begin{center}
\begin{tabular}{cccccccccc}
\hline\noalign{\smallskip}
\hline\noalign{\smallskip}
Array & Project & Frequency & Bandwidth & Date & Time  & FWHM, p.a.  &   rms  &  $\theta_{\rm LAS}$  \\
      &  &  (MHz)    &   (MHz)   &      &  (min) & ($^{\prime \prime} \times^{\prime \prime}$, $^{\circ}$)\phantom{00} & ($\mu$Jy b$^{-1}$) &  ($^{\prime}$) \\
\noalign{\smallskip}
\hline\noalign{\smallskip}
GMRT      &  $05VKK01$      &   333   &   32           &   2004 Apr 24    & 67    & $14\times12$, $29$   & 550   &   $\sim 30$  \\
GMRT      &  $30_{-}065$     &   607   &   33    &   2016 Jul 23    & 325   &  $8\times4$, 14       &  35   &   $\sim 17$ \\
GMRT      &  $30_{-}065$     &   1300 &   33   &    2016 Aug 20   & 185   & $4\times2$, 2         &  35   &   $\sim 7$ \\
VLA$-$BnA$^a$ &  AC696          & 1435   &    50   & 2003 Oct 02      & 200   & $8.3\times4.0$, 3     & 35    &   $\sim 2$    \\
VLA$-$DnC$^a$ &  AC696          & 1435   &    50    & 2004 May 30      & 260   & $43\times30$, 58  & 50    &   $\sim 16$  \\

\hline{\smallskip} 
\end{tabular}
\end{center}
\label{tab:obs}
{\bf Notes.} Column 1: radio telescope. Column 2: project code. Columns 3--5: observation frequency, bandwidth and date. 
Column 6: effective time on source (after flagging). Column 7 and 8: full width at half-maximum (FWHM) and position angle 
(p.a.) of the synthesized beam and rms noise level ($1\sigma$) in images made using a Briggs robustness parameter of 0. 
Column 9: largest angular scale detectable by the observation.

$^a$ VLA archival observations from \cite{2019ApJ...880...70G}. 
\end{table*}

\subsection{GMRT 607 and 1300 MHz observations}

We observed A\,3444 on July 24 and August 20 2016 using the GMRT at 
the frequencies of 607 MHz and 1300 MHz (project 30$_{-}$065) for a total of
8.5 hours and 5.6 hours, respectively (including time overheads for calibration). 
The visibilities were acquired in spectral-line observing mode using 
the GMRT software backend with a total observing bandwidth of 33 MHz, 
subdivided in 512 channels, and both RR and LL polarizations. 

The data were reduced using the 
Astronomical Image Processing System
\citep[AIPS\footnote{http://www.aips.nrao.edu.},][]{2003ASSL..285..109G}.
We used the task RFLAG to excise visibilities 
affected by radio frequency interference (RFI), followed by manual flagging to remove 
residual bad data. Gain and bandpass calibrations were applied using the primary 
calibrators 3C147 and 3C286 at 607 MHz and 3C286 at 1300 MHz. 
The sources 1018-333 and 1526-138, observed several times during the observations, 
were used to calibrate the data in phase. A number of phase self-calibration 
cycles were applied to the target visibilities. Non-coplanar effects were taken 
into account using wide-field imaging by decomposing the primary beam area
into $\sim$80 smaller facets at 607 MHz and $\sim 50$ facets at 1300 MHz.

The final images at 607 MHz were obtained in AIPS using multi-scale imaging.
The final self-calibrated data set at 1300 MHz was first converted into
 a measurement set using the Common Astronomy Software Applications \citep[CASA\footnote{https://casa.nrao.edu.}, version 5.1][]{2022PASP..134k4501C}
and then imaged
using the multi-scale deconvolution option in WSClean \citep{2014MNRAS.444..606O, 2017MNRAS.471..301O}.
The root mean square (rms) sensitivity level ($1\sigma$) achieved in the images at full resolution,
obtained setting the Briggs robustness weighting to 0 \citep{1995PhDT.......238B},
is 35 $\mu$Jy beam$^{-1}$ both at 607 MHz and 1300 MHz (Tab.~\ref{tab:obs}). 

We also produced images with lower angular resolution increasing the robustness parameter and/or
applying tapers to the $uv$ data during
the imaging process. The noise in these images is 50 $\mu$Jy beam$^{-1}$ at 607 MHz ($20^{\prime\prime}$
resolution) and 43 $\mu$Jy beam$^{-1}$ at 1300 MHz ($15^{\prime\prime}$ resolution). 

Finally, for flux density measurements, we corrected the final images for the \gmrt\ primary beam 
response\footnote{http://www.ncra.tifr.res.in:8081/{\textasciitilde}ngk/
\\
primarybeam/beam.html.} 
using PBCOR in AIPS. All flux densities are given on the \cite{2017ApJS..230....7P}
wide-band scale. Residual amplitude errors are estimated to be within 
$5\%$ at both frequencies \citep[e.g.,][]{2004ApJ...612..974C}.


\begin{table*}
\caption{uGMRT radio observation containing MS\,1455.0+2232.}
\begin{center}
\begin{tabular}{cccccccc}
\hline\noalign{\smallskip}
\hline\noalign{\smallskip}
Project & Phase center & Distance & Frequency & Bandwidth &  Time  & FWHM, p.a.  &   rms  \\
        &  (h m s, $^{\circ}$ $^{\prime}$ $^{\prime\prime}$ ) & ($^{\prime}$) &  (MHz)    &   (MHz)   &     (min) & ($^{\prime \prime} \times^{\prime \prime}$, $^{\circ}$)\phantom{00} & ($\mu$Jy b$^{-1}$) \\
\noalign{\smallskip}
\hline\noalign{\smallskip}
 38$_{-}$010 & 14 57 35.13, +22 32 01 & 12.3 & 383 & 200 &  262 & 7.1$\times$6.1, 22 & 24\\
 \hline{\smallskip} 
\end{tabular}
\end{center}
\label{tab:obs2}
{\bf Notes.} Column 1: project code. Columns 2--3: phase center of the observation and distance of MS\,1455.0+2232 from the phase center. Columns 4--6: central frequency, bandwidth and time on source. Columns 7--8: FWHM and p.a. of the beam and rms noise level ($1\sigma$) in images made using a Briggs robustness parameter of 0.
\end{table*}


\subsection{Archival GMRT data at 333 MHz}

We analyzed archival GMRT observations of A\,3444 at 333 MHz.
The cluster was observed in April 2004 as part of the GMRT Cluster Key Project 
(05VKK01) for a total of $\sim 1$ hour on source. The data were recorded 
using the old hardware correlator and both the upper and lower side bands 
(USB and LSB), providing a total observing bandwidth of 32 MHz. 
The default spectral-line mode was used with 128 frequency channels per band, 
each of width 125 kHz. 3C286 was observed as flux density and bandpass calibrator. 
We calibrated the USB and LSB data sets individually in AIPS. A combination 
of RFLAG and manual flagging was used to remove RFI and bad data. After 
bandpass calibration and a priori amplitude calibration using the \cite{2017ApJS..230....7P} flux density scale, 
a number of phase-only self-calibration cycles and imaging were carried 
out for each data set. Wide-field imaging was implemented in each step of 
the data reduction, with 25 facets covering the primary beam area. 
The final self-calibrated USB and LSB data sets was then converted into
measurement sets using CASA 5.1 
and imaged together using joint deconvolution in WSClean. 
The rms noise in the final image at full resolution 
($14^{\prime\prime}\times12^{\prime\prime}$ for a robust of 0) 
is $550$ $\mu$Jy beam$^{-1}$.
Residual amplitude errors are estimated to be $<10\%$ \citep[e.g.,][]{2004ApJ...612..974C}.

\section{uGMRT radio observations of MS\,1455.0+2232}\label{sec:obs2}

We reduced and imaged an archival uGMRT observation in Band 3 
(250-500 MHz; project 38$_{-}$010) containing MS\,1455.0+2232 
at $12^{\prime}.3$ from the phase center (the FWHM of the GMRT primary beam in Band 3 is $75^{\prime}$).
The observation was made on 2020 Jul 29, for a total of $\sim 260$ minutes on 
target (excluding calibration overheads; Tab.~\ref{tab:obs2}). 

We used the Source Peeling and Atmospheric Modeling \citep[SPAM;][]{2009A&A...501.1185I}
pipeline\footnote{http://www.intema.nl/doku.php?id=huibintemaspampipeline} to process the uGMRT data, adopting a standard calibration scheme that consists of bandpass and complex gain calibration and direction-independent self-calibration, followed by direction-dependent self-calibration. The flux density scale was set using 3C\,286 and \cite{2012MNRAS.423L..30S}. 
The wide-band dataset was first divided into six narrower sub-bands, each 33.3 MHz wide.
Each sub-band was then individually processed by the pipeline. 
The SPAM self-calibrated visibilities were converted into measurement sets using CASA and finally imaged together using joint-channel and multi-scale deconvolution in WSClean. For the imaging, we applied different weighting schemes (from uniform weights to a robust of $+0.2$) and $uv$ tapers.

Correction for the \gmrt\ primary-beam response was applied using the task PBCOR in AIPS and the primary-beam shape parameters given 
in a December 2018 GMRT memo\footnote{http://www.ncra.tifr.res.in/ncra/gmrt/gmrt-users/
observing-help/ugmrt-primary-beam-shape}. The systematic amplitude uncertainty 
was assumed to be 15$\%$.

%
%
\begin{figure*}
\gridline{\fig{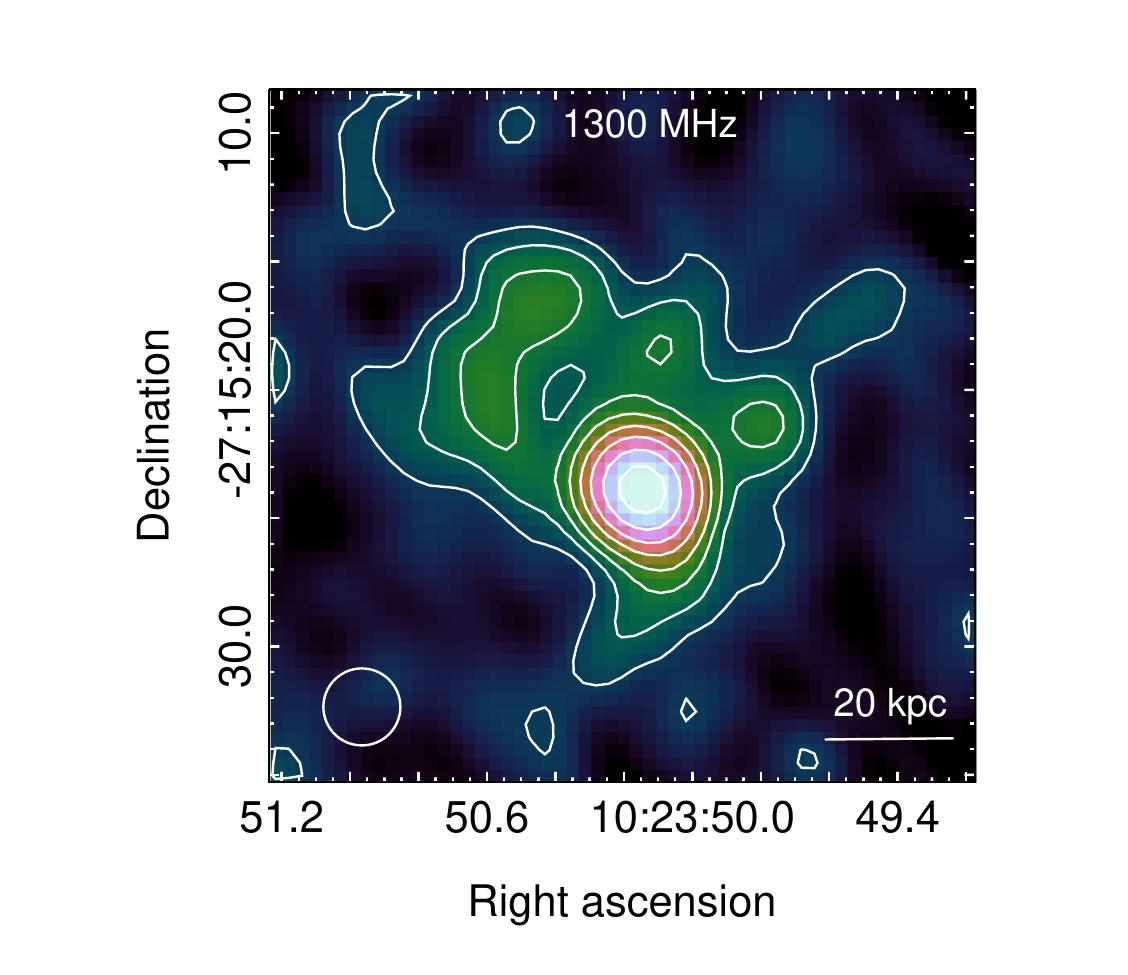}{0.5\textwidth}{(a)}
\hspace{-0.8cm}
          \fig{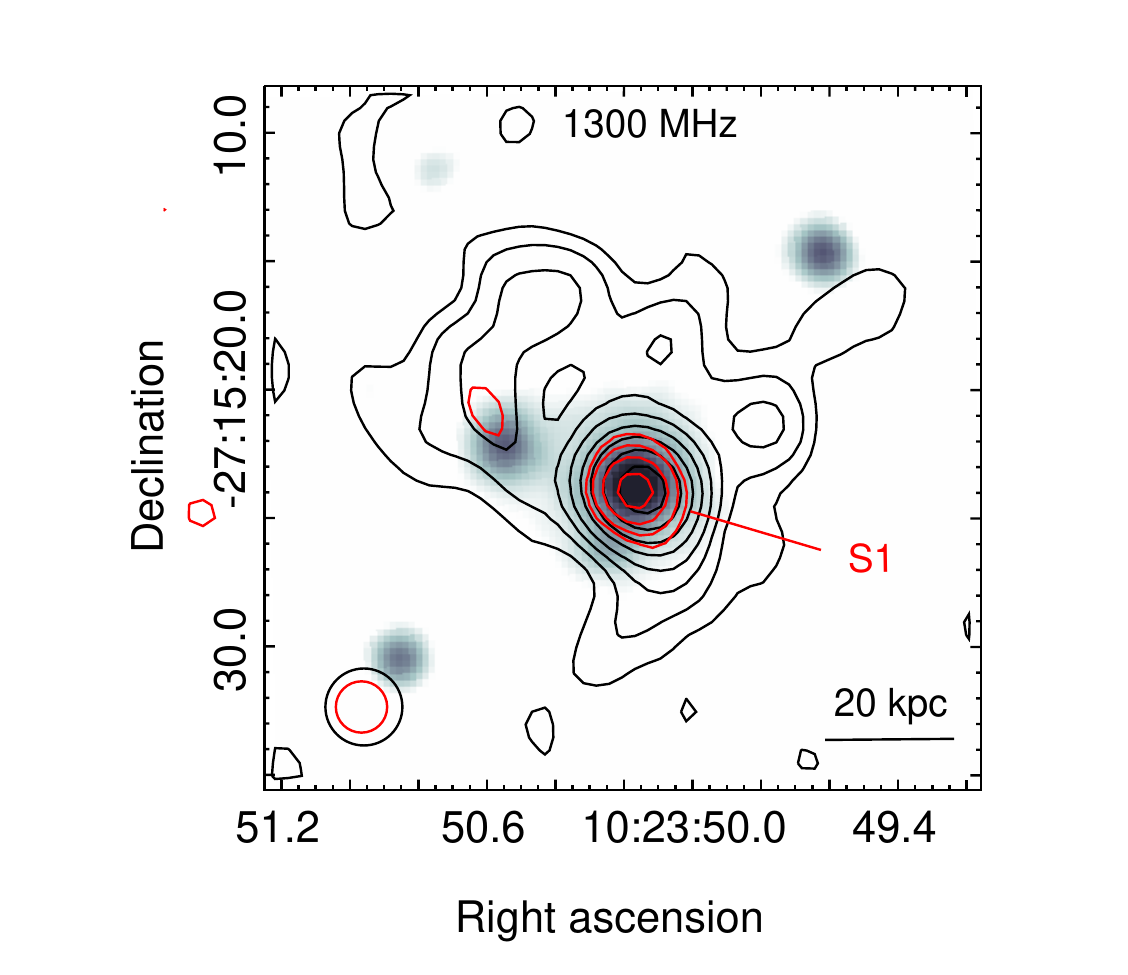}{0.5\textwidth}{(b)}}
\vspace{-0.5cm}\gridline{\hspace{0.8cm}\fig{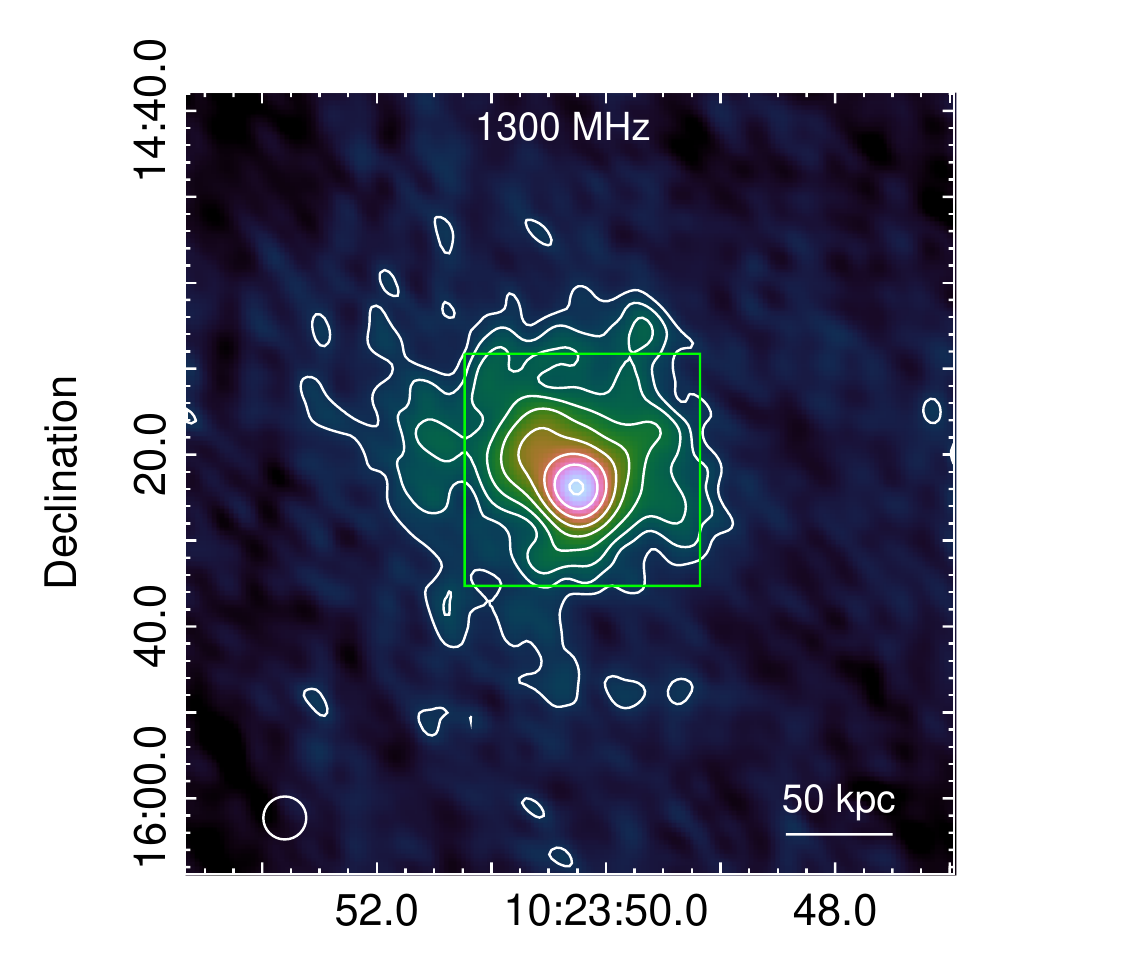}{0.48\textwidth}{(c)}
\fig{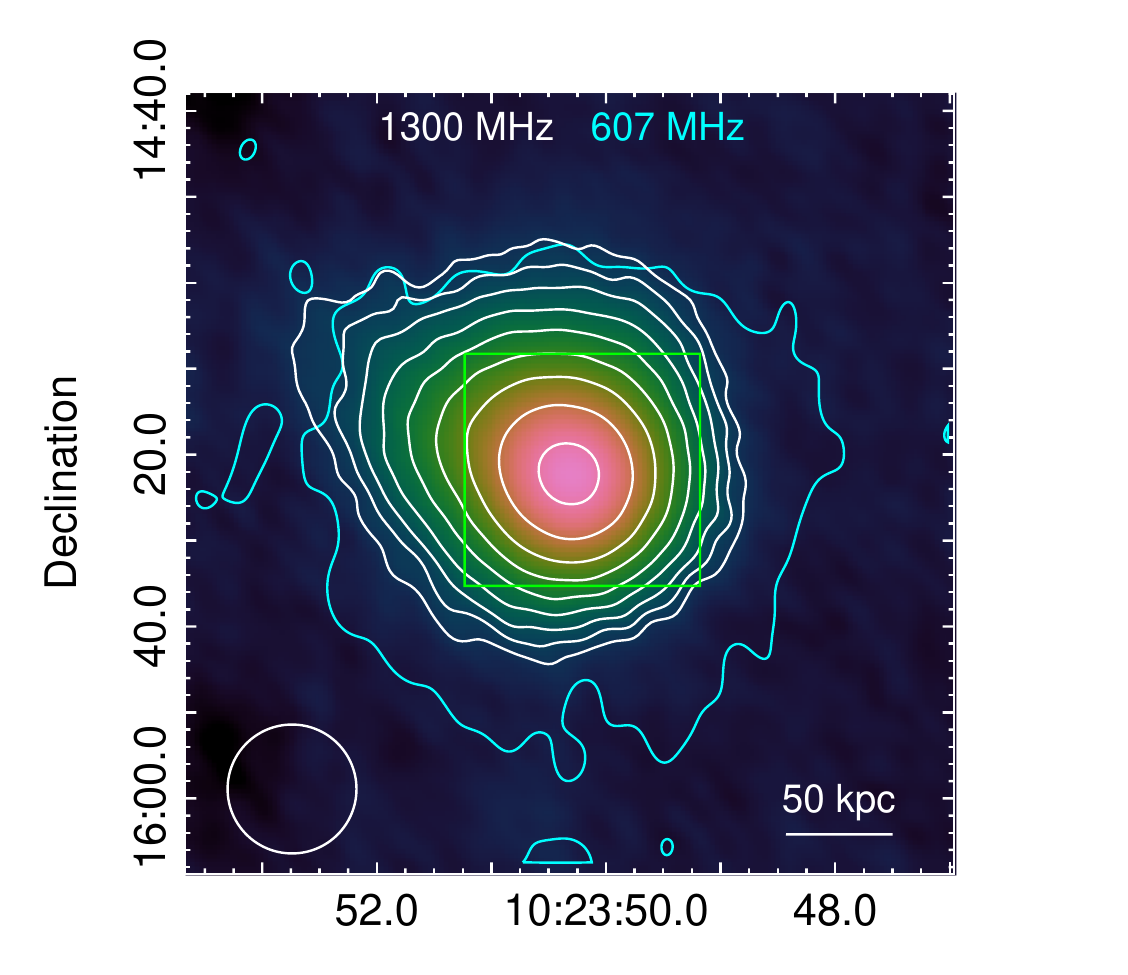}{0.48\textwidth}{(d)}}
          \caption{A\,3444. (a) GMRT 1300 MHz image with robust 0.5 and 
$3^{\prime\prime}$ beam. The noise is $1\sigma=35$ $\mu$Jy beam$^{-1}$. 
Contours are spaced by a factor of $\sqrt{2}$ from $0.1$ mJy beam$^{-1}$. 
No levels at $-0.1$ mJy  beam$^{-1}$ are present. 
(b) GMRT 1300 MHz contours (black) from (a), overlaid on the optical $z$-band Pan-STARRS-1 image. Red contours are from a $2^{\prime\prime}$-resolution image at 1300 MHz made with pure uniform weighting, and are spaced by a factor of 2 from $0.1$ mJy beam$^{-1}$. The compact source S1 is associated with the BCG. (c, d) GMRT 1300 MHz
  images (colors and white contours) with robust 2 and $5^{\prime\prime}$ and $15^{\prime\prime}$ circular beams, respectively. The noise levels 
  are $1\sigma=43$ $\mu$Jy beam$^{-1}$ and $70$ $\mu$Jy beam$^{-1}$.
  Contours are spaced by a factor of $\sqrt{2}$ from $+3\sigma$.
  No levels at $-3\sigma$ mJy beam$^{-1}$ are present. The 
  green box marks the region covered by panels (a) and (b). The lowest 
  contour of the minihalo at 607 MHz (from Fig.~\ref{fig:mh_610_327}(a)) is 
  reported in cyan for a visual comparison.}
\label{fig:mh_1.28}
\end{figure*}
%
%

\section{Radio images of A\,3444}\label{sec:images}

We present our new GMRT image (contours and colors) at 1300 MHz in Figure \ref{fig:mh_1.28}(a)
The image has been obtained using a robust of 0.5 and
restored with a $3^{\prime\prime}$ circular beam. In Figure \ref{fig:mh_1.28}(b), we 
overlay the same radio contours (black) on the optical Pan-STARRS-1\footnote{Panoramic Survey 
Telescope and Rapid Response System \citep{2016arXiv161205560C}.} $z$-band image. In red, 
we overlay contours from a 1300 MHz image made using pure uniform weighting and 
with a higher angular resolution of $2^{\prime\prime}$. A compact source, labelled S1, is 
associated with the brightest cluster galaxy (BCG). At this sensitivity level and resolution, no clear jets/lobes are detected from S1. 
Its $<2^{\prime\prime}$ angular size implies a physical extent of less than 8 kpc.

%
%
\begin{figure*}
\gridline{\fig{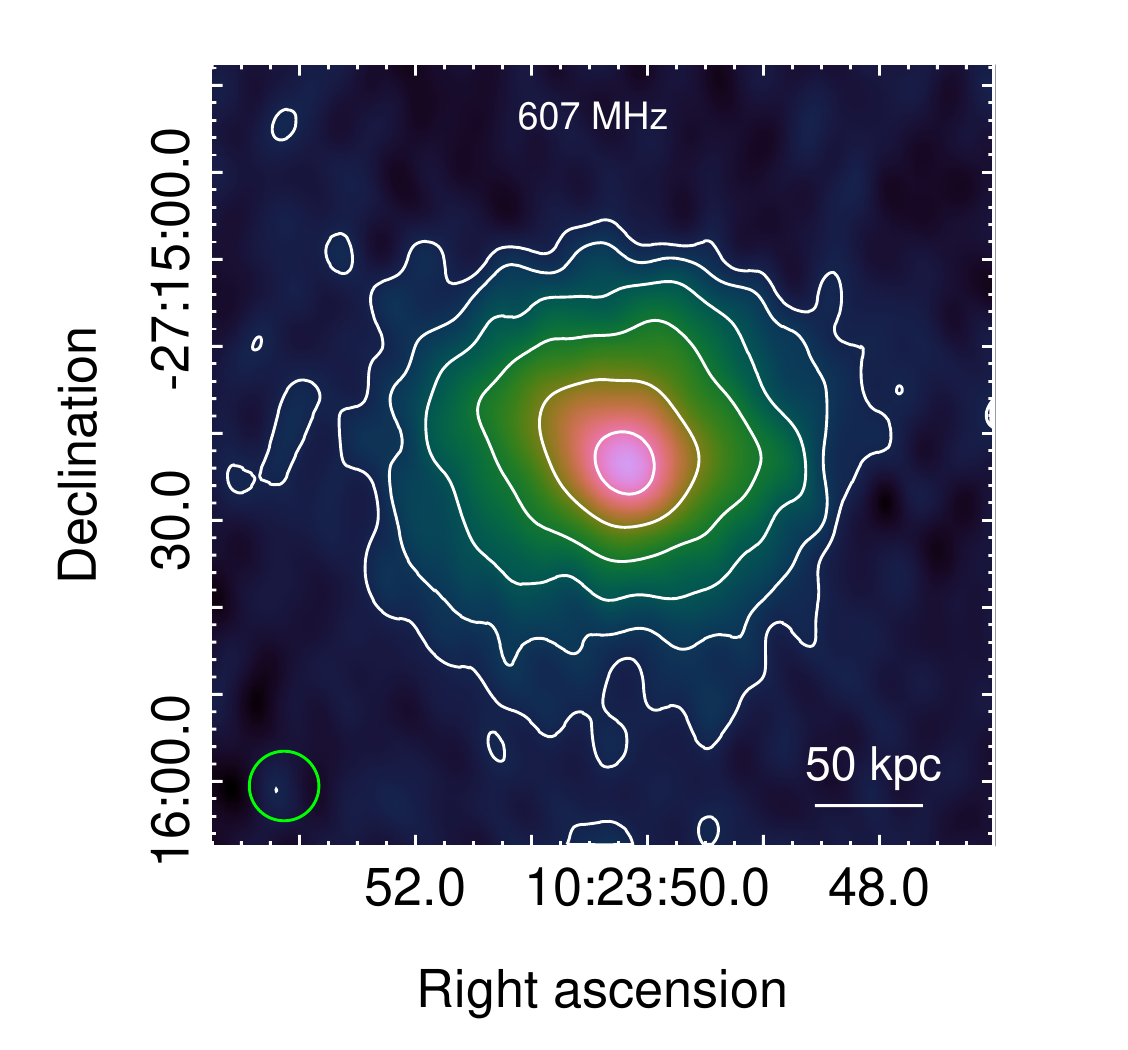}{0.5\textwidth}{(a)}
\hspace{-0.8cm}
\fig{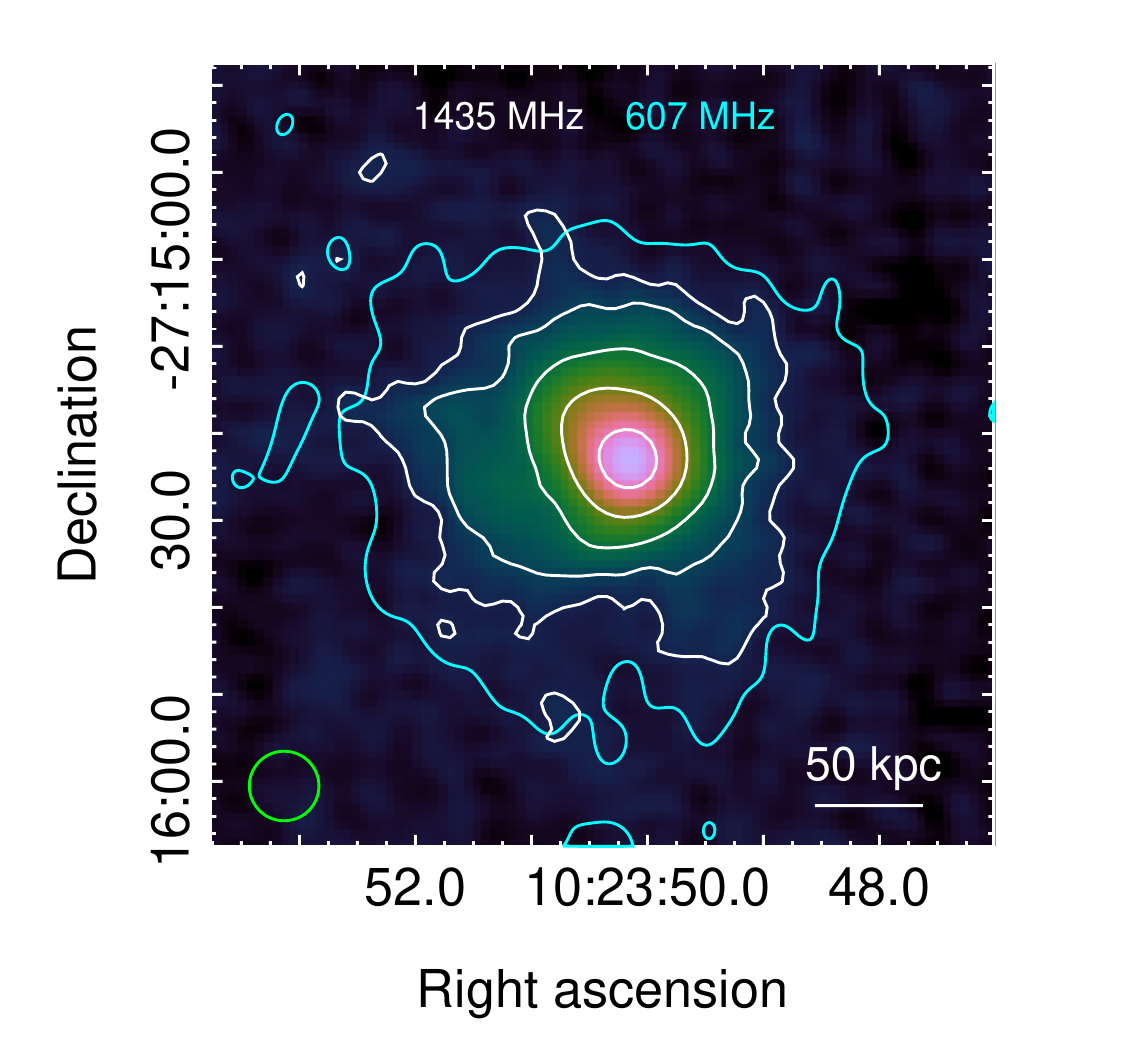}{0.5\textwidth}{(b)}}          
\vspace{-0.5cm}\gridline{\fig{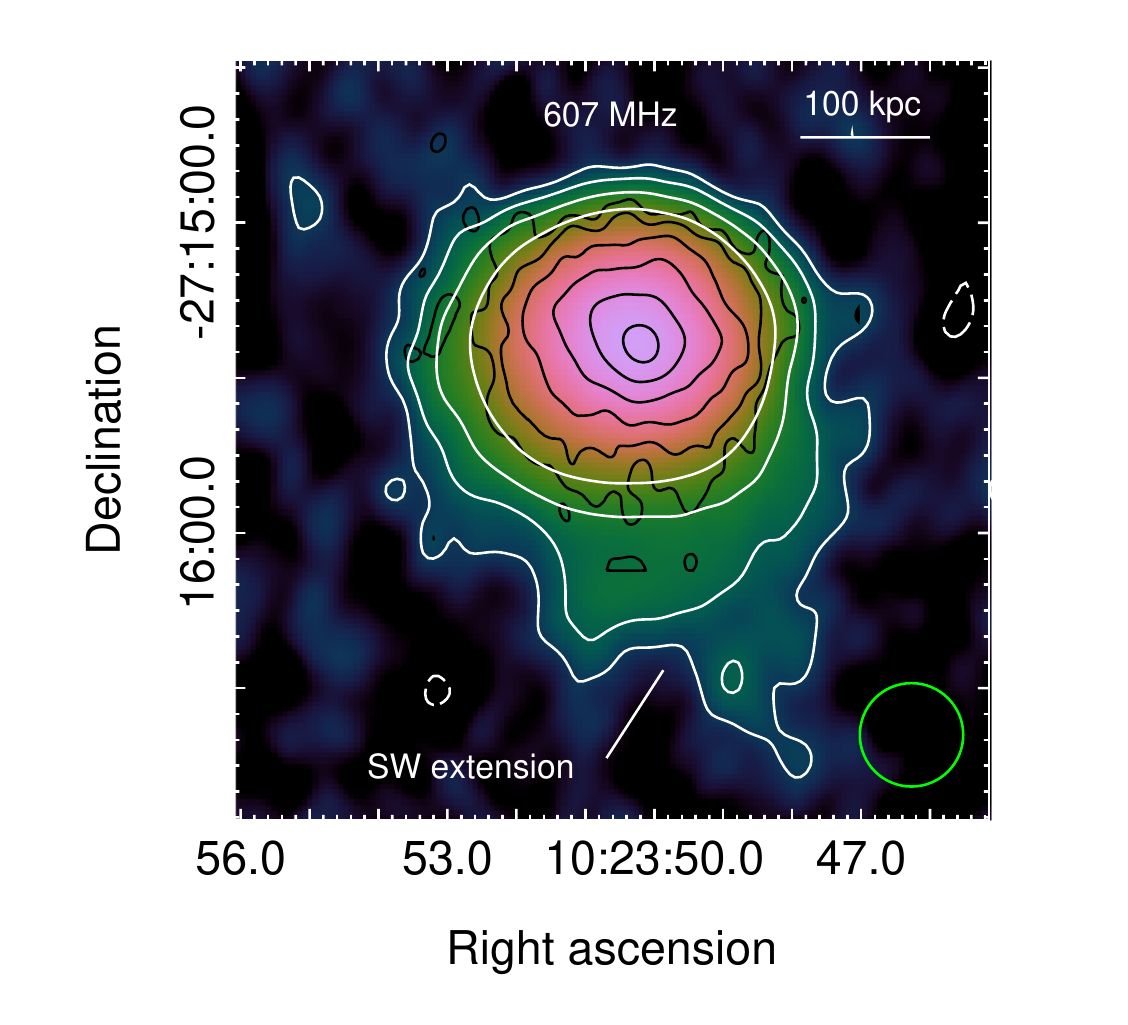}{0.52\textwidth}{(c)}
\hspace{-0.8cm}
\fig{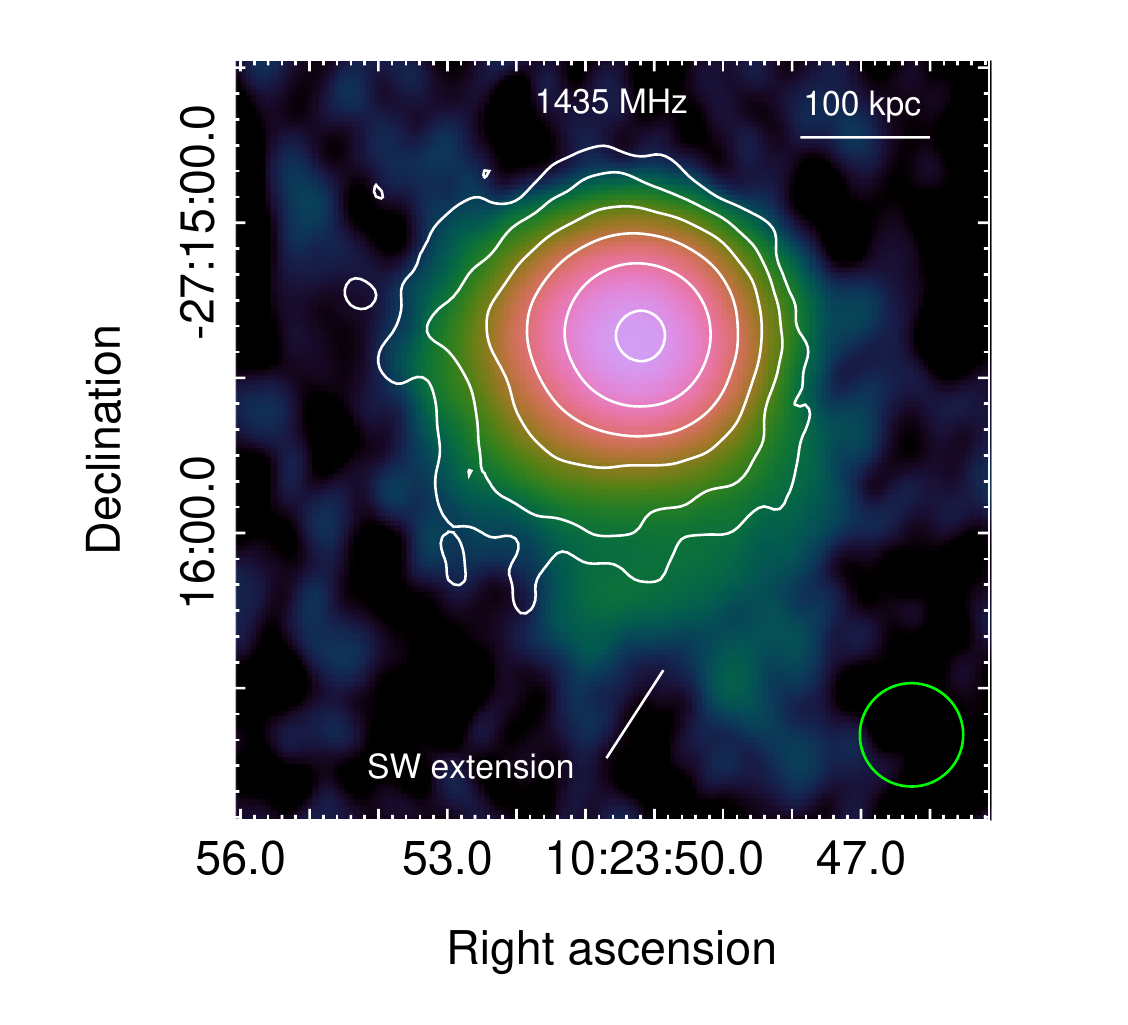}{0.52\textwidth}{(d)}}
\caption{A\,3444. (a) GMRT 607 MHz image (colors and contours) and (b) VLA DnC+BnA image at 1435 MHz (colors and white contours). Both images are restored with a $8^{\prime\prime}$ circular beam (green circle). 
Contours are spaced by a factor of 2 from $+3\sigma$, 
with $1\sigma=37$ $\mu$Jy beam$^{-1}$ and $1\sigma=40$ $\mu$Jy beam$^{-1}$, respectively. No levels at $-3\sigma$ mJy beam$^{-1}$ are present. 
In (b), the lowest contour at 607 MHz is reported in cyan. 
(c) 607 MHz image (colors and white contours) with a $20^{\prime\prime}$ beam (green circle) and $1\sigma=50$ $\mu$Jy beam$^{-1}$. White contours are 0.15, 0.3, 0.6 and 1.2 mJy beam$^{-1}$. Contours at $-0.15$ mJy beam$^{-1}$ are shown as dashed. Black contours are from panel {\em a}. (d) 607 MHz color image, from (c), with VLA 1435 MHz contours at $20^{\prime\prime}$ resolution overlaid in white. Contours are spaced by a factor of 2 from $+3\sigma=150$ $\mu$Jy beam$^{-1}$. No levels at $-3\sigma$ mJy beam$^{-1}$ are present.}
\label{fig:mh_610_327}
\end{figure*}
%
%

%
%
\begin{figure}
\centering
\includegraphics[width=9.2cm]{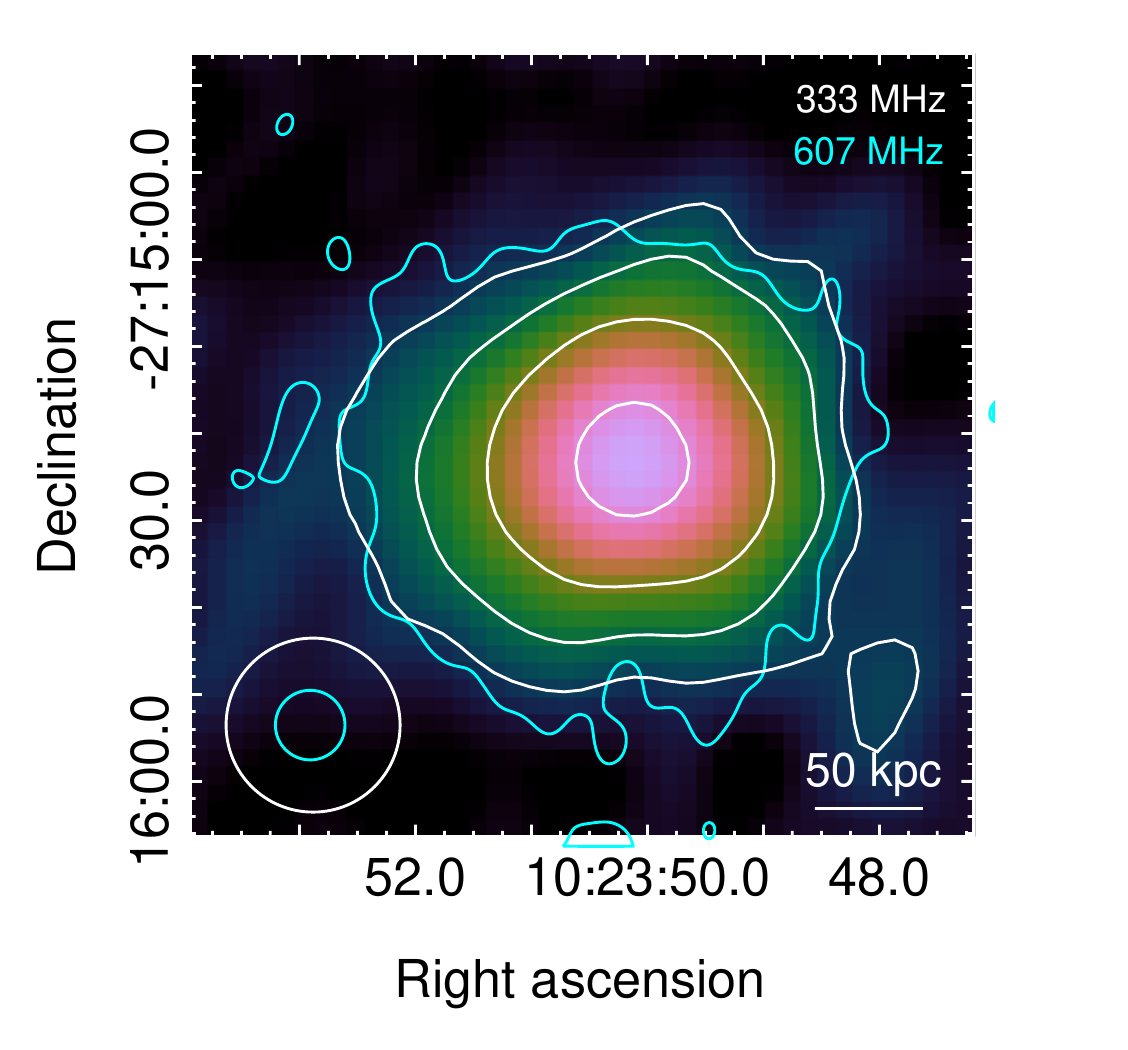}
\smallskip
\caption{A\,3444. GMRT image at 333 MHz (colors and white contours) 
with a $20^{\prime\prime}$ beam (white circle). The noise 
is $1\sigma=800$ $\mu$Jy beam$^{-1}$ and contours are spaced by a factor of 
2 from $2.5$ mJy beam$^{-1}$. No levels at $-3\sigma$ mJy beam$^{-1}$ are present. The lowest contour at 607 MHz (from Fig.~\ref{fig:mh_610_327}(a)) is reported in cyan. The cyan circle shows the $8^{\prime\prime}$ beam at 607 MHz. }
\label{fig:mh_327}
\end{figure}
%
%

%
%
\begin{figure*}
\gridline{\fig{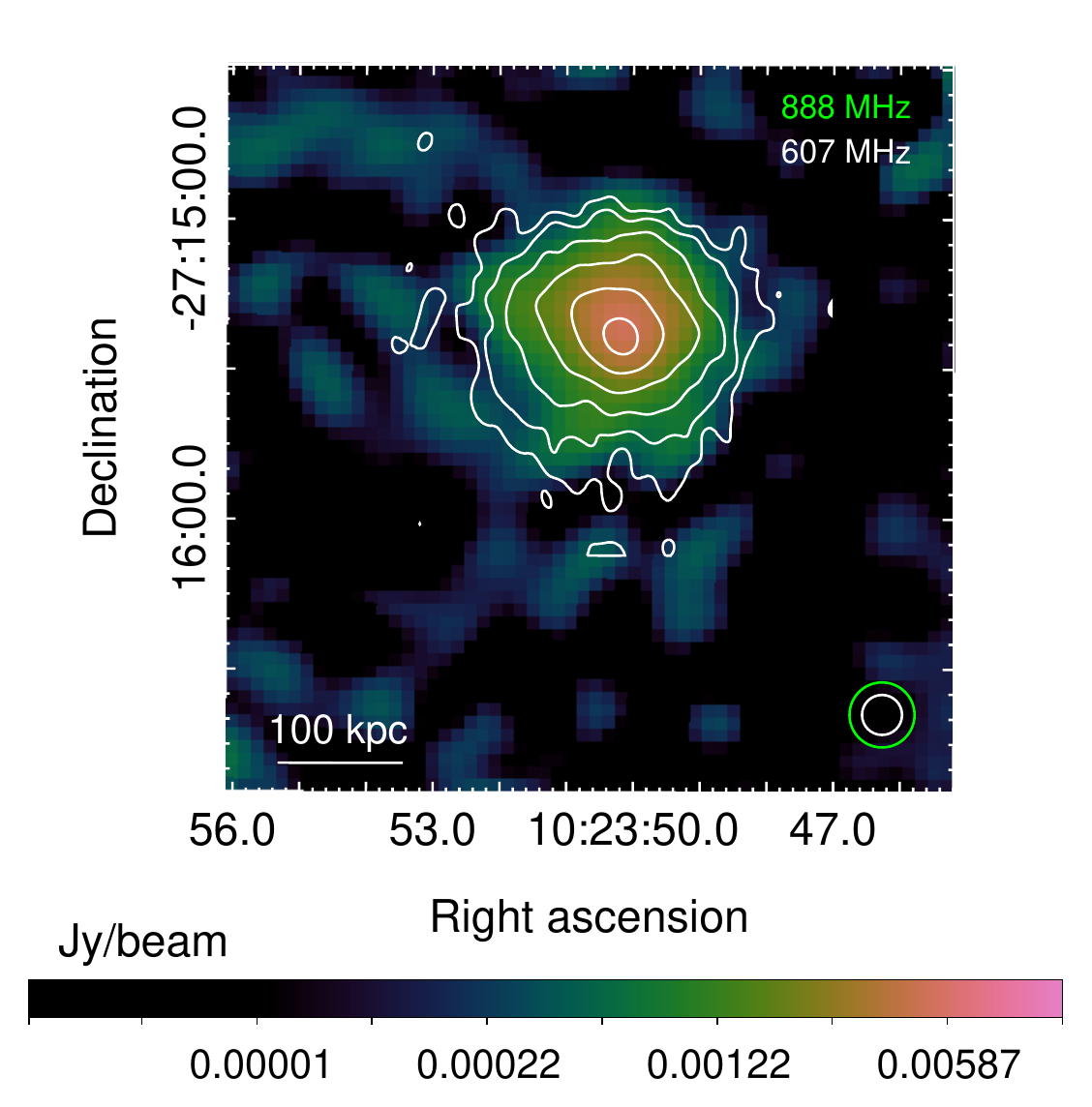}{0.5\textwidth}{(a)}
\hspace{-0.8cm}
\fig{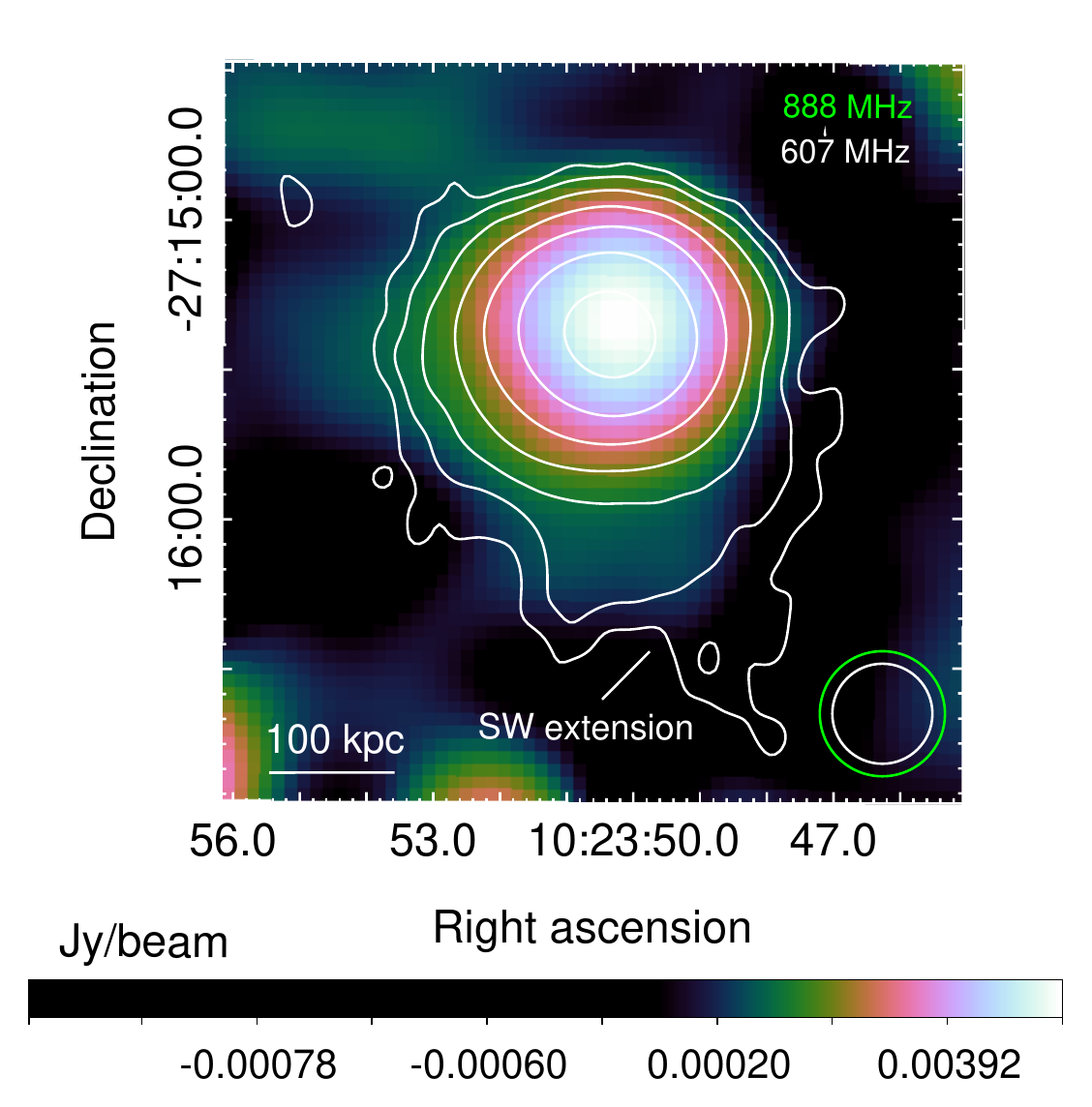}{0.5\textwidth}{(b)}}       
\caption{A\,3444. RACS images at 888 MHz at $13^{\prime\prime}.6\times12^{\prime\prime}.9$ resolution (a) 
and $25^{\prime\prime}$ resolution (b). The rms is $1\sigma=160$ $\mu$Jy beam$^{-1}$ and 200 $\mu$Jy beam$^{-1}$, respectively. The GMRT 607 MHz high-- and low--resolution contours from Fig.~\ref{fig:mh_610_327} are overlaid in white. 
Green and white circles show the RACS and GMRT beams.} 
\label{fig:mh_racs}
\end{figure*}
%
%

The compact source is enshrouded by diffuse emission from the larger-scale minihalo.
In Figures \ref{fig:mh_1.28}(c) and \ref{fig:mh_1.28}(d), we show two 
lower-resolution images at 1300 MHz obtained with a robustness parameter of 2 and a 
beam of $5^{\prime\prime}$ and $15^{\prime\prime}$, respectively. 
The green box marks the region covered by panel (a).
For a visual comparison, in (d) we also overlay 
in cyan the $+3\sigma=0.1$ mJy beam$^{-1}$
isocontour of our new image at 607 MHz, which is 
presented in Figure \ref{fig:mh_610_327}(a), along with a VLA image at 
1435 MHz (b). At 607 MHz, the minihalo reaches a radial distance of 
$r\sim 120$ kpc from the cluster center. The minihalo is also detected on a similar scale in the 
GMRT images at 333 MHz. We show one of these images, at $20^{\prime\prime}$ resolution, in Fig.~\ref{fig:mh_327}. 

We made images with a coarser angular resolution to highlight the
diffuse emission and show a pair of them in 
Figures \ref{fig:mh_610_327}(c) and \ref{fig:mh_610_327}(d) with 
a beam of $20^{\prime\prime}$. These images were obtained using a 
common minimum $uv$ baseline ($0.4$ k$\lambda$) 
and by tapering down baselines longer than 20 k$\lambda$. The minihalo does 
not appear to grow significantly in size in any direction but toward South West. 
Along this direction, the 607 MHz image reveals a "tail" of fainter emission 
(labelled SW extension) that extends well beyond the average minihalo 
radius of 120 kpc, reaching a distance of $\sim 380$ kpc from the center. 
This tail of emission is not detected in the images at 333 MHz and 1435 
MHz because of their lower sensitivity, but it is well detected in the new 
MeerKAT images at 1.28 GHz \citep{2023MNRAS.520.4410T}. 
Due to the higher sensitivity ($1\sigma \sim 9$ $\mu$Jy beam$^{-1}$ 
for a beam of $7^{\prime\prime}$), the MeerKAT images trace the 
emission from this tail out to $r\sim 400$ kpc.

\section{Radio flux density of A\,3444}

We summarize the radio properties of S1 (BCG) and of the surrounding minihalo in Table \ref{tab:flux}.
The flux density of S1 at 1300 MHz and 1435 MHz was measured on our highest-resolution images by fitting the source with a Gaussian model (task JMFIT in AIPS). At 607 MHz, we report the JMFIT peak flux density measured within a $r=4^{\prime\prime}$=16 kpc central region on images obtained after 
cutting the innermost $10$ k$\lambda$ region of the $uv$ plane to 
suppress the surrounding larger-scale emission from the minihalo.
The flux density of S1 at 333 MHz, where it is not possible to separate well the 
point source from the minihalo due to the lower angular resolution, is instead 
extrapolated using the 607-1435 
MHz best-fit spectral index ($\alpha_{\rm fit, \, S1}=0.6\pm0.2$, as calculated 
in Section \ref{sec:spectrum}).

To measure the flux density of the minihalo, we first made images at all frequencies using 
a common minimum projected baseline in the $uv$ plane of $0.4$ k$\lambda$. This spacing 
corresponds to a nominal largest detectable scale of $\sim 10^{\prime}$, which is significantly 
larger than the average area covered by the minihalo emission ($\sim 1^{\prime}$ in diameter). 
We then measured the total flux density on these images using a circular region centered 
on the cluster center and with a radius $r=40^{\prime\prime}$ (160 kpc) that encompasses 
the $+3\sigma$ iso-contour at 607 MHz.
Finally, we subtracted the flux density of S1 from the total emission and obtained a measurement 
of the minihalo total flux density, reported in Table \ref{tab:flux} as {\em MH, total}. 

Even though the BCG emission is unresolved in our highest-resolution images (Fig.~1(b)), we also provide measurements of the minihalo emission after excision of a larger region around S1 (i.e., in the interval $12^{\prime\prime} < r < 40^{\prime\prime}$) that avoids any possible contamination of the BCG into the immediately surrounding structure of the minihalo. These measurements are 
reported in Table \ref{tab:flux} as {\em MH, outer}.
We also provide flux density values for the inner $4^{\prime\prime} < r < 12^{\prime\prime}$ region of the minihalo ($16$ kpc $ < r < 50$ kpc), 
referred to as {\em MH, inner}.

\begin{table*}
\caption{Properties of the BCG and minihalo in A\,3444}
\begin{center}
\begin{tabular}{ccccccccccc}
\hline\noalign{\smallskip}
\hline\noalign{\smallskip}
Source & region & $S_{\rm 333 \, MHz}$ & $S_{\rm 607 \, MHz}$ & $S_{\rm 1300 \, MHz}$ & $S_{\rm 1435 \, MHz}$ &    $\alpha_{\rm fit}$    &   r \\
       &   (kpc)     &(mJy)              & (mJy)             & (mJy)              & (mJy)               &              & (kpc) \\
\noalign{\smallskip}
\hline\noalign{\smallskip}
S1 (BCG)    & $<16$ & $3.5\pm0.9^{a}$   & $2.4\pm0.2$    & $1.47\pm0.07$ & $1.41\pm 0.07$ & $0.6\pm0.2$  & $<8$   \\
{\em MH, total} & $16-160$ & $55\pm6$  & $29\pm3$   & $13.0\pm0.8$  & $12.1\pm0.8$ & $1.0\pm0.1$    &   $120^b$ \\
{\em MH, outer} & $50-160$ & $35\pm4$  & $18\pm2$   & $6.1\pm0.5$   & $5.9\pm0.5$  & $1.3\pm0.1$    &   $120^{b}$ \\
{\em MH, inner} & $16-50$ & $20\pm2$  & $11\pm1$   & $6.9\pm0.6$   & $6.2\pm0.5$  & $0.8\pm0.1$    &   $50$ \\
\hline{\smallskip}
\end{tabular}
\end{center}
\label{tab:flux}
{\bf Notes.} Column 1: source/component. Column 2: radius of the region used to measure the flux density (see also circular regions in Fig.~\ref{fig:spectrum}{\em a}). Columns 3--6: flux densities. Column 6: best-fit spectral index (calculated in Sec.~\ref{sec:spectrum}). Column 7: linear size (radius) measured on Figs. \ref{fig:mh_1.28}{\em b} (S1) and \ref{fig:mh_610_327}{\em a} (minihalo). 
\\
$^a$ Extrapolated from 607 MHz using $\alpha_{\rm fit}$. 
$^b$ The diffuse emission extends to $r\sim 380$ kpc at 607 MHz (SW extension in Fig.~\ref{fig:mh_610_327}{\em c}).
\end{table*}

\section{Flux density of A\,3444 from radio surveys}\label{sec:low}

We inspected images containing A\,3444 from available radio surveys, including the VLA Sky Survey \citep[VLASS,][]{2020PASP..132c5001L} at 3 GHz, the VLA Low-frequency Sky Survey Redux \citep[VLSSr,][]{2014MNRAS.440..327L} at 74 MHz, the TIFR GMRT Sky Survey Alternative Data Release \citep[TGSS-ADR,][]{2017A&A...598A..78I} at 150 MHz, the Galactic and Extragalactic All-Sky MWA survey \citep[GLEAM,][]{2017MNRAS.464.1146H} at 200 MHz, and the Rapid ASKAP Continuum Survey \cite[RACS,][]{2020PASA...37...48M, 2021PASA...38...58H} at 888 MHz.

\subsection{VLA Sky Survey at 3 GHz}

 In the VLASS Epoch 2.2 quick-look image, we found a faint, compact source at the position of the BCG. A Gaussian fit to the source gives a peak flux density of $1.05\pm0.21$ mJy beam$^{-1}$ (this value was corrected for a low-flux density bias of $10\%$ that affects faint sources in quick-look images from VLASS 1.2 onward, as reported in the NRAO Quick-Look Image web page\footnote{https://science.nrao.edu/vlass/data-access/vlass-epoch-1-quick-look-users-guide}; a total uncertainty of $20\%$ was assumed).
The extended emission from the minihalo is not detected in the VLASS image due to the high angular resolution ($\sim 2^{\prime\prime}$), noise level ($1\sigma=130$ $\mu$Jy beam$^{-1}$) and sparse $uv$ coverage at short baselines of the VLA in its B configuration.

\subsection{Low-frequency radio surveys}

A radio source is detected at the cluster center in the 74 MHz VLSSr image at $3\sigma$ level 
($1\sigma=80$ mJy beam$^{-1}$) and in the GLEAM image at 200 MHz at $16\sigma$ ($1\sigma=6.3$ mJy beam$^{-1}$). In both images (not shown here), the radio structure is unresolved at a resolution of $75^{\prime\prime}$ and $136^{\prime\prime}\times128^{\prime\prime}$, respectively. A $16\sigma$, marginally extended source is 
visible in the TGSS-ADR image at 150 MHz (not shown here), that has a smaller beam of $36^{\prime\prime}\times25^{\prime\prime}$ and a local noise of $1\sigma=3.5$ mJy beam$^{-1}$. 

We also obtained an image from the first RACS data release (DR1) 
in the lowest band (888 MHz; RACS-low), convolved to a resolution of $25^{\prime\prime}$, and a higher-resolution image ($13^{\prime\prime}$) 
of the A\,3444 field \citep{2020PASA...37...48M} from the CSIRO ASKAP Science Data Archive (CASDA). 
Both are shown in colors in Fig.~\ref{fig:mh_racs}. The morphology of the minihalo agrees well with the structure detected at 607 MHz (contours).
The brightest part of the SW extension is marginally detected in the RACS-low 
image (at $\sim 2\sigma$).

%
%
\begin{figure*}
\gridline{\fig{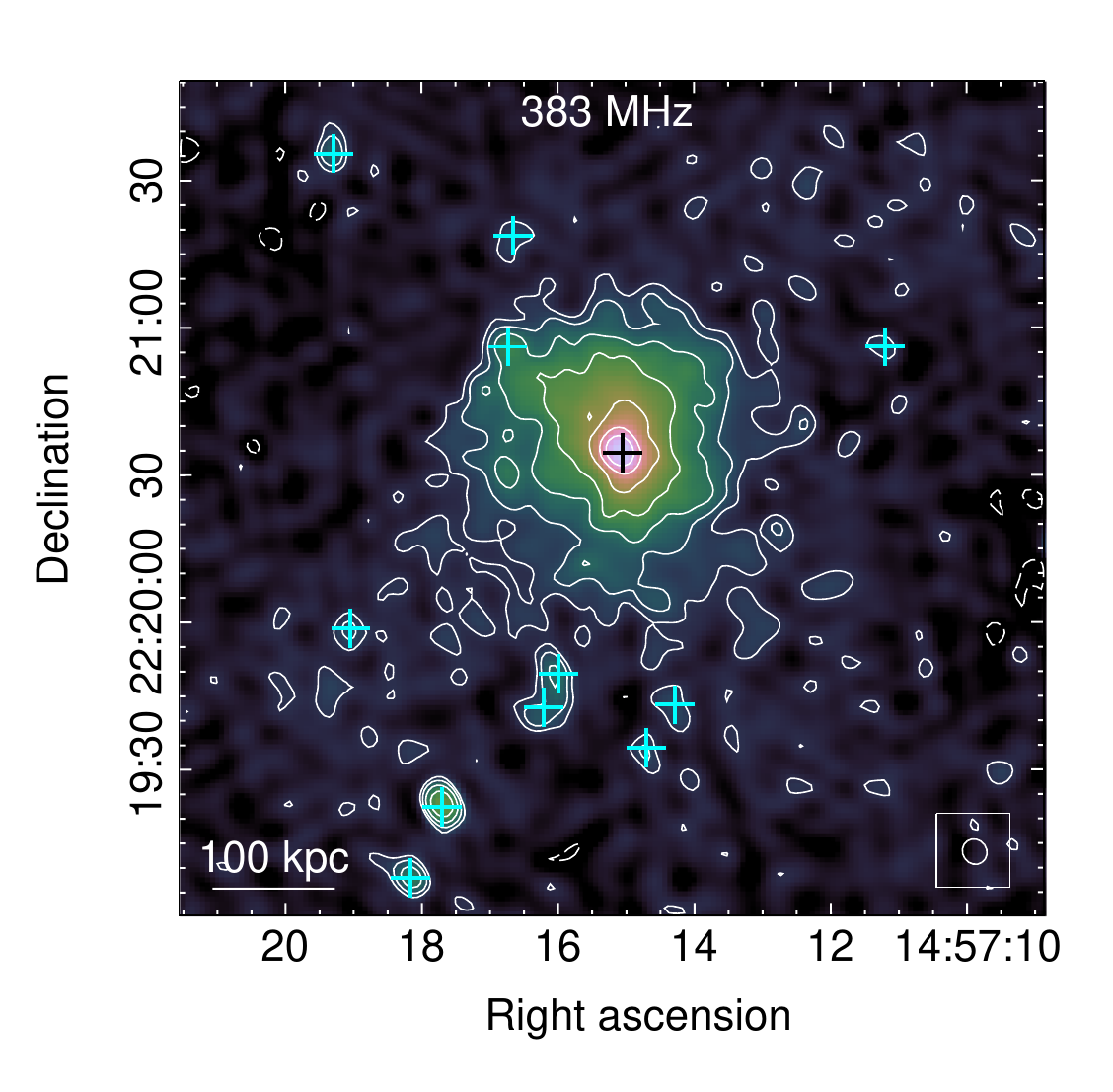}{0.5\textwidth}{(a)}
\hspace{-0.8cm}
          \fig{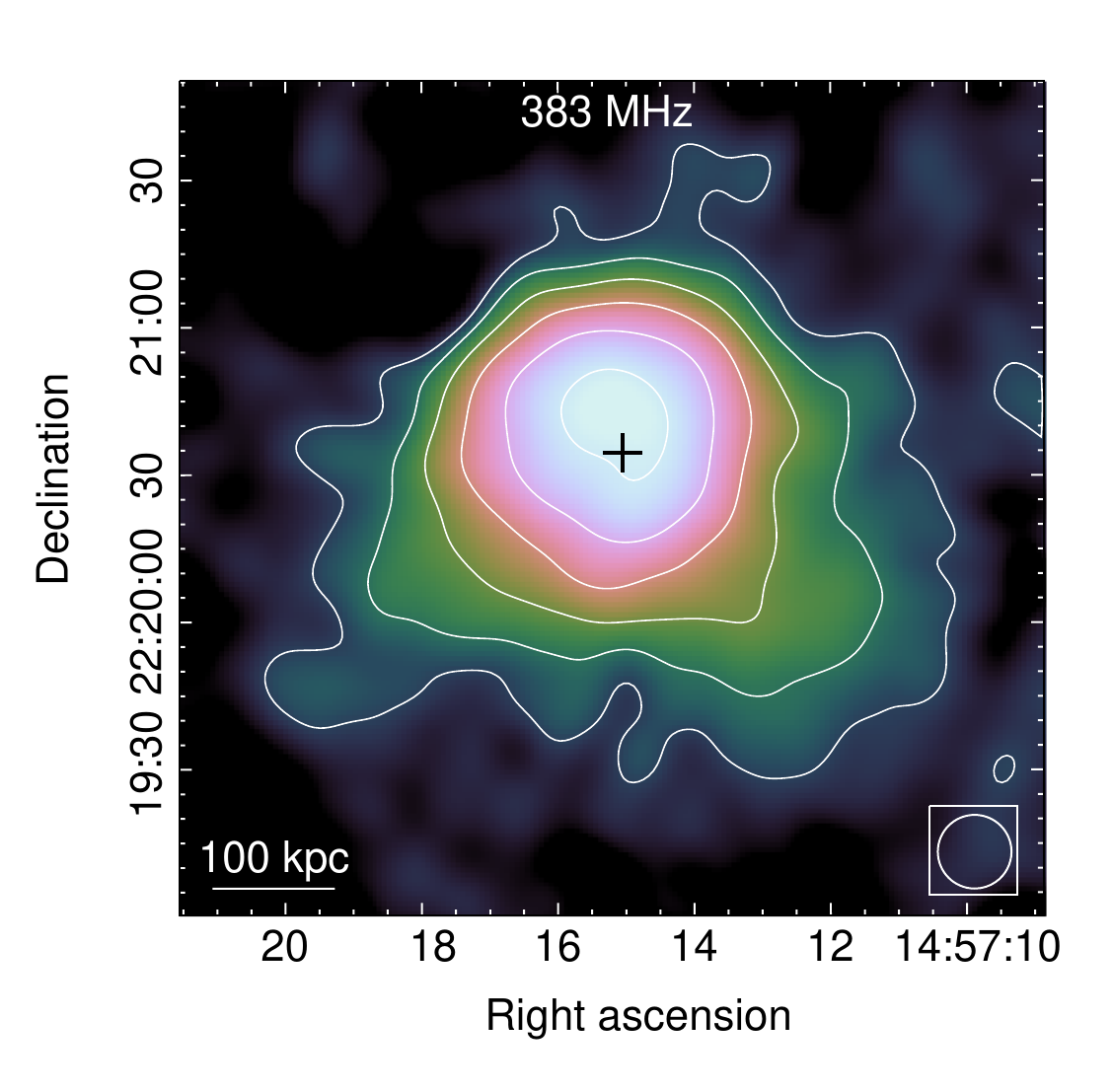}{0.5\textwidth}{(b)}}
\caption{MS\,1455.0+2232. (a) uGMRT 338 MHz high-resolution image 
with uniform weights. The beam is $4.5^{\prime\prime}\times5.5^{\prime\prime}$ (ellipse in the corner) and $1\sigma=25$ $\mu$Jy beam$^{-1}$. 
Contours are spaced by a factor of 2 from $0.07$ mJy beam$^{-1}$. 
Negative levels at $-0.07$ mJy  beam$^{-1}$ 
are shown as dashed contours. The black cross marks the compact source 
at the BCG. Cyan cross show the location of other discrete radio sources. 
(b) uGMRT 338 MHz low-resolution MHz image of the diffuse emission, 
after subtraction of discrete sources. The image was made using robust $+0.2$, a Gaussian taper of $10^{\prime\prime}$, and restored with a beam of $15^{\prime\prime}$ (also shown in the corner). Contours are spaced by a factor of 2 from $+3\sigma=0.2$ mJy beam$^{-1}$. No negative contours at $-0.2$ mJy beam$^{-1}$ 
are present. For reference, the position of the BCG (which was subtracted out) 
is shown by the black cross.}
\label{fig:mh_ms1455}
\end{figure*}
%

\begin{table}
\caption{Flux density of A\,3444 from surveys}
\begin{center}
\begin{tabular}{cccccccccc}
\hline\noalign{\smallskip}
\hline\noalign{\smallskip}
Survey & $\nu$ & $S_{\nu}$ \\
       & (MHz) & (mJy)   \\
\noalign{\smallskip}
\hline\noalign{\smallskip}
VLSSr & 74  & $261\pm40^{a}$  \\
TGSS-ADR   & 150 & $97\pm10^{a}$ \\
GLEAM      & 200 & $132\pm10^{a}$ \\
RACS-low  & 888 & $17\pm2$ \\ 
\hline{\smallskip}
\end{tabular}
\end{center}
\label{tab:flux1}
      {\bf Notes.} Flux density of the A\,3444 minihalo from low-frequency radio surveys measured
      within $r=40^{\prime\prime}$ kpc.
      The 200 MHz value is from the GLEAM catalog. All values have been re-scaled to the \cite{2017ApJS..230....7P} flux density scale adopted in this paper.
      
      $^a$ Includes S1. 
\end{table}

In Table ~\ref{tab:flux1}, we summarize the total flux density measured on the survey images within the same $r=40^{\prime\prime}$ kpc central region used for the minihalo total emission in Table \ref{tab:flux}.
The 200 MHz value is from the GLEAM catalog. All flux densities have been re-scaled to the \cite{2017ApJS..230....7P} scale adopted in this paper, using the appropriate factors listed in \cite{2017ApJS..230....7P}\footnote{Differences between the \cite{2017ApJS..230....7P} scale and the original scales 
are less than 2$\%$.}.
The 74, 150 and 200 MHz values include both the minihalo and central radio galaxy S1. However, based on its spectral index ($\alpha_{\rm fit}=0.6$, Sec.~\ref{sec:spectrum}) the contribution of S1 at these frequencies is expected to be only a few $\%$. At 888 MHz, the total (MH+S1) flux density is $18.9\pm1.9$ mJy, of which $1.9\pm0.2$ mJy are expected to be from S1, leaving $17\pm2$ mJy in minihalo emission.

\section{MS\,1455.0+2232: radio images and flux density at 383 MHz}\label{sec:images}

We present uGMRT images of MS\,1455.0+2232 
at the central frequency of 383 MHz 
in Fig.~\ref{fig:mh_ms1455}. In (a) we show
a high-resolution image, made with uniform weights, and mark the position 
of the compact source at the BCG (black cross) and other discrete radio
sources in the region identified above $3\sigma=0.075$ mJy beam$^{-1}$ (cyan crosses). 
In (b) we show a lower-resolution image of the
diffuse emission after subtraction of the clean components associated 
with discrete sources from the $uv$ data. The image was obtained 
adopting a robust of $+0.2$, a $10^{\prime\prime}$ 
Gaussian $uv$ taper, and restored with a beam of $15^{\prime\prime}$. The diffuse emission spans
a total area of $\sim 145^{\prime\prime}\times 110^{\prime\prime}$ ($\sim 580$ kpc $\times$ 400 kpc), which is significantly larger than measured in previous 
GMRT and VLA images \citep[][]{2008A&A...484..327V,2019ApJ...880...70G}, and 
in good agreement with the new MeerKAT and LOFAR images of \cite{2022MNRAS.512.4210R}. 

Using Fig.~\ref{fig:mh_ms1455}(a), we measure a 383 MHz flux density of $9.7\pm1.5$ mJy for the BCG, which agrees well with the integrated radio 
spectrum of the source presented by \cite{2022MNRAS.512.4210R}.
From Fig.~\ref{fig:mh_ms1455}(b), we measure a total flux density in
diffuse emission of $60\pm9$ mJy within the $3\sigma=0.2$ mJy beam$^{-1}$ contour.
On a similar scale, \cite{2022MNRAS.512.4210R} measured a total of 
$18.7\pm0.9$ at 1283 MHz and $154.5\pm15.6$ mJy at 145 MHz, and computed a spectral index of $\alpha=0.97\pm0.05$.
Our measurement at 383 MHz is consistent with this 
spectral index: we find $0.97\pm0.19$ in the 145-383 MHz interval, and $0.96\pm0.13$ between
383 MHz and 1283 MHz.

%
%
\begin{figure*}
\gridline{\fig{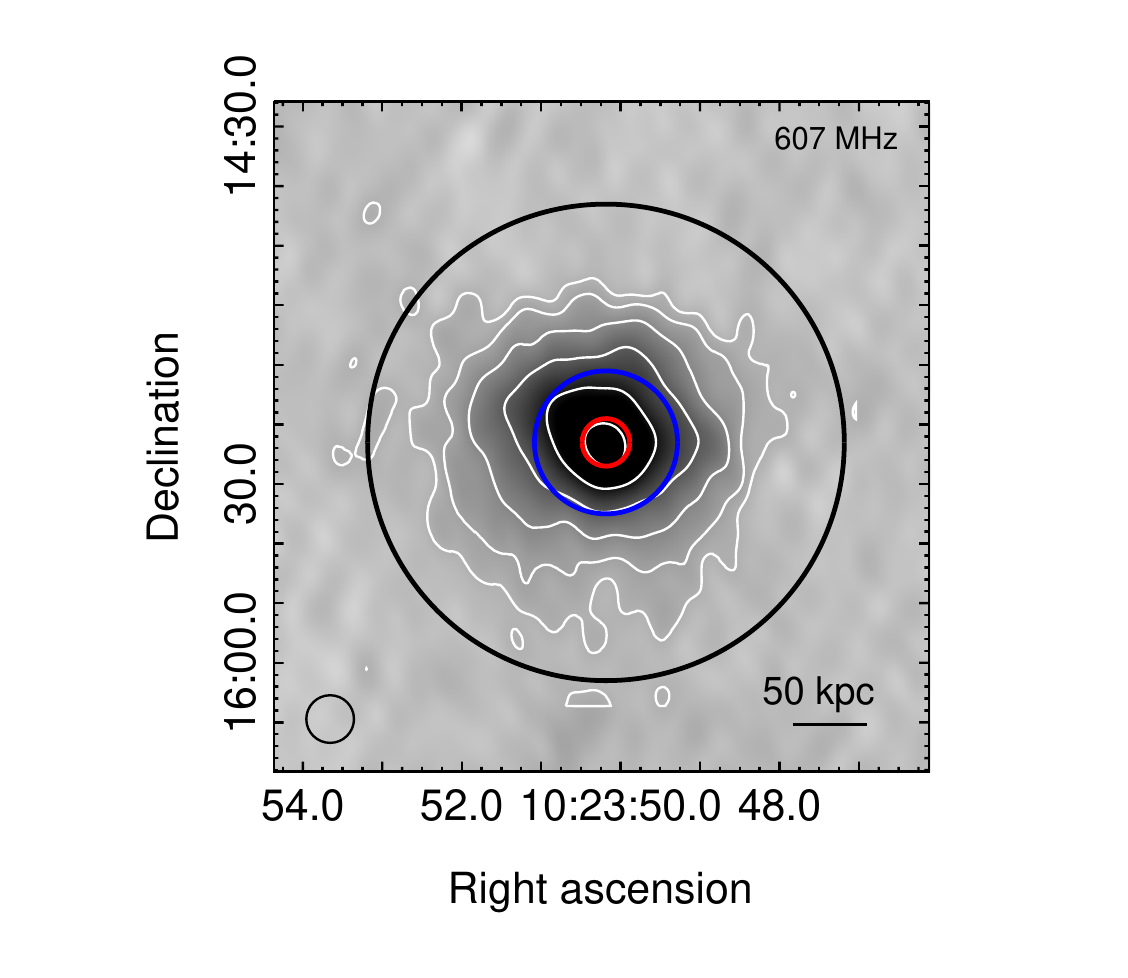}{0.52\textwidth}{(a)}
\hspace{-0.8cm}
          \fig{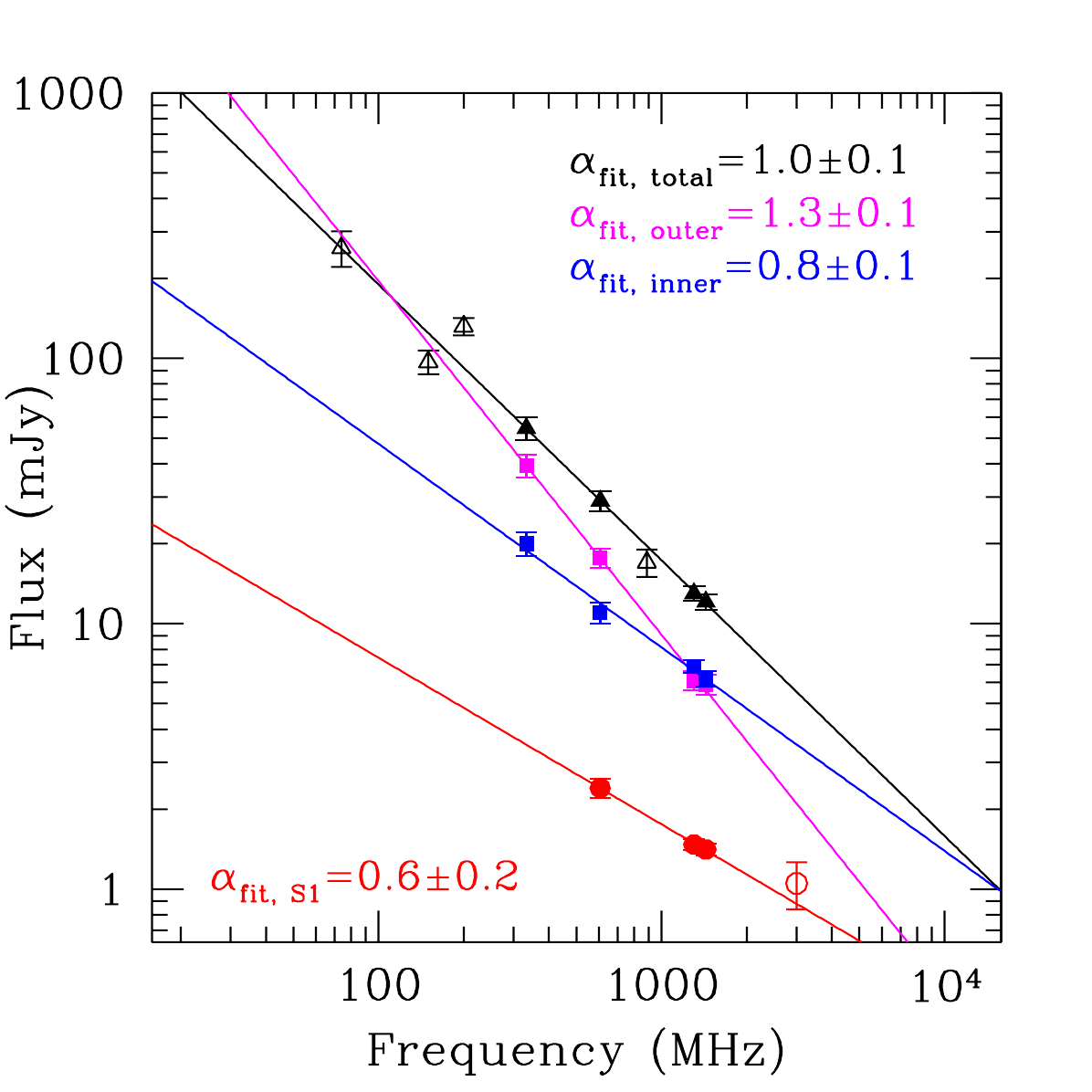}{0.45\textwidth}{(b)}}
\caption{A\,3444. (a) Regions used to measure the flux densities in Tab.~\ref{tab:flux} of S1 (red circle), 
{\em MH, total} (between the black and red circles), {\em MH, inner} (between the red and blue circles) and
{\em MH, outer} (between the blue and black circles). The red circle has $r=4^{\prime\prime} = 16$ kpc, the blue circle
has $r=12^{\prime\prime} = 50$ kpc, and the black circle has $r=40^{\prime\prime}$=160 kpc. The image is from Fig.~\ref{fig:mh_610_327}(a). 
 (b) Radio spectra of S1 (red), {\em MH, total} (black filled triangles), {\em MH, outer} (magenta) and {\em MH, inner }(blue), computed using the regions in (a) and the flux densities in Tab.~\ref{tab:flux}. The empty red circle is the VLASS 2.2 flux density of S1. White empty triangles are from the radio surveys (Tab.~\ref{tab:flux1}); all, but RACS, include S1 (its contribution is estimated to be a few $\%$). The solid lines are power-law fits (with slopes) to the data using only the filled data points.}
\label{fig:spectrum}
\end{figure*}
%
%

%
%
\begin{figure*}
\gridline{\fig{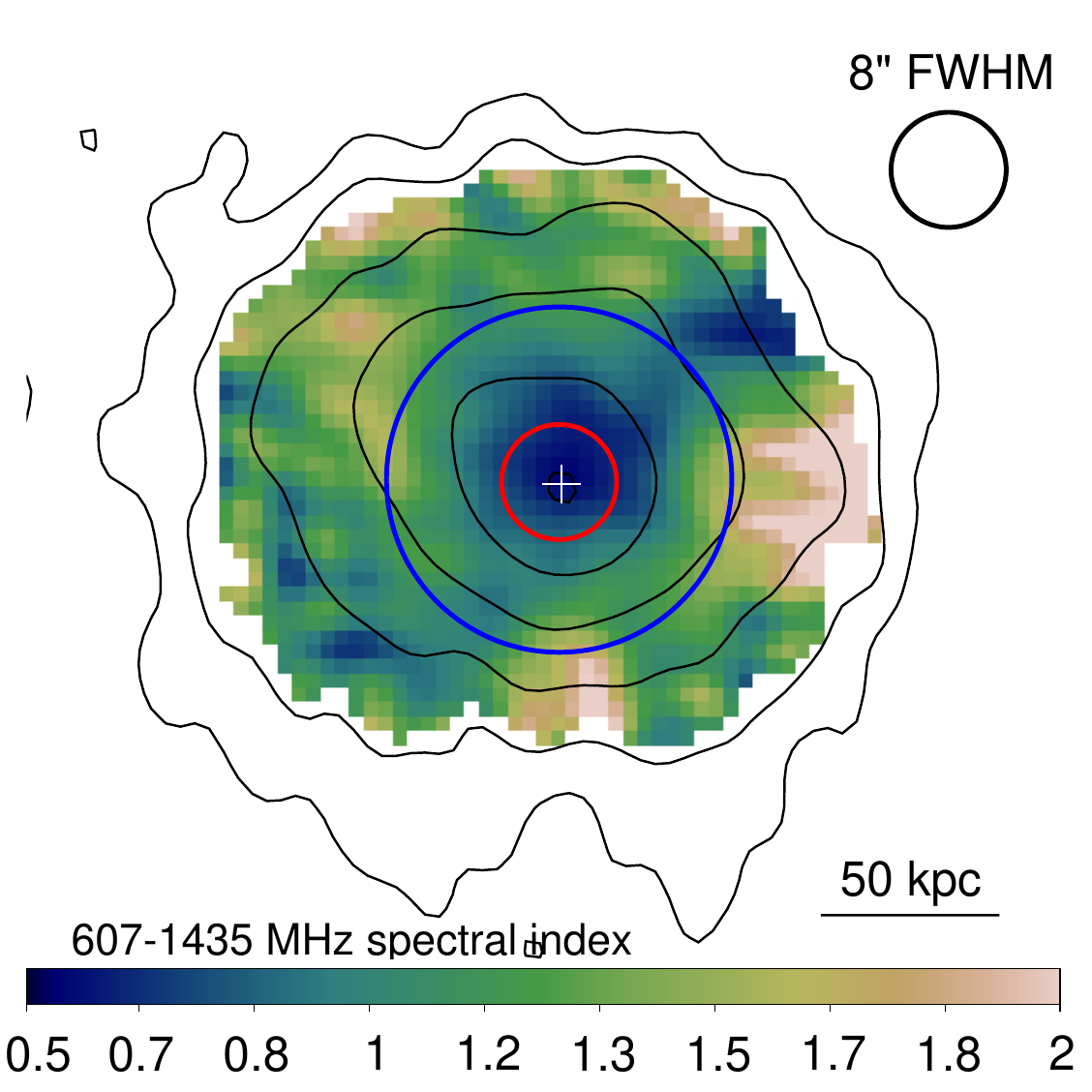}{0.4\textwidth}{(a)}
\hspace{-0.8cm}
          \fig{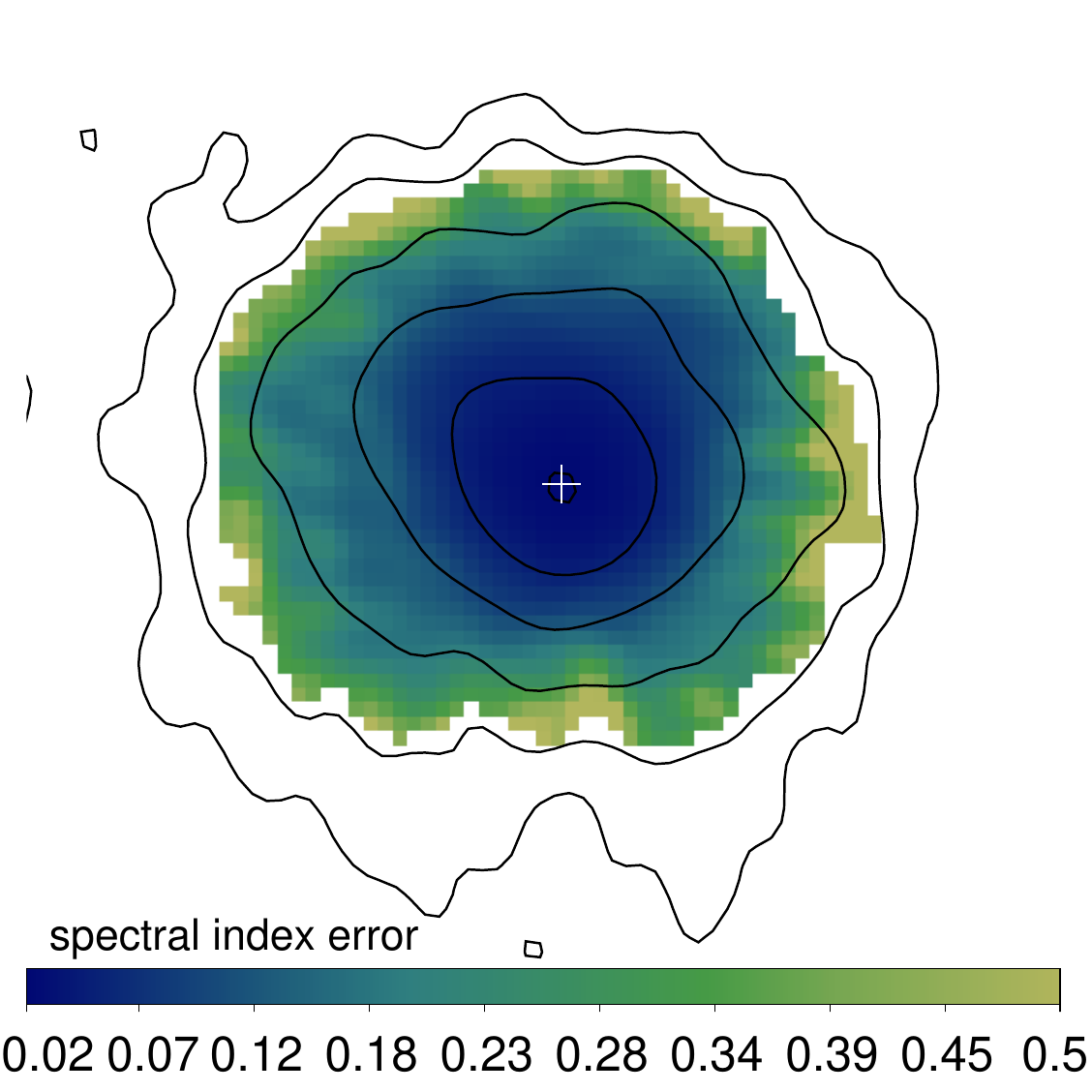}{0.4\textwidth}{(b)}}
\caption{A\,3444. Color-scale image of the spectral index distribution 
between 607 MHz and 1435 MHz (a) and associated error map (b),
computed from primary beam corrected images with noise of $1\sigma=40$ $\mu$Jy beam$^{-1}$, 
same $uv$ range, pixel size and circular beam 
of $8^{\prime\prime}$ (black circle). The spectral index 
was calculated in each pixel where the surface brightness is above the 
$3\sigma$ level in both images. Overlaid are the 607 MHz contours, 
spaced by a factor of 2 from $3\sigma=0.12$ mJy beam$^{-1}$. 
The white cross marks the cluster center. The red and blue circles have
$r=4^{\prime\prime}$ (S1) and $r=12^{\prime\prime}$ ({\em MH, inner};  
see also Fig.~\ref{fig:spectrum}(a)).}
\label{fig:spix}
\end{figure*}
%
%

%
%
\begin{figure*}
\gridline{\fig{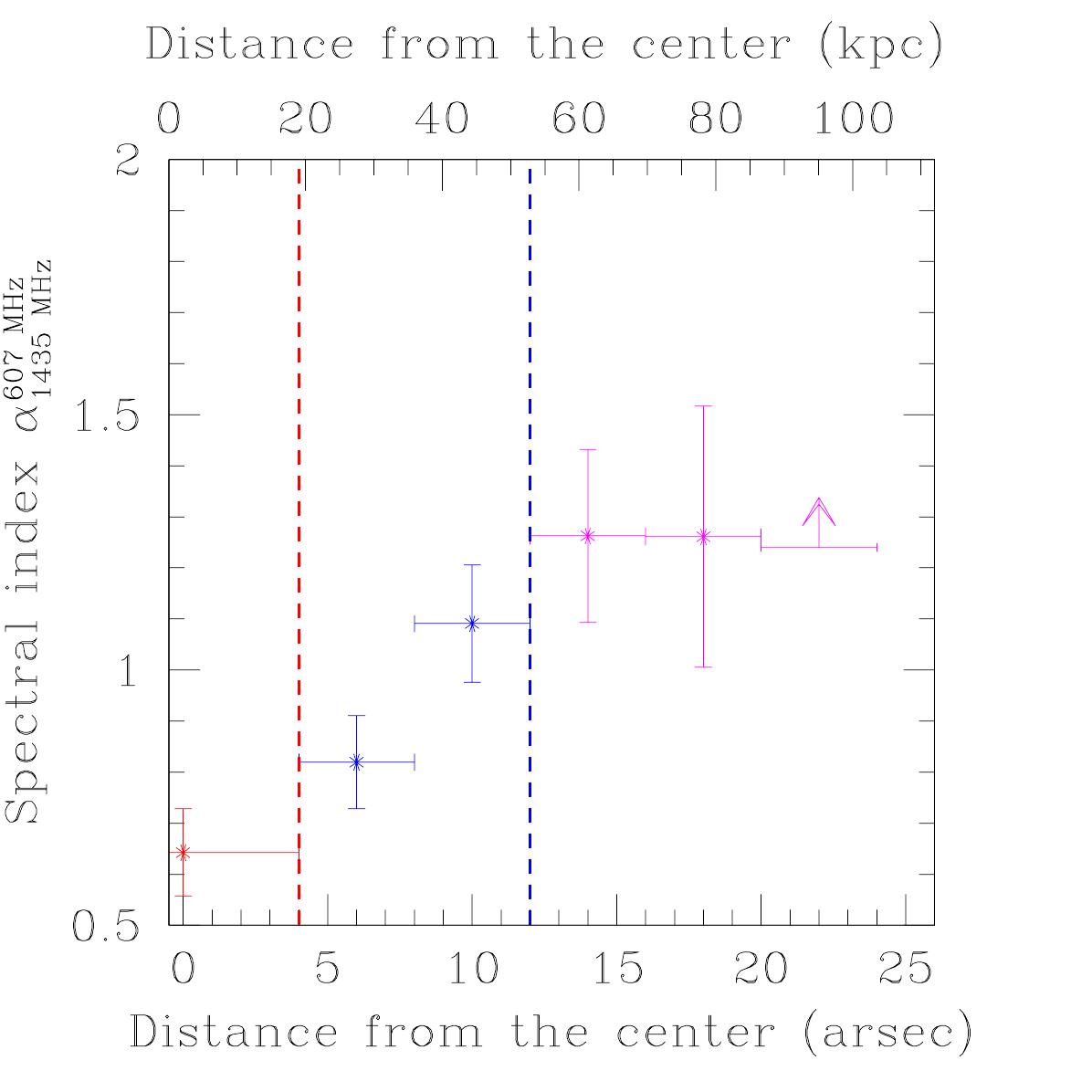}{0.41\textwidth}{(a)}
\hspace{-0.8cm}
          \fig{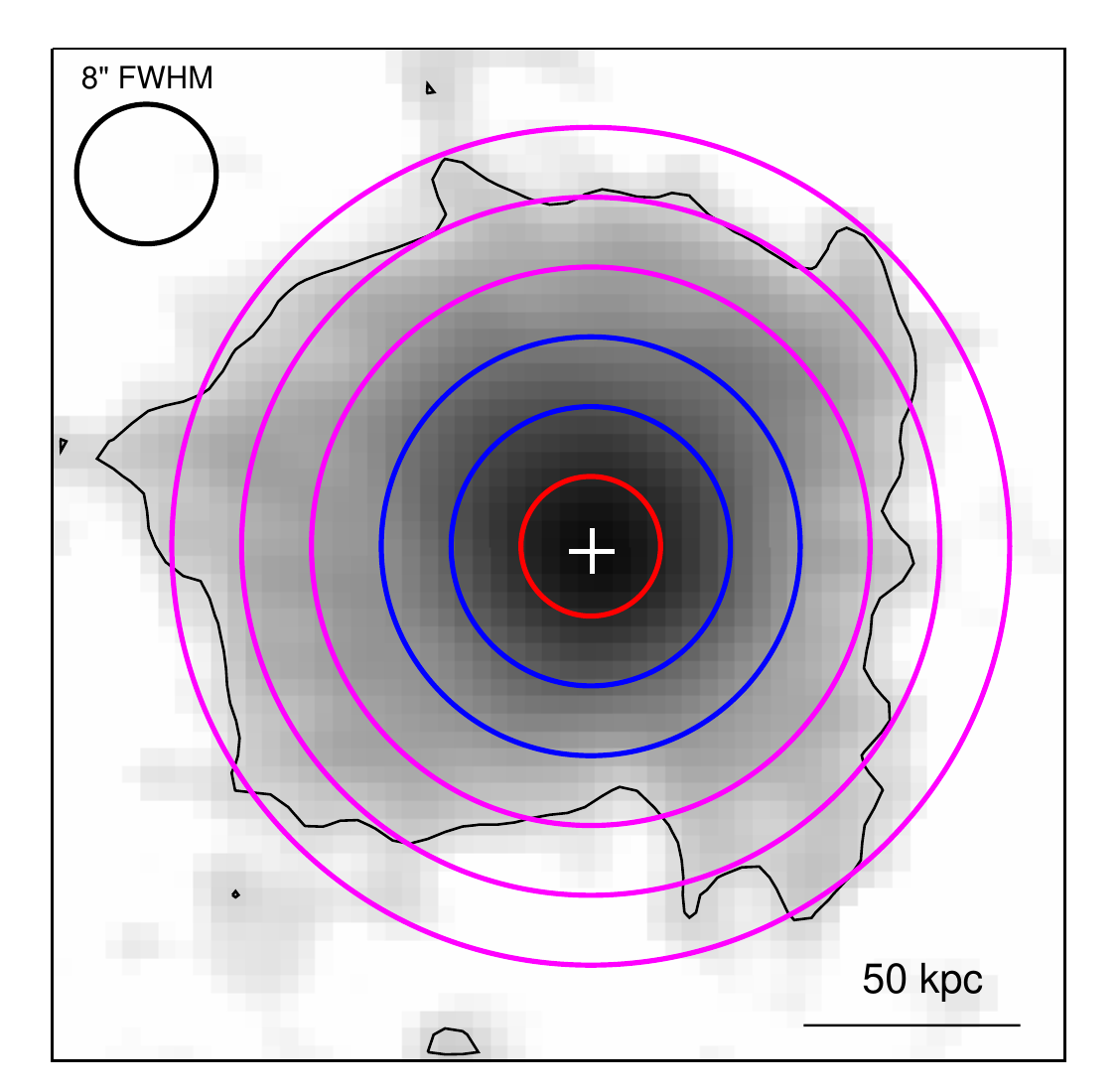}{0.4\textwidth}{(b)}}
\caption{A\,3444. (a) Spectral index between 607 MHz and 1435 MHz as a function of the distance
  from the cluster center, derived using the regions shown (b), centered on the cluster center 
  (white cross) and with a radial step of 4$^{\prime\prime}$. The red and blue dashed 
  lines mark $r=4^{\prime\prime}$ (S1) and $r=12^{\prime\prime}$ ({\em MH, inner}; see also Fig.~\ref{fig:spectrum}). The radio image is at 1435 MHz (Fig.~2(b)) and the   black contour is at $+3\sigma=40$ $\mu$Jy beam$^{-1}$.}
\label{fig:spix_prof}
\end{figure*}
%
%

\section{Spectral analysis of A\,3444}

\subsection{Integrated radio spectra}\label{sec:spectrum}

In Figure \ref{fig:spectrum}(b), we show the radio spectra of the 
BCG (S1, red) and diffuse emission (black, magenta and blue) based on 
the flux densities in Tables \ref{tab:flux} and \ref{tab:flux1} and 
the regions in Fig.~\ref{fig:spectrum}(a).

The spectrum of S1 has a power-law best-fit spectral index of $\alpha_{\rm fit,\, S1}= 0.6\pm0.2$. The empty red circle is the VLASS 2.2 data point 
at 3 GHz. Due to the preliminary nature of the VLASS 2.2 quick-look images, 
we did not include this point in the spectral fitting. It is, however, in agreement with the best fit within the errors. 

The total spectrum of the minihalo (black filled triangles) is described by a power law in the 333--1435 MHz range, with a steep slope of $\alpha_{\rm fit, \, total}=1.0\pm0.1$. Its outer region ($>50$ kpc; magenta squares) has a steeper spectrum with $\alpha_{\rm fit, \, outer}=1.3\pm0.1$, whereas its inner region in blue (i.e., between the red and blue circles in Fig.~\ref{fig:spectrum}(a)) has $\alpha_{\rm fit, \, inner}=0.8\pm0.1$.

The best fits in Figure \ref{fig:spectrum}(b) were computed using only the pointed observations presented in this paper (filled points). Empty triangles are the data points from the radio surveys (Table \ref{tab:flux1}). Despite the different angular resolution, sensitivity and uncertainties on the BCG contribution, the survey points show marginal scatter and appear in general agreement with the overall spectrum of the minihalo.

We note that \cite{2023MNRAS.520.4410T} measured a MeerKAT in-band spectral index of $\alpha=1.5\pm0.4$ for the minihalo, which is steeper than the spectral index we measure here ($\alpha_{\rm fit, \, total}=1.0\pm0.1$), even though it is still consistent within the $1\sigma$ errors. For the spectral analysis, \cite{2023MNRAS.520.4410T} used a region enclosed by the $5\sigma$ contour of each MeerKAT sub-image. This region is larger (by $\sim 10^{\prime\prime}$) than our extraction 
region (black circle in Fig.~6) and includes the brightest part of 
the SW extension, which may contain emission with a steeper 
spectrum \citep[see also Sect.~\ref{sec:sw_sp}.]{2023MNRAS.520.4410T}.
Therefore, it is unclear whether the steeper spectral index measured by MeerKAT reflects an actual steepening of the integrated radio spectrum of the minihalo at higher frequencies (the highest-frequency end of the MeerKAT band is 1.7 GHz) and/or whether the steeper measurement is affected by the contribution of the SW extension.

\subsection{Radio spectral index distribution}\label{sec:sp_index_map}

In Figure~\ref{fig:spix}, we show a color image of the spectral index 
distribution between 607 MHz and 1435 MHz (a) and associated
error map (b), obtained by comparing a pair of primary-beam corrected 
images made with the same interval of projected baselines in the $uv$ plane 
($0.4-50$ k$\lambda$) and a circular restoring beam of $8^{\prime\prime}$. 
The two images have a similar noise of $\sim 40 \, \mu$Jy beam$^{-1}$. 
Contours at 607 MHz are overlaid on both panels to provide a reference for 
the source morphology. The white cross marks the cluster center and the red 
and blue circles are the same as in Fig.~\ref{fig:spectrum}(a). 

The flattest spot in the center ($\alpha \sim 0.6$) coincides with the
region occupied by S1, associated with the BCG, and is consistent with
the slope of its integrated spectrum in Fig.~\ref{fig:spectrum}(b). 
The surrounding diffuse emission has a spectral index that ranges from 
$\alpha \sim 0.8-1.2$ within the central 50 kpc (blue circle) to 
$\sim 1.2-1.5$ at larger radii, and becomes even steeper in some 
of the outermost regions, though with large uncertainties ($>0.4$). 
This behavior suggests a steepening of the minihalo spectral 
index with increasing distance from the center.

To investigate this possible steepening, we obtained a spectral index 
radial profile (Fig.~\ref{fig:spix_prof}(a)), by extracting the 607 MHz 
and 1435 MHz flux density in annuli centered on the cluster center 
(Fig.~\ref{fig:spix_prof}(b)) and then computing the corresponding spectral 
indices. We used the same $8^{\prime\prime}$-resolution images used to 
derive the spectral index map and a radial step of $4^{\prime\prime}$ for 
the annuli. The red and blue vertical dashed lines mark $r=4^{\prime\prime}$ 
(S1) and $r=12^{\prime\prime}$ (inner region). The profile confirms the trend 
hinted at by the spectral index map, showing a gradual radial increase 
in spectral index within the diffuse emission, 
with $\alpha=0.8\pm0.1$ to $\alpha=1.3\pm0.3$ going from $r \sim 20$ kpc to $\sim 80$ kpc.

\subsection{Spectral index of the SW extension}\label{sec:sw_sp} 
  
For the SW extension detected in the low-resolution image at 607 
MHz (Fig.~2(c)), we were not able to derive a spectral index 
image or a radial profile of the spectral index because the signal 
is mostly below the $3\sigma$ level at all other frequencies. For this 
reason, we only obtained an estimate of its overall spectral index 
by measuring the total flux density at 607 MHz and 1435 MHz within a 
sector containing the SW extension and ranging from $\sim 100$ kpc to 
$\sim 380$ kpc from the cluster center. We found a spectral index of $\alpha \sim 1.4$.
This value is very uncertain, but it indicates that the SW tail may be slightly 
steeper than the minihalo emission in the outer $r\sim 50-100$ kpc range 
($\alpha \sim 1.3$; Fig.~8a). A steeper spectral index in the SW tail 
is also suggested by the MeeKAT in-band spectral index image 
presented by \cite{2023MNRAS.520.4410T}.

%
%
\begin{figure*}[h!]
\gridline{\fig{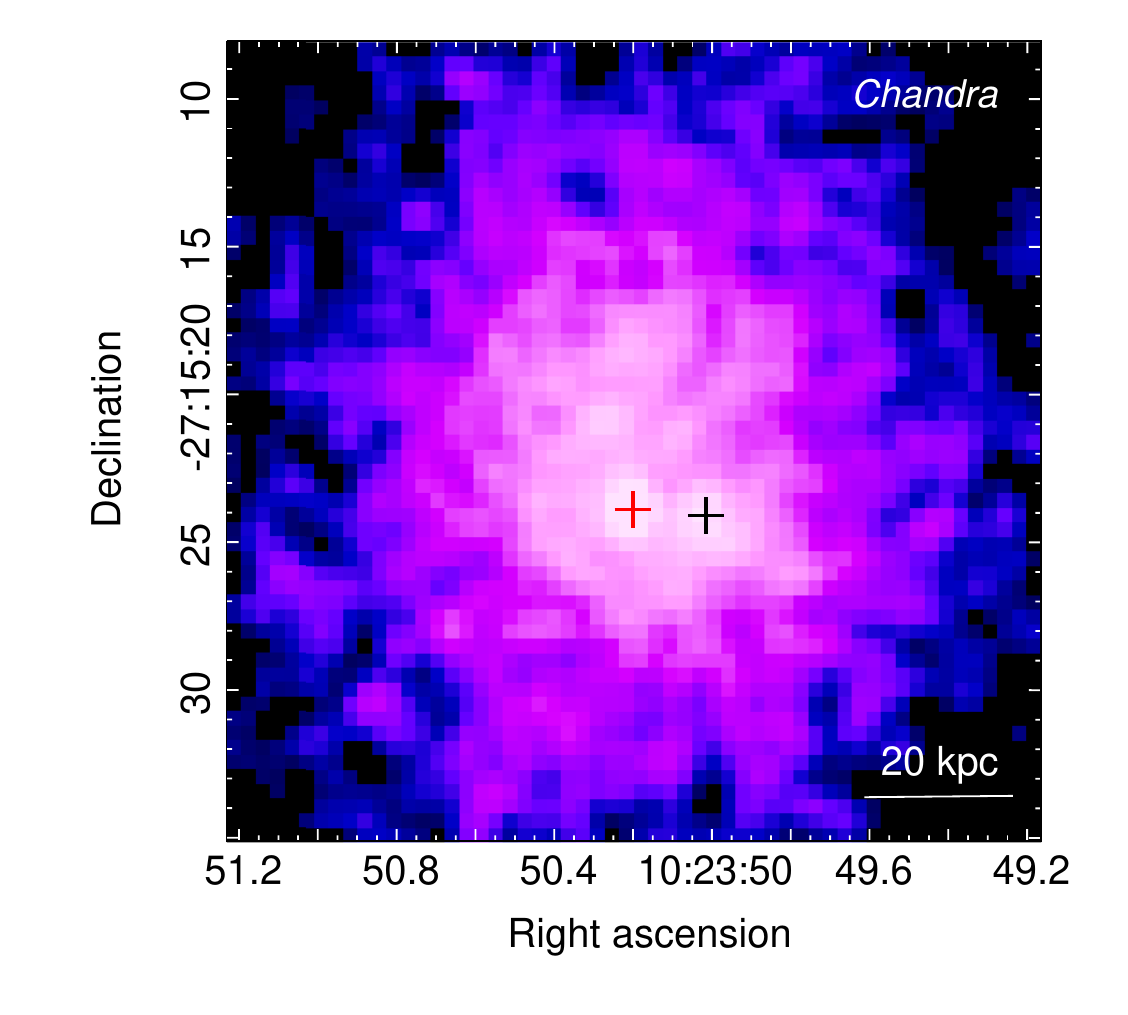}{0.5\textwidth}{(a)}
\hspace{-0.8cm}
          \fig{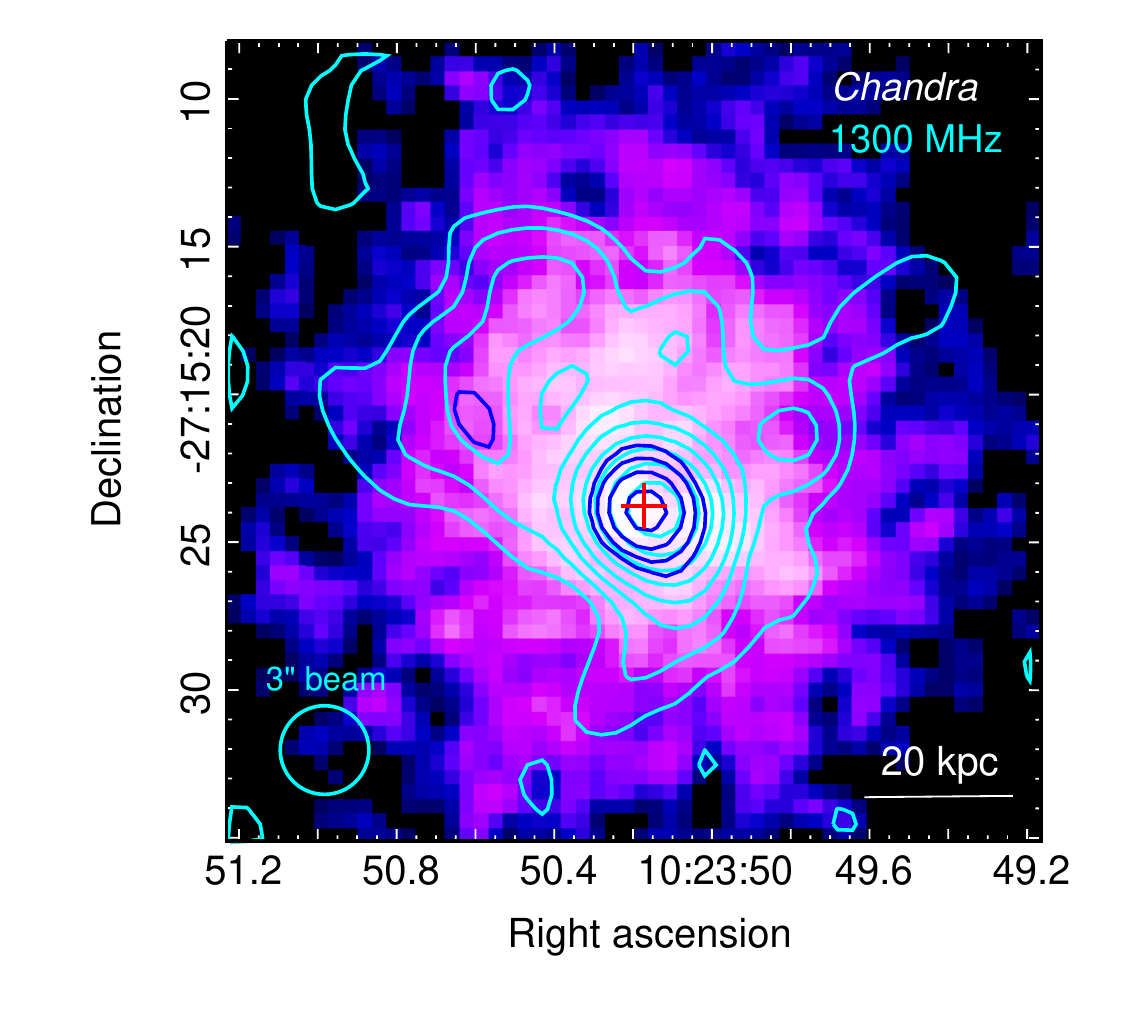}{0.5\textwidth}{(b)}}
\caption{A\,3444. (a) {\em Chandra} X-ray image in the 0.5--2.5 keV energy band,
smoothed with a Gaussian with $\sigma=1''$. The background is subtracted and the image divided by the exposure map. The red cross marks the position of an X-ray point source coincident with the BCG. The black cross is the peak of the extended X-ray emission, which is offset from the BCG \citep[see also][]{2012MNRAS.421.3409H}. (b)  
  The same {\em Chandra} image with GMRT 1300 MHz contours from Fig.~1(b), with $3^{\prime\prime}$ (cyan) and $2^{\prime\prime}$ (blue) resolution.}
\label{fig:core}
\end{figure*}
%
%

\section{ X-ray analysis: cold fronts in the core of A\,3444}\label{sec:chandra}

MS\,1455.0+2232 exhibits a pair of prominent sloshing cold fronts in its cool core  in the {\em Chandra}\/ image \citep{2001astro.ph..8476M,2008ApJ...675L...9M}. 
Here we search for sloshing signatures in A3444 using the archival {\em Chandra}\/ ACIS-S data (OBSID 9400, 36 ks).
We refer to \cite{2017ApJ...841...71G} for details on the data reduction 
and image preparation. All radial distances $r$\/ are from the peak of the X-ray extended emission in the {\em Chandra} 0.5--2.5 keV image (10h23m50.0s, $-27^{\circ}$15$^{\prime}$24.1$^{\prime\prime}$).

In Figure \ref{fig:core}, we show a {\em Chandra}\/ image of the central $r\sim$ 50 kpc region of the cluster with GMRT 1300 MHz contours overlaid from Fig.\ref{fig:mh_1.28}(b). 
No obvious X-ray cavities, associated with the extended radio emission, 
are visible in this inner region. The red cross marks the position of a weak
X-ray point source that is coincident with the BCG, suggesting the presence 
of an AGN \citep{2012MNRAS.421.3409H}. The peak of the extended X-ray emission is marked by a black cross. This peak is offset 
by $\sim 2^{\prime\prime}.5$ ($\sim 10$ kpc) from the BCG \citep{2012MNRAS.421.3409H}. A\,3444 is one of the few clusters in which such a significant offset of the gas peak 
from the optical galaxy has been observed \cite[e.g.,][]{2010ApJ...710.1776J, 2012MNRAS.421.3409H, 2016MNRAS.460.1758H, 2016MNRAS.460.2752W,2019ApJ...885..111P, 2021ApJ...911...66P, 2022ApJ...934...49G}. These offsets 
may be a transient phenomenon caused by sloshing motions of the ICM in the cluster core, which can temporarily displace the densest and coolest gas from the optical nucleus.

%
%
\begin{figure*}
\gridline{\fig{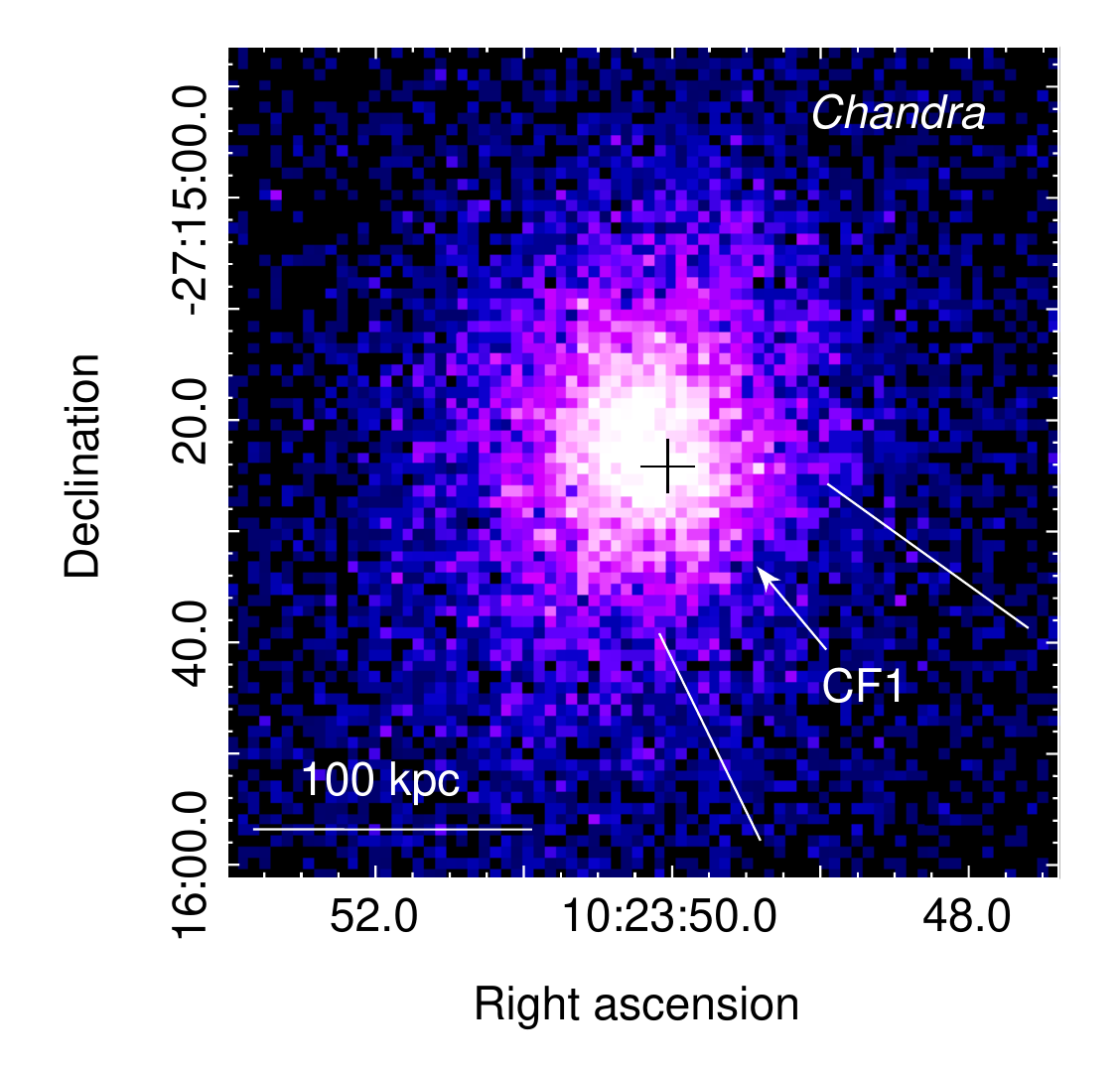}{0.45\textwidth}{(a)}
         \fig{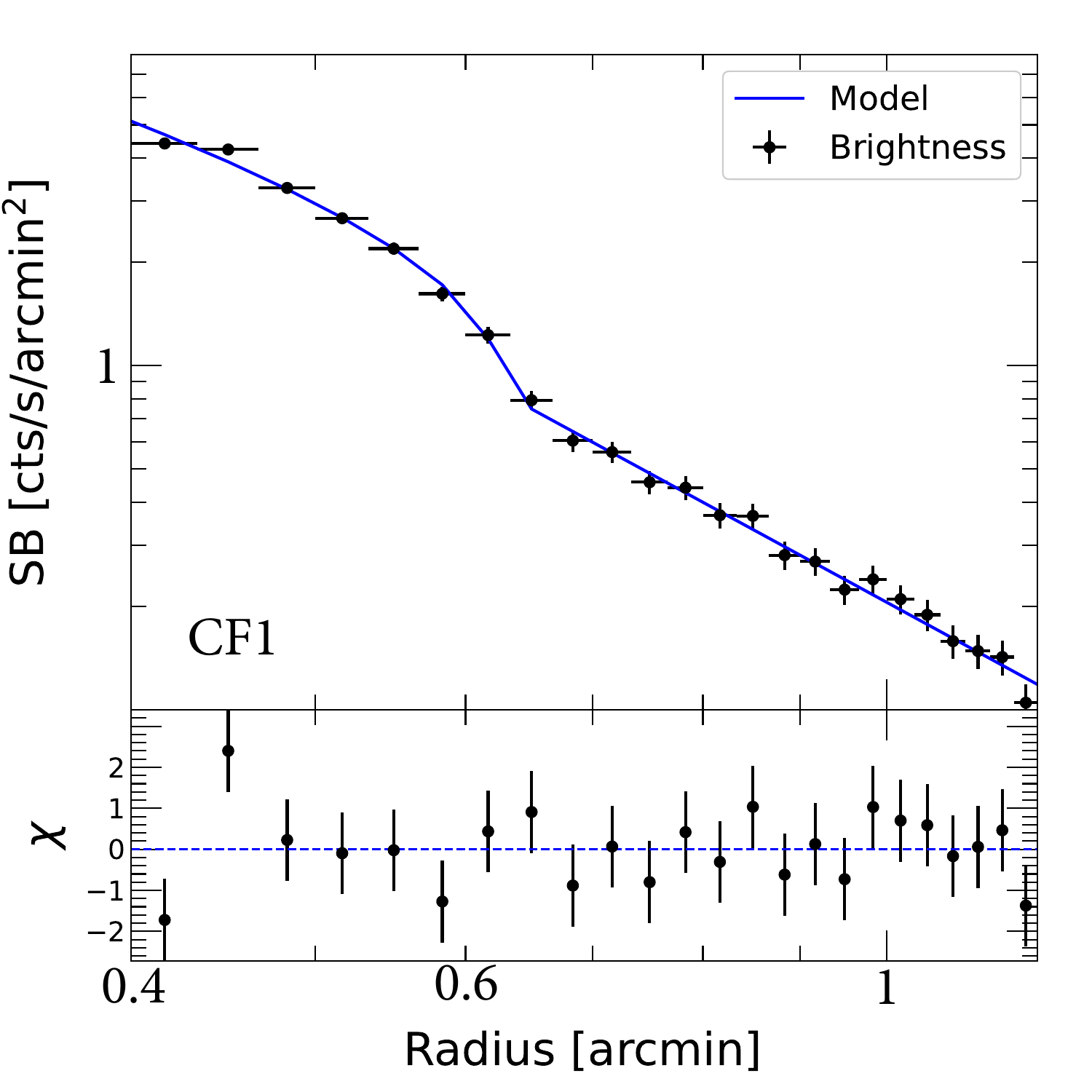}{0.45\textwidth}{(b)}}
\gridline{\fig{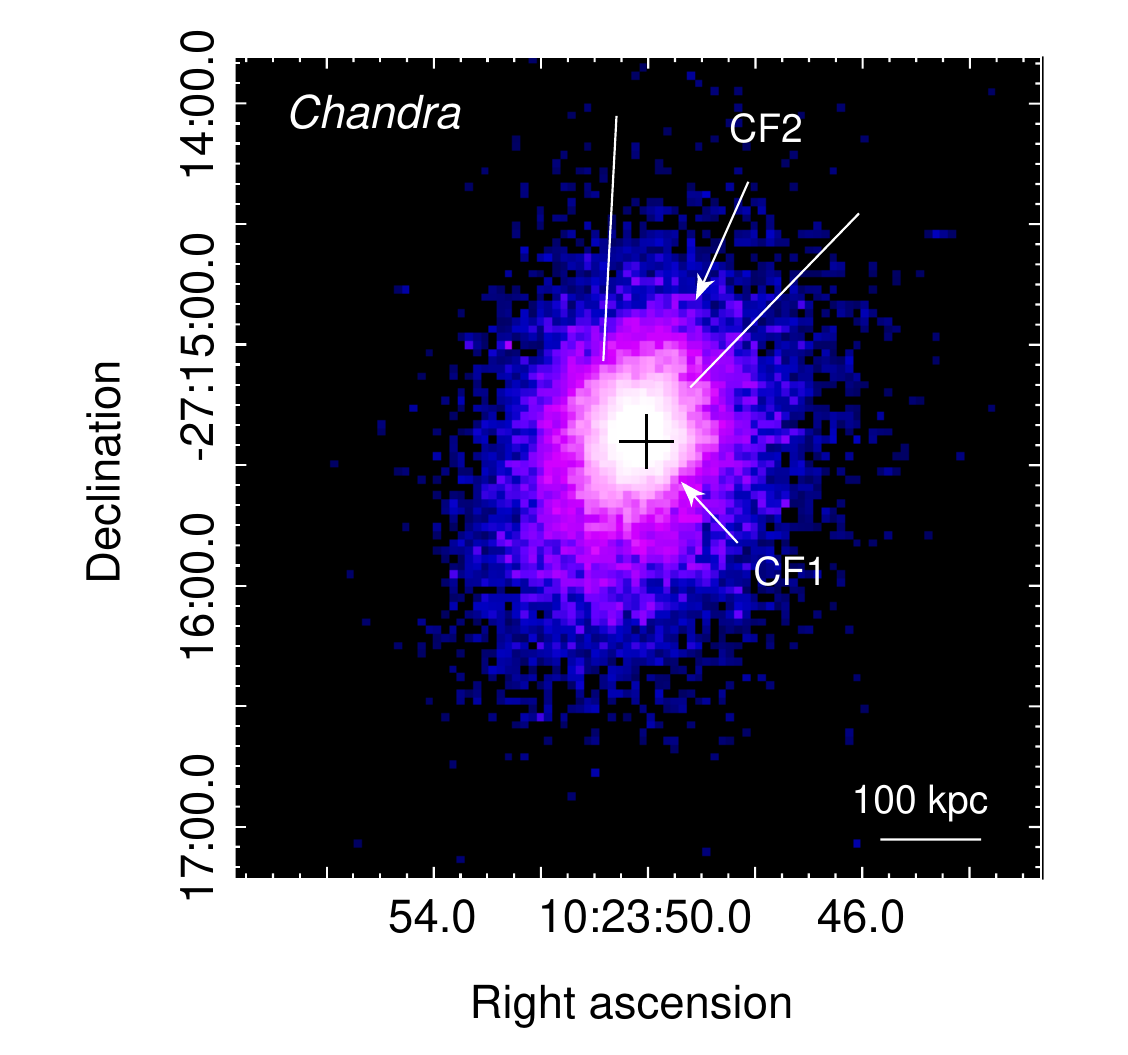}{0.48\textwidth}{(c)}
          \fig{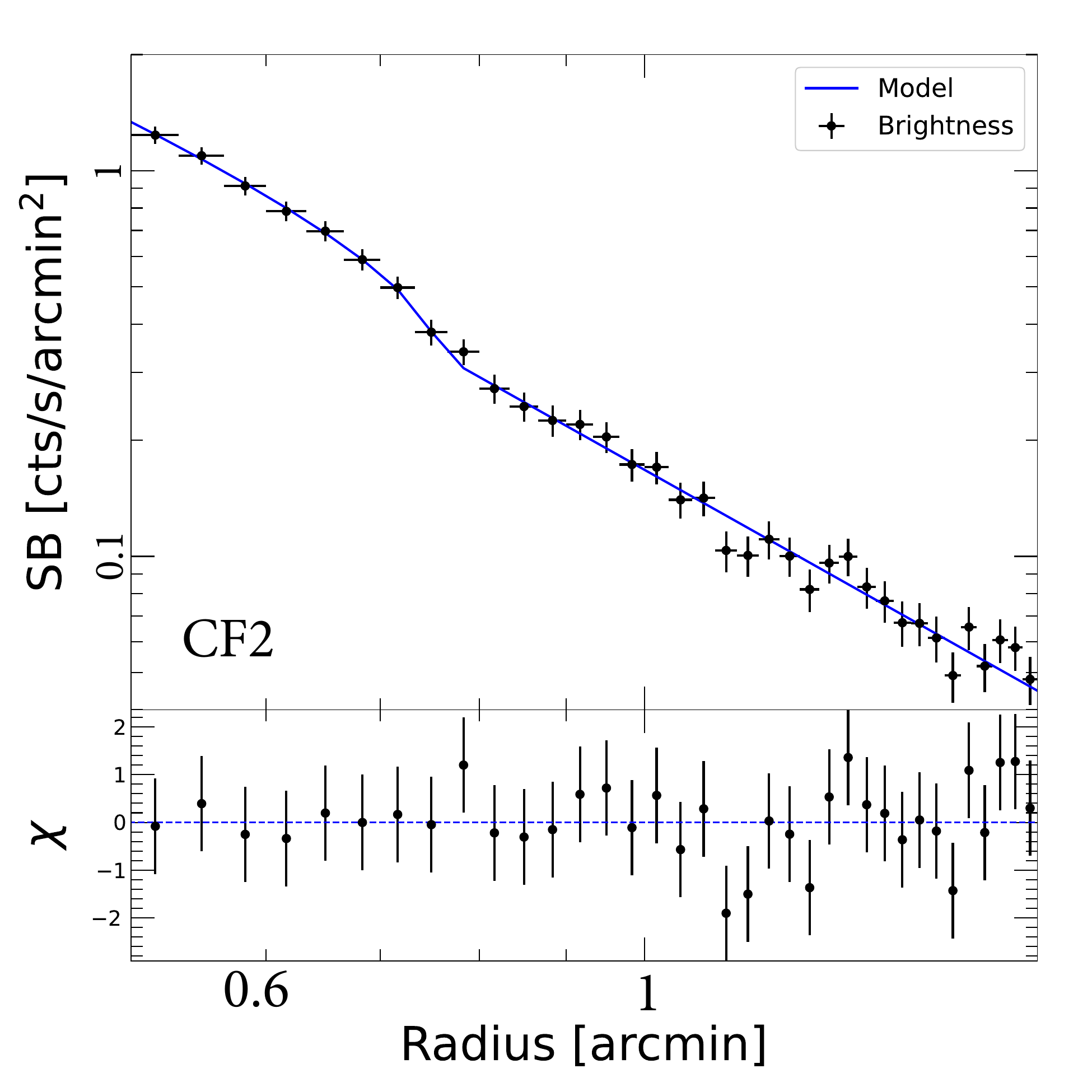}{0.45\textwidth}{(d)}}      
\caption{A\,3444. {\em Chandra} X-ray images in the 0.5--2.5 keV (a) and 0.5--4.0 keV (c) bands, background subtracted, divided by the exposure map and binned by 2 (with the resulting $1^{\prime\prime}$ pixel size). The cross is the X-ray peak. Arrows mark the positions of the cold fronts. White lines show the sectors used to extract the surface brightness profiles in (b) and (d). The best-fitting projected broken power-law models for the density profiles are shown in blue. The $\chi^{2}$/d.o.f. of the fits are 19.2/19 (CF1) and 20.5/31 (CF2).}
\label{fig:chandra}
\end{figure*}
%
%

%
%
\begin{figure*}
\gridline{\fig{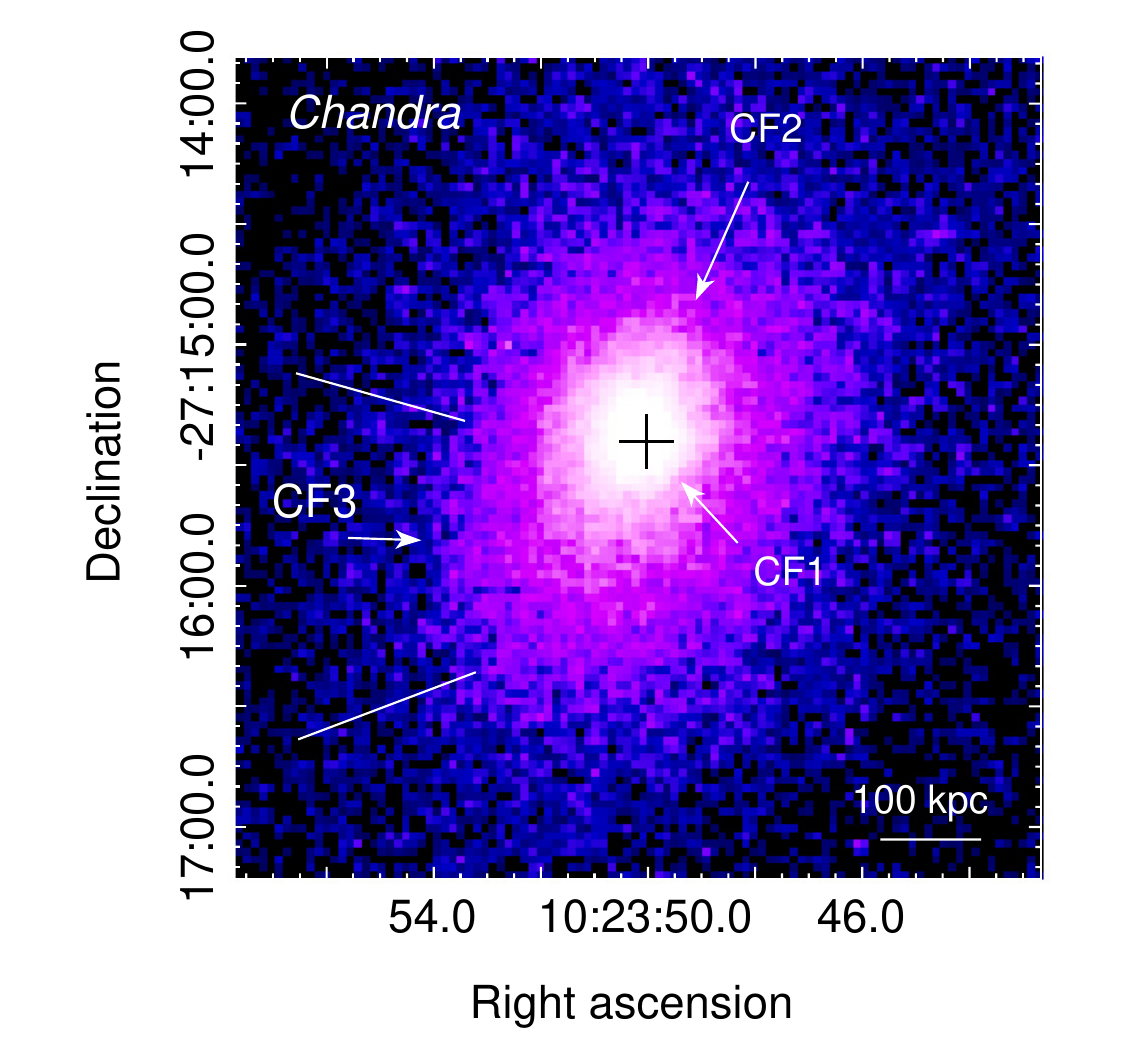}{0.48\textwidth}{(a)}
         \fig{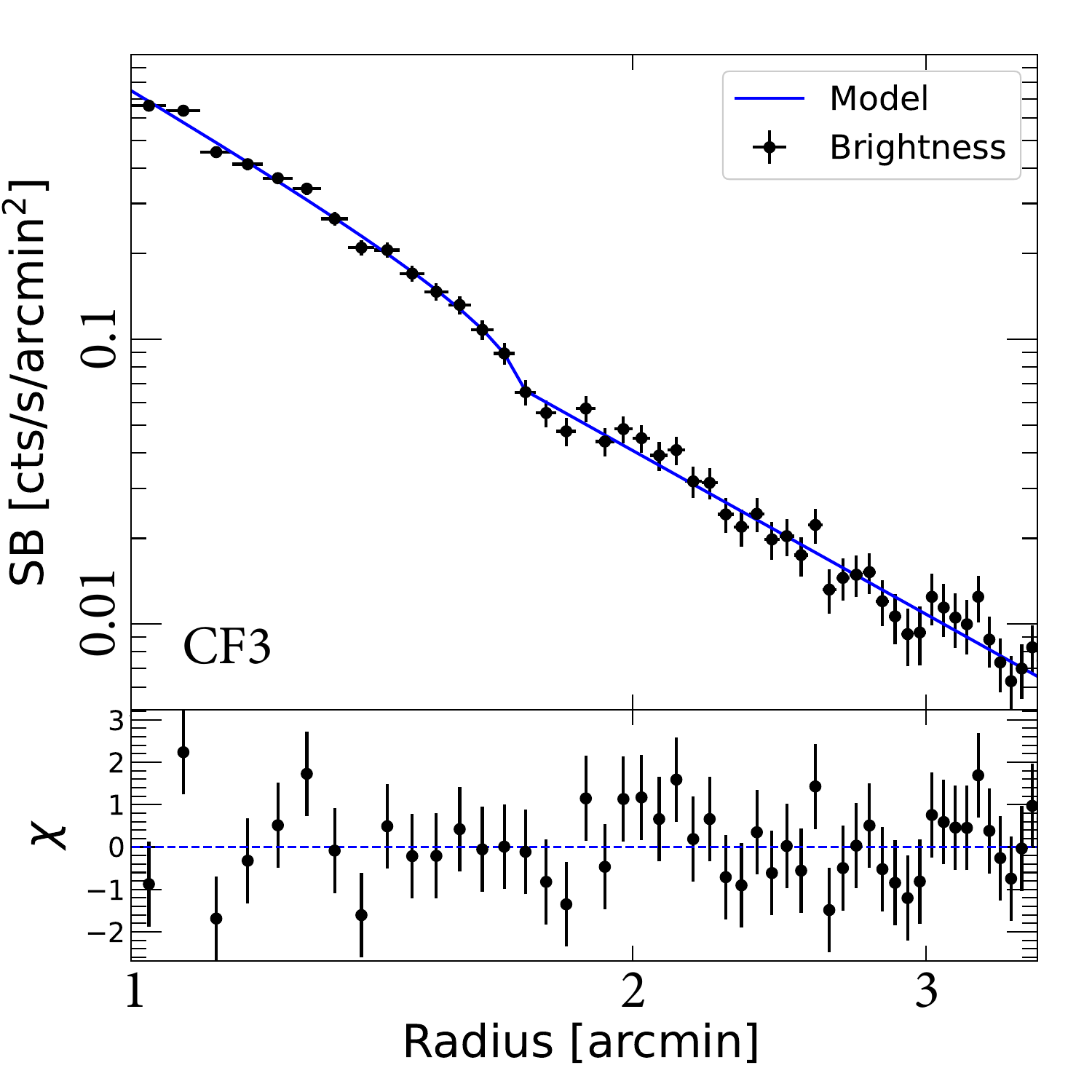}{0.45\textwidth}{(b)}}
     \gridline{\fig{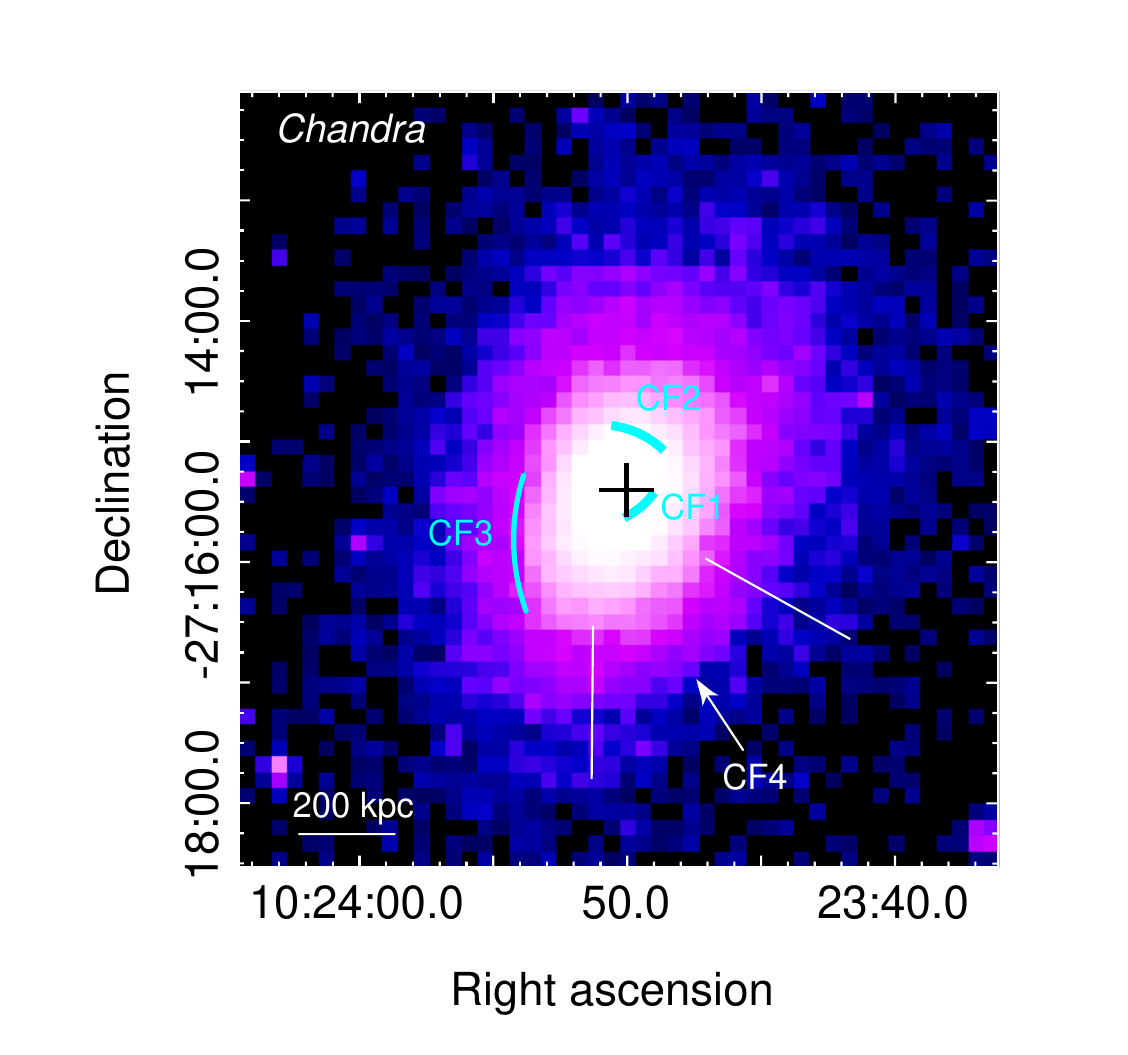}{0.48\textwidth}{(c)}
         \fig{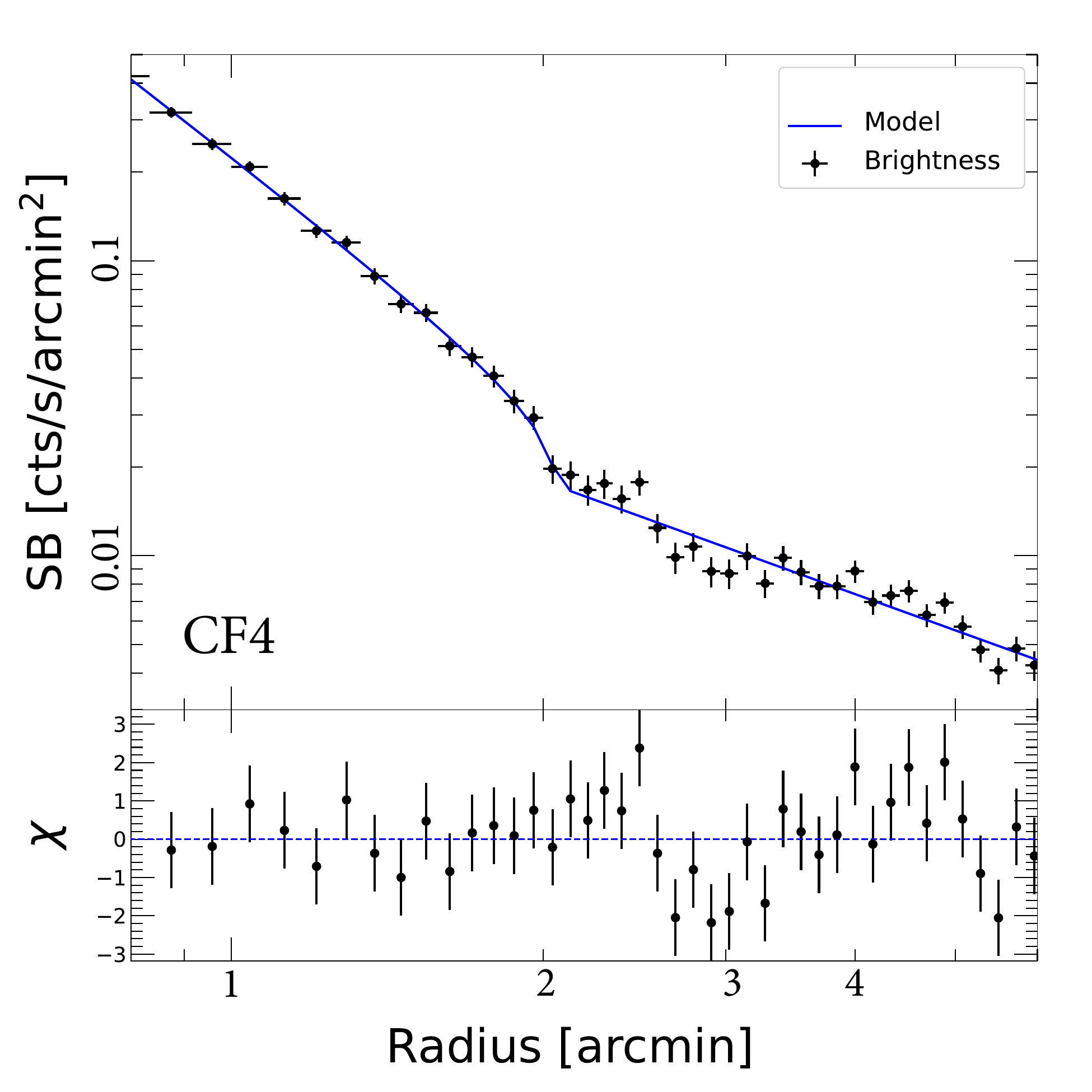}{0.45\textwidth}{(d)}}    
\caption{A\,3444. {\em Chandra} X-ray images in the 0.5--4.0 keV band, background subtracted, divided by the exposure map and binned by 2 (a) and 16 (c) (with the resulting $1^{\prime\prime}$ and $8^{\prime\prime}$ pixel sizes). The cross is the X-ray peak. Arrows and arcs mark the positions of the cold fronts. White lines show the sectors used to extract the surface brightness profiles in (b) and (d). The best-fitting projected broken power-law models for the density profiles are shown in blue. The $\chi^{2}$/d.o.f. of the fits are 41.1/45 (CF3) and 69/57 (CF4).}
\label{fig:chandra2}
\end{figure*}
%
%

%
%
\begin{figure}
\centering
\includegraphics[width=8cm]{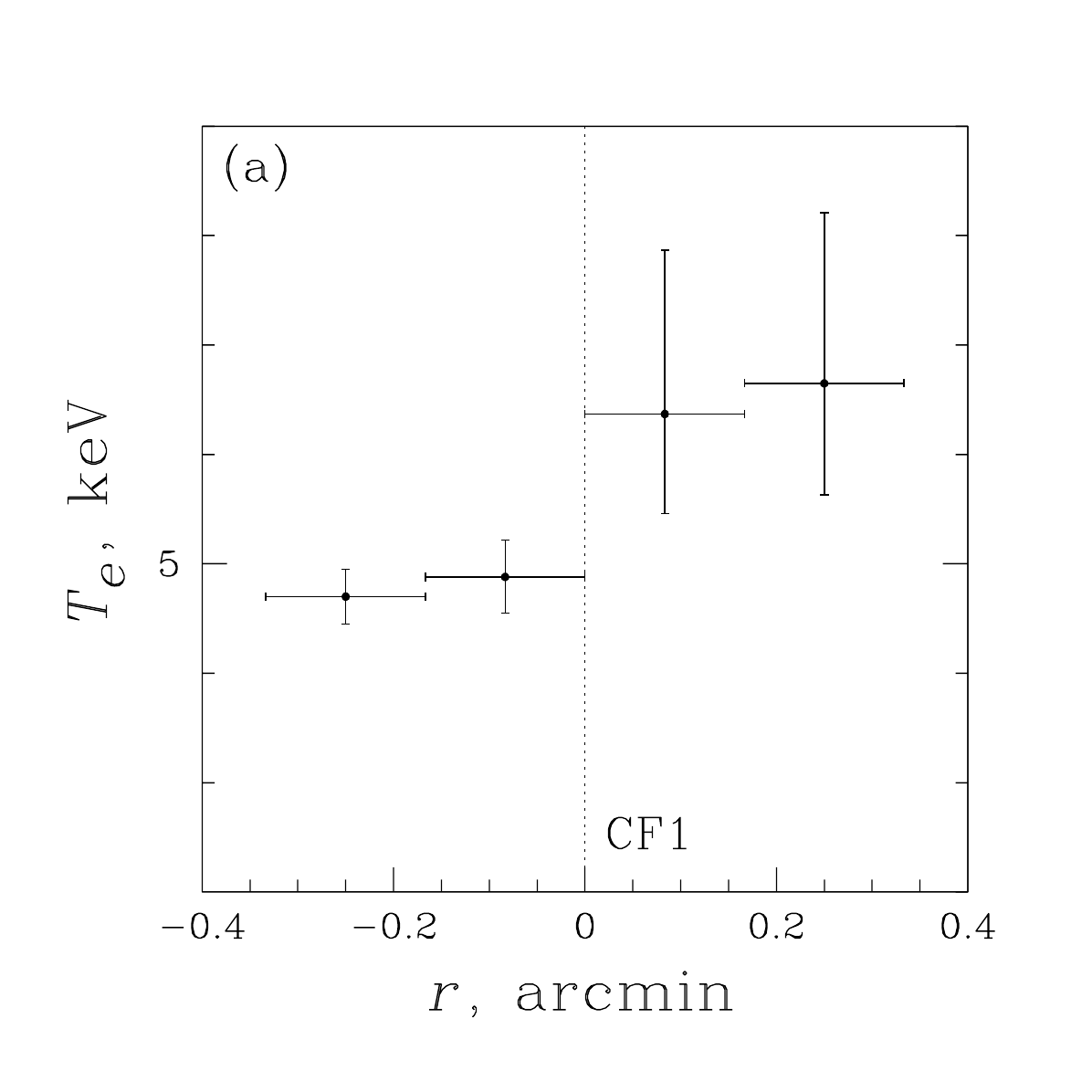}
\includegraphics[width=8cm]{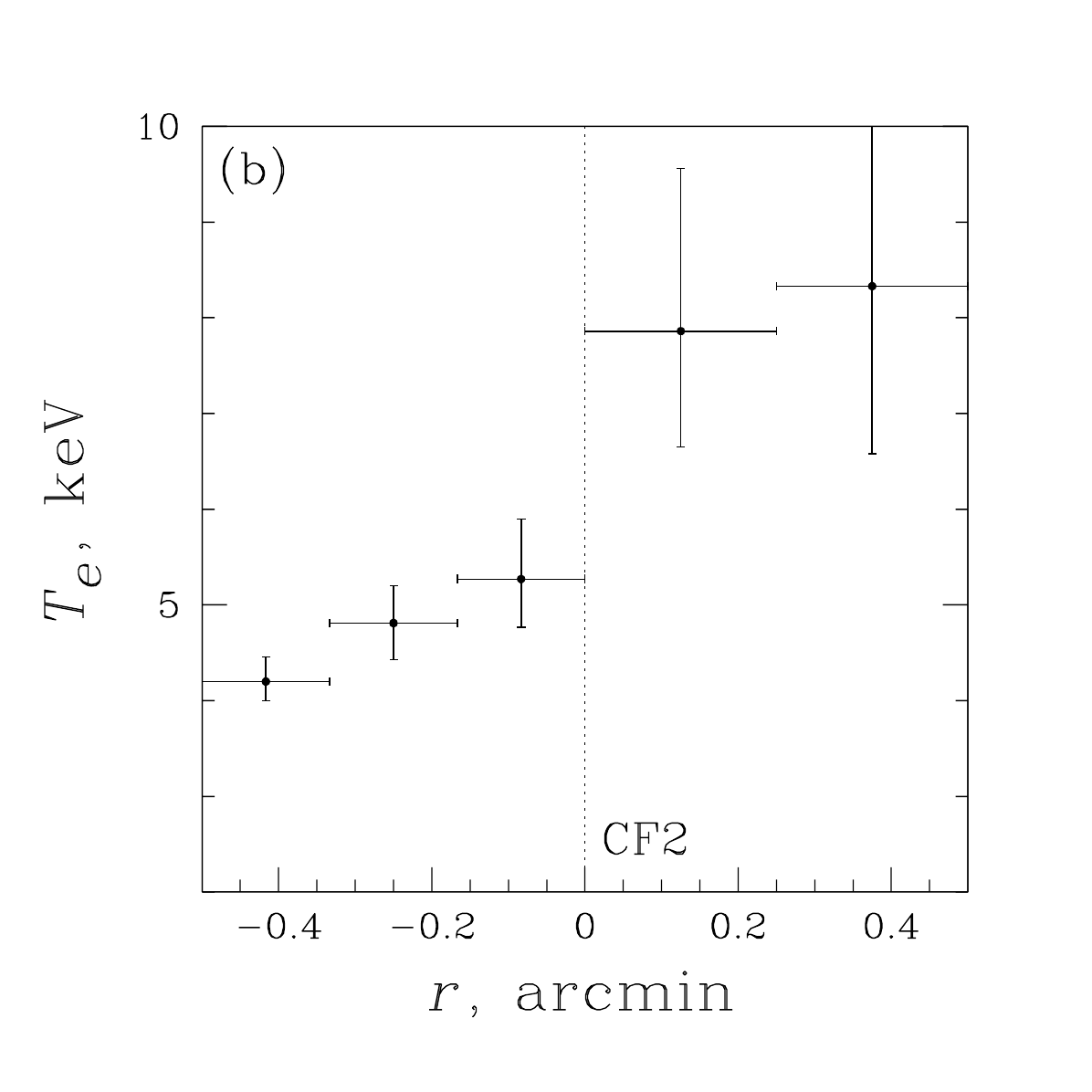}
\smallskip
\caption{A\,3444. Radial projcted temperature profiles across CF1 (a) and CF2 (b)
measured within the sectors shown in Fig.~\ref{fig:chandra}.
The zero of the x-axis is at the front radius (dotted lines). 
Error bars are $1\sigma$.} 
\label{fig:tchandra}
\end{figure}
%
%

In Figures \ref{fig:chandra} and \ref{fig:chandra2}, 
we present {\em Chandra} images showing the ICM emission on a larger scale. 
We see three surface brightness edges in these images. One (CF1) is located $\sim 60$ kpc southwest of the X-ray peak, 
CF2 is located at $r\sim 120$ kpc, and CF3 is at $r\sim 230$ kpc. These edges appear to define a spiral-shaped pattern, similar to the structure seen in numerical simulations of gas sloshing in cluster cores \citep[e.g.,][]{2006ApJ...650..102A}
and often observed in the X-ray images of cool cores with cold 
fronts \citep[e.g.,][for reviews]{2007PhR...443....1M,2022hxga.book...93Z}.
An additional edge (CF4) is located well outside of the core, at $r\sim 400$ kpc, suggesting that the ICM is sloshing on a cluster-wide scale.

We used {\tt pyproffit} \citep{2020OJAp....3E..12E} to extract and model 
the surface brightness profiles across these edges using a source-subtracted 
X-ray image in the 0.5--2.5 keV band and corresponding exposure map. 
To account for the detector and sky background, we also 
included a re-projected, normalized blank-sky background image and the 
ACIS readout artifact \citep{2000ApJ...541..542M}.
To compute the X-ray surface brightness profiles, we used the sectors 
indicated in Figs.~\ref{fig:chandra}(a, c) and \ref{fig:chandra2}(a, c). 
Each profile is centered on the center of curvature of the front 
(note that these centers 
are not coincident with the cluster X-ray peak). The  profiles 
are shown in Figs.~\ref{fig:chandra}(b, d) and \ref{fig:chandra2}(b, d).
We modeled these profiles 
assuming spherical symmetry and a broken power-law density 
profile in 3D \citep[e.g.,][]{2007PhR...443....1M}, with the power-law slopes 
and the position and amplitude of the density jump as free parameters.
We found that the surface brightness discontinuities at the fronts are well
described by density jumps with factors $C_{\rm CF1}=1.6\pm0.1$,  $C_{\rm CF2}=1.4\pm0.1$, $C_{\rm CF3}=1.4\pm0.1$, and $C_{\rm CF4}=1.8\pm0.1$, 
as given by the best-fit broken power-law models shown in Figs.~\ref{fig:chandra}(b, d) and \ref{fig:chandra2}(b, d).

We measured gas temperature profiles across these edges (Fig.~\ref{fig:tchandra}) 
by fitting spectra as described in \cite{2017ApJ...841...71G}. 
Even though statistical errors are large, panels (a) and (b) show that 
the temperature increases moving outward across the surface brightness edge, 
indicating that both CF1 and CF2 are cold fronts. For CF3 and CF4, 
the statistics is too low to constrain the temperature variation across 
this front; however, their temperature profiles (not shown) are consistent 
with cold fronts, as is the large-scale spiral pattern formed by CF1, CF2, CF3 and CF4.

\section{Discussion}

In a number of clusters, radio minihalos have been found to be confined within the X-ray cold fronts. The likely origin of those X-ray fronts is gas sloshing, which would drive turbulence inside the cores and orient the ICM magnetic fields along the fronts. This would naturally explain the confinement of the radio emission within those fronts, if the synchrotron-emitting electrons are reaccelerated by turbulence \citep{2013ApJ...762...78Z}. 
MS\,1455.0+2232 was one of the first examples of the confinement of the radio emission by cold fronts \citep{2008ApJ...675L...9M}. 
However, deeper radio images of a few minihalos, 
including MS\,1455.0+2232, have revealed diffuse emission that ``leaks out'' beyond the cold fronts \citep[e.g.,][]{2017MNRAS.469.3872G,2018MNRAS.478.2234S,2019A&A...622A..24S,2021MNRAS.508.3995B, 2022MNRAS.512.4210R}, in apparent tension with the above picture. Below we see if there is indeed a tension.

%
%

\begin{figure*}
\gridline{\fig{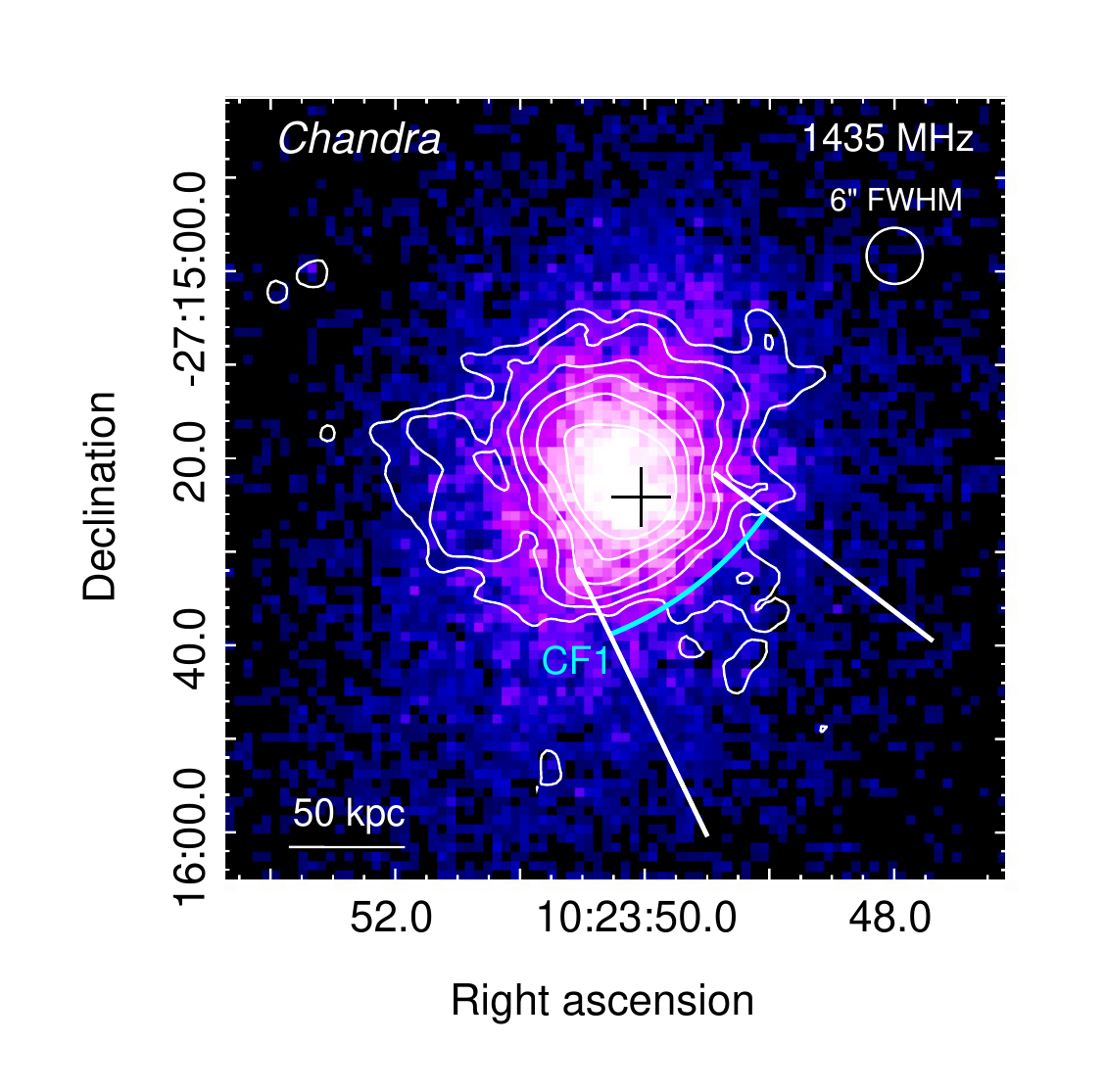}{0.45\textwidth}{(a)}
         \hspace{-1.5cm} \fig{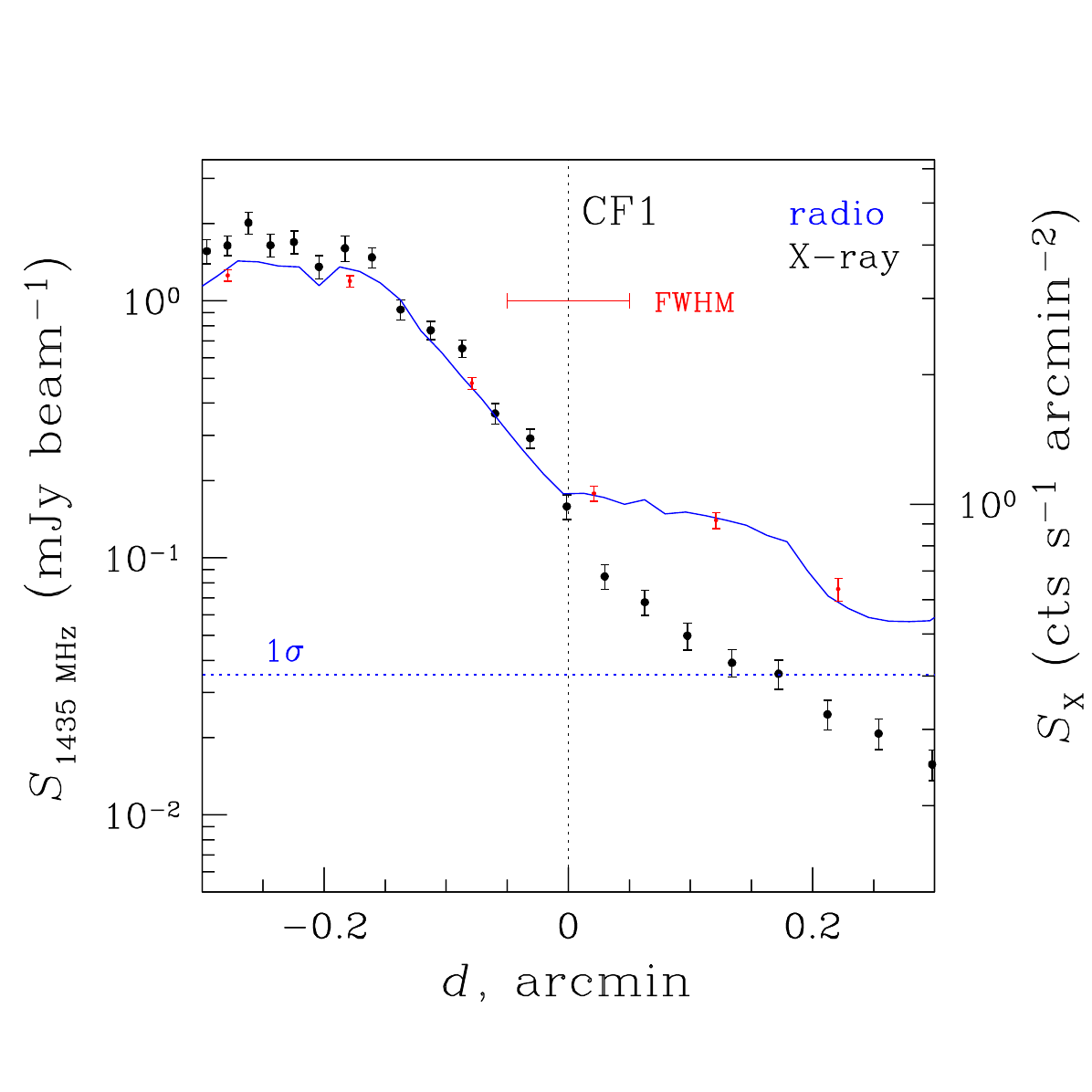}{0.48\textwidth}{(b)}}
\vspace{-1cm}
\gridline{\fig{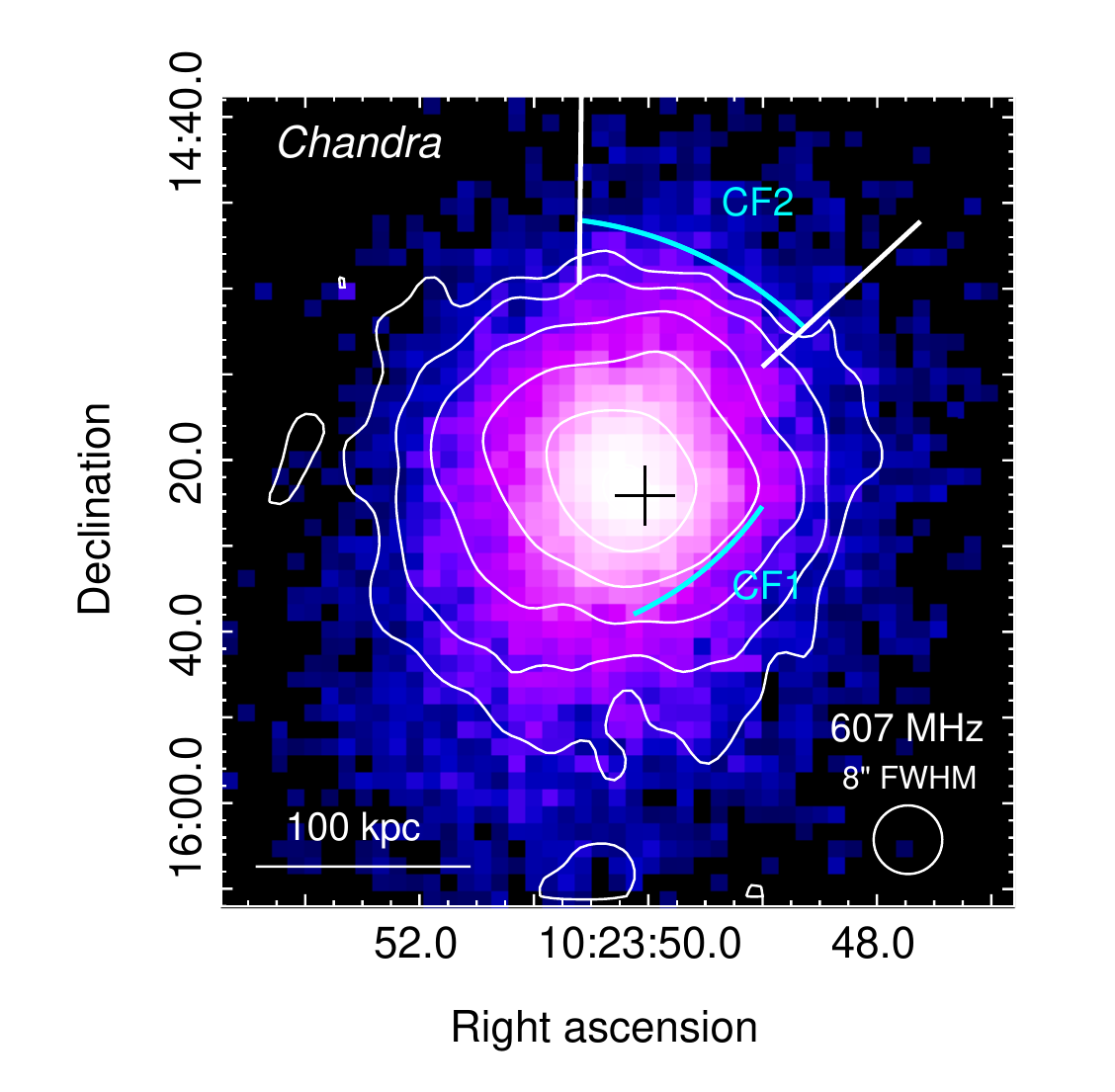}{0.45\textwidth}{(c)}
          \hspace{-1.5cm}\fig{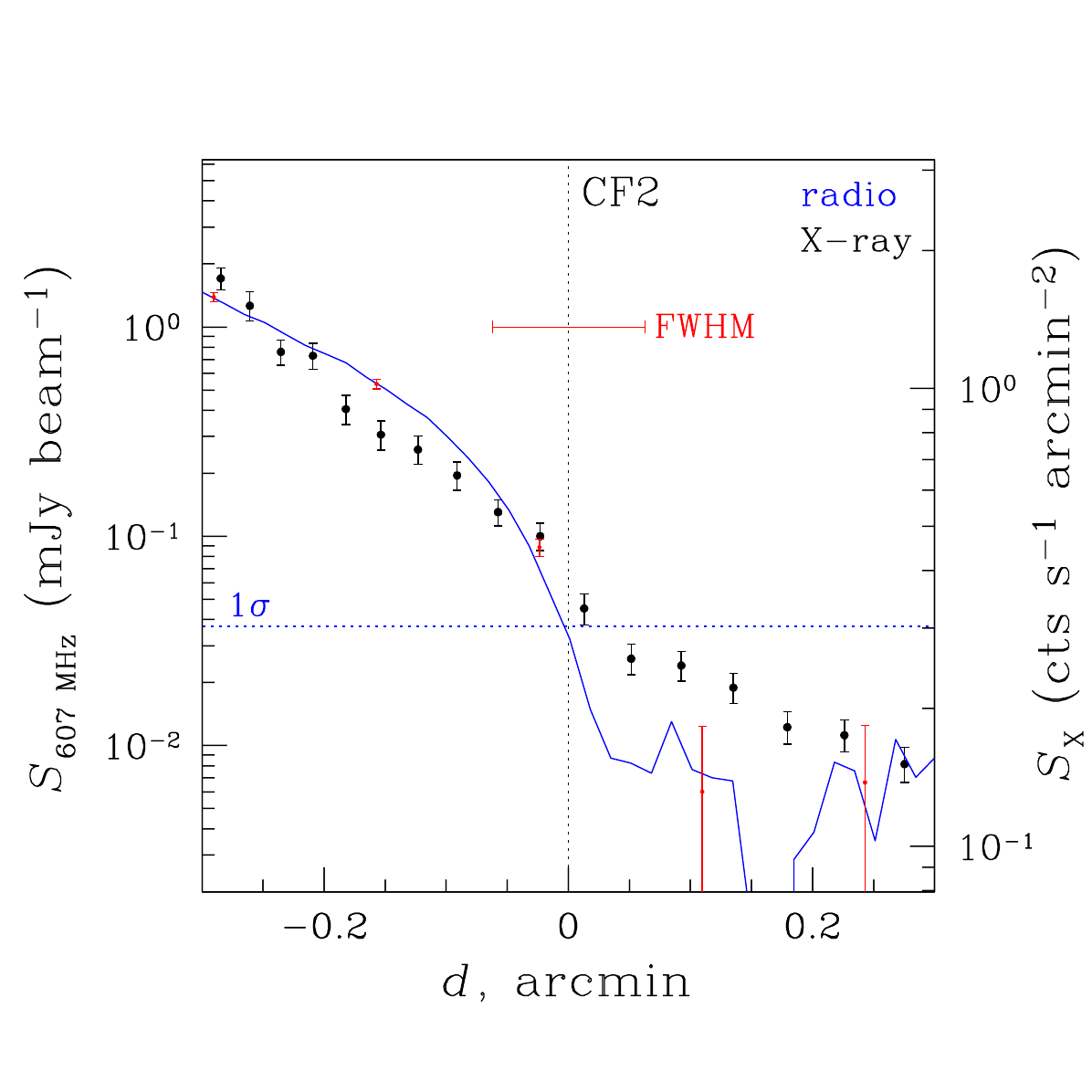}{0.48\textwidth}{(d)}}
             \caption{A\,3444. (a, c) Radio contours overlaid on the {\em Chandra} X-ray images (from Fig.~\ref{fig:chandra}). VLA contours at 1435 MHz are spaced by a factor $\sqrt{2}$ starting from $3\sigma=0.12$ mJy beam$^{-1}$ ($6^{\prime\prime}$ beam). Contours at 607 MHz are from Fig.~\ref{fig:chandra2}(a). The cross marks the X-ray peak. Cyan arcs mark the cold fronts CF1, CF2 at the best-fit radii from Figs.~\ref{fig:chandra} and \ref{fig:chandra2}. The radio (blue, red) and X-ray (black) brightness profiles, extracted in sectors marked by white lines, are shown in (b) and (d). The x-axis zero is at the cold front radius. The blue profiles use radial bins of $1^{\prime\prime}$. The red points are radial bins as wide as the FWHM ($6^{\prime\prime}$ and $8^{\prime\prime}$). Blue horizontal dotted lines indicate the noise level of the radio images.}
\label{fig:radioprof1}
\end{figure*}

%
%
%
\begin{figure*}
\gridline{\fig{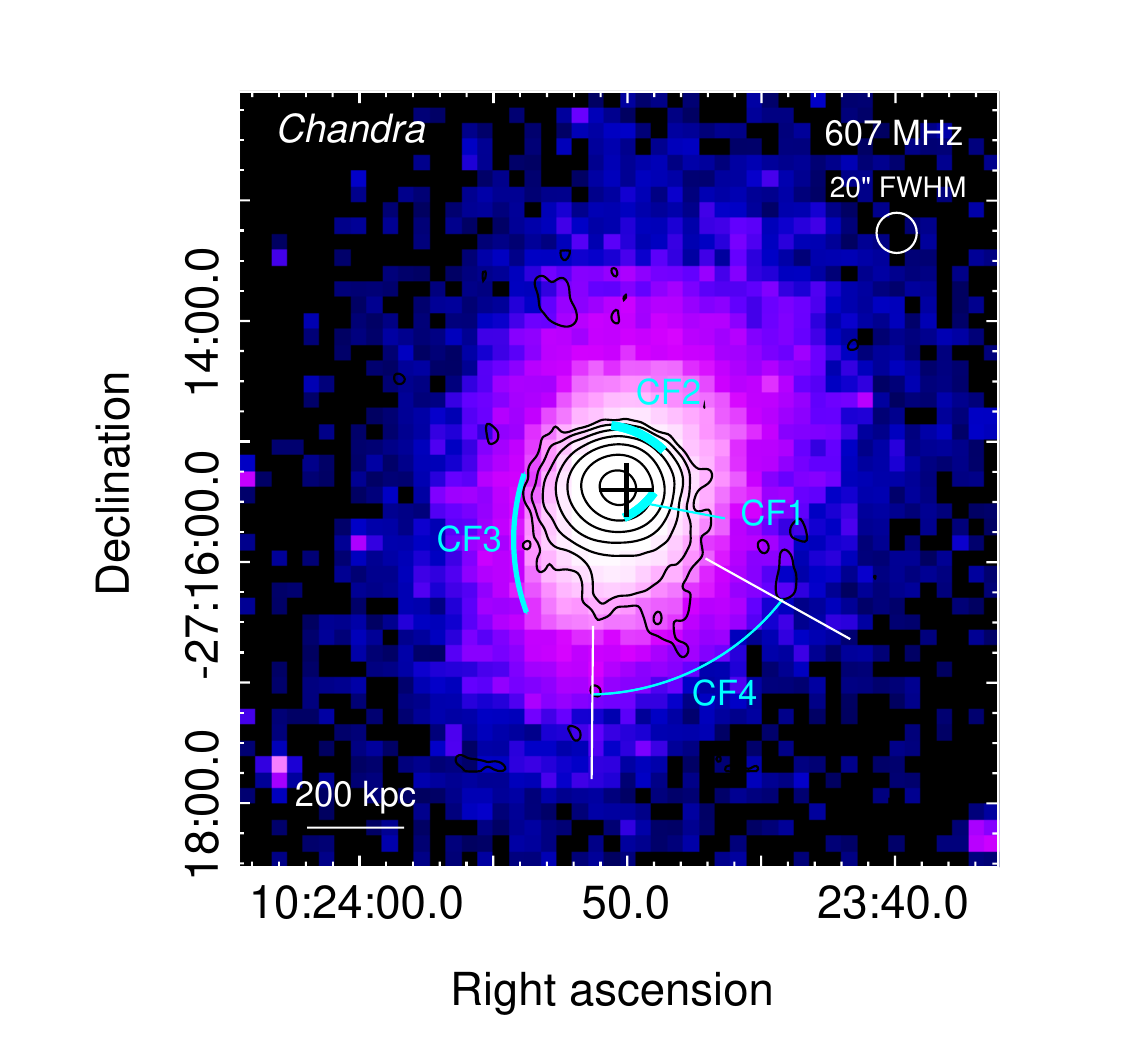}{0.5\textwidth}{(a)}
         \fig{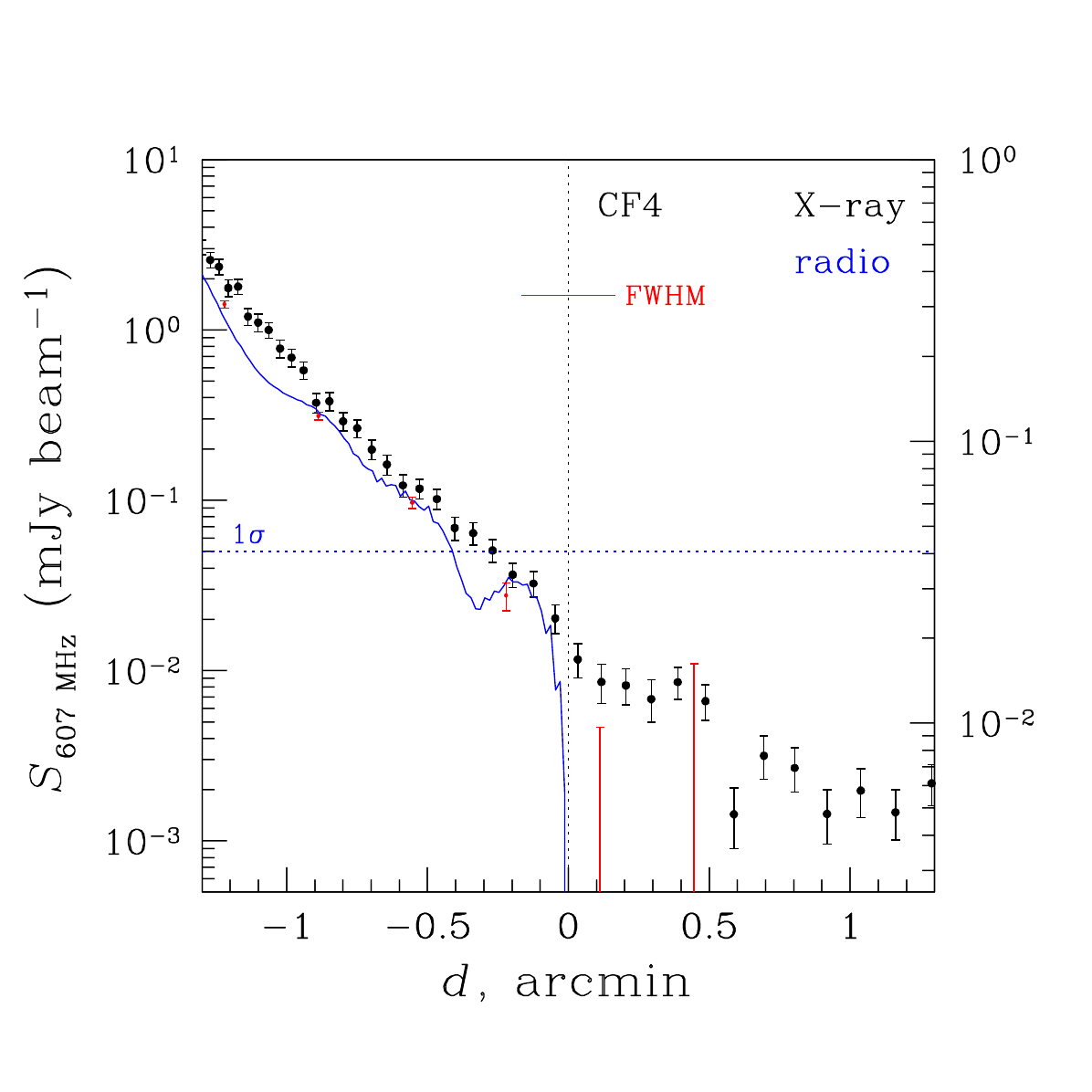}{0.5\textwidth}{(b)}}
    \caption{A\,3444. (a) {\em Chandra} X-ray image from Fig.~\ref{fig:chandra2}(c) with radio contours from Fig.~\ref{fig:mh_610_327}(c). The black cross is the X-ray peak. Arcs mark the cold fronts at the best-fit radii from Figs.~\ref{fig:chandra} and Fig.~\ref{fig:chandra2}. The radio (blue, red) and X-ray (black) brightness profiles, extracted in the sector marked by white lines, are shown in (b). The x-axis zero is at the cold front radius. The blue profiles use radial bins of $1^{\prime\prime}$. The red points are radial bins as wide as the FWHM ($20^{\prime\prime}$). The blue horizontal dotted line indicates the noise level of the radio images.}
\label{fig:radioprof2}
\end{figure*}

%
%

\subsection{Large-scale sloshing in A\,3444}

In Section \ref{sec:chandra}, we have identified three sloshing cold fronts 
in the {\em Chandra} images of A\,3444, at $r=60$, 120, and 230 kpc. In addition, we have found an outer, prominent X-ray surface brightness edge at $r\sim 400$ kpc. This edge is likely an old, large-scale sloshing cold front that has risen outwards. Numerical simulations of gas sloshing show that, as they age, cold fronts should propagate radially from the center into the outer regions of the cluster and
should be long-lived \citep[e.g.,][]{2006ApJ...650..102A,2010ApJ...717..908Z,2023arXiv231009422B}.
Large-scale sloshing cold fronts have been in fact found in several clusters, 
including extreme cases, such as the Perseus cluster, A\,2142, and RXJ\,2014.8-2430, 
in which they have been detected out to $\sim 1$ Mpc from the center \citep{2012ApJ...757..182S,2013A&A...556A..44R, 2014MNRAS.441L..31W, 2018NatAs...2..292W}. 

In Figures ~\ref{fig:radioprof1} and \ref{fig:radioprof2}, we compare the radio emission to the position of the X-ray cold fronts in A\,3444, marked by cyan arcs. 
In Fig.~\ref{fig:radioprof1}(a), we overlay contours at 1435 MHz of the innermost region of the minihalo, and in (b) we compare the X-ray and radio brightness profiles extracted in the sector containing CF1. The radio profile shows a clear edge at the position of CF1, suggesting a connection to the cold front. Beyond this edge, 
however, fainter radio emission is detected toward the region occupied by the SW radio extension. 

In Fig.~\ref{fig:radioprof1}(c),  we overlay the 607 MHz contours 
on the X-ray image. In the northern region of the core, the minihalo 
is edged by CF2 at $r\sim 120$ kpc. Panel (d) shows 
the X-ray/radio profiles across this front. The radio emission drops 
abruptly at the cold front location. Beyond the front, we do not detect any 
significant radio emission, at least at the sensitivity level of our 607 MHz 
image ($1\sigma=37$ $\mu$Jy beam$^{-1}$) that allows us to detect emission at 
least 2 orders of magnitude below the radio peak. We note that the largest 
detectable scale of our 607 MHz dataset is $\sim 10^{\prime}$, i.e., 10 times 
larger than the minihalo extent, thus the drop in radio brightness at the front 
is not caused by a limit in the scale that can be reconstructed by the 
interferometric data. This indicates that the minihalo is confined here by the cold front.  

In Fig.~\ref{fig:radioprof2}(a), we show a {\em Chandra} image with the low-resolution contours at 607 MHz, showing the SW radio extension beyond CF1. 
In (b) we compare the X-ray and radio profiles extracted within the sector containing the radio extension and CF4. The radio profile traces remarkably well the X-ray profile inside the cold front. The radio signal falls below the $1\sigma$ level before CF4, however the radio emission appears to reach the cold front radius in the deeper MeerKAT images of \cite{2023MNRAS.520.4410T}. This suggests that the origin of the 
entire diffuse radio emission seen in A\,3444 is related to large-scale sloshing 
motions of the ICM.

\begin{figure*}
\gridline{\fig{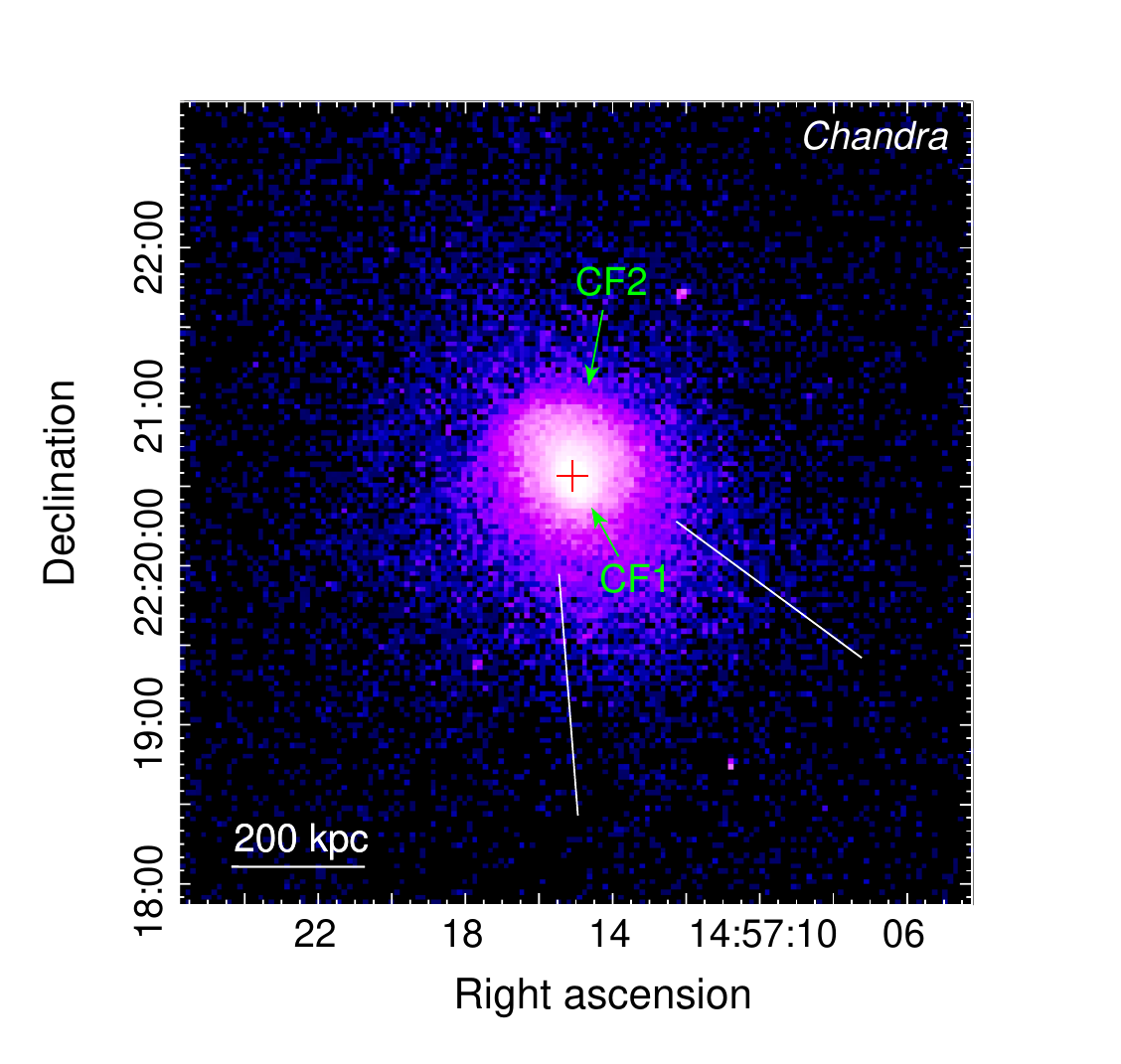}{0.5\textwidth}{(a)}
          \fig{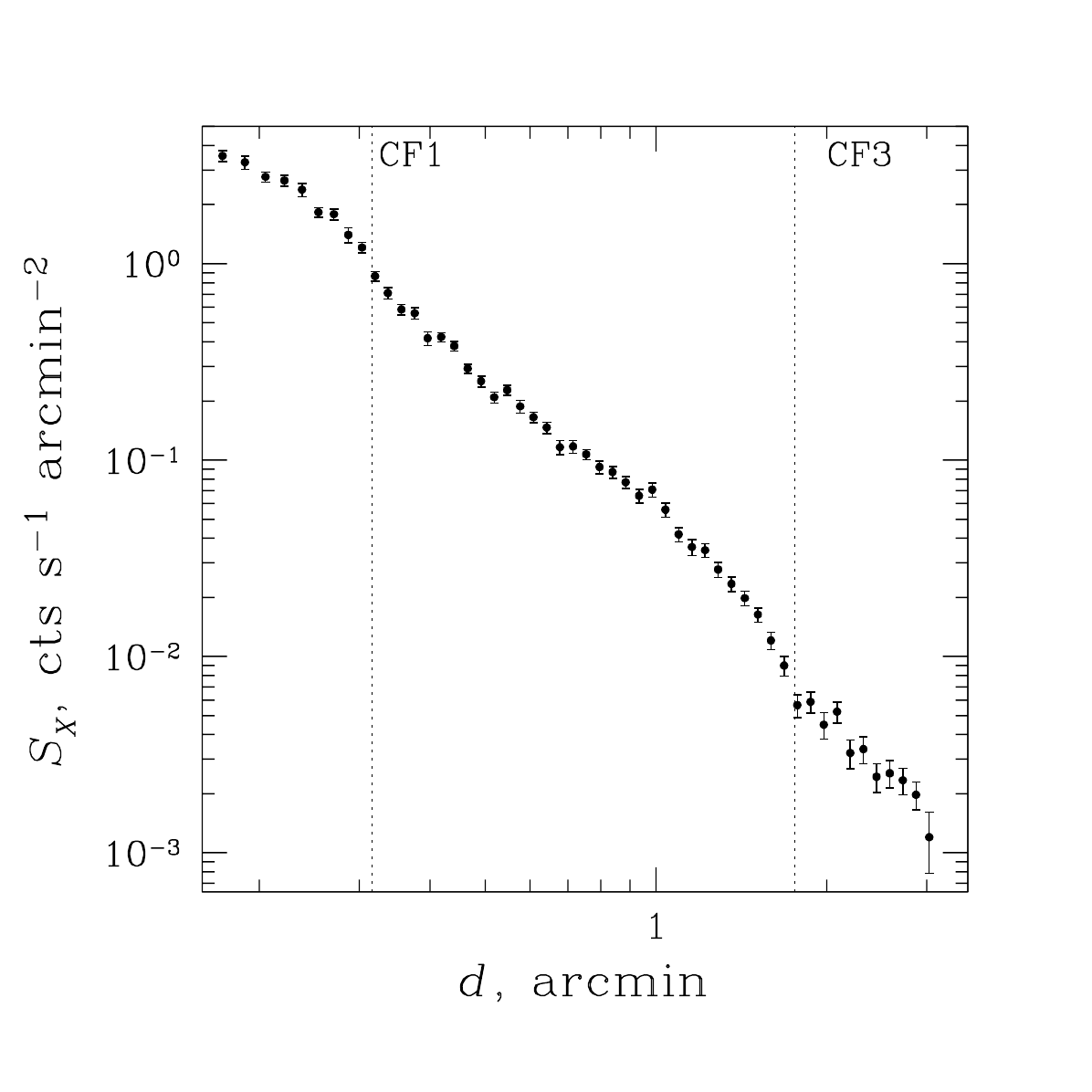}{0.485\textwidth}{(b)}}
          \vspace{-0.5cm}
\gridline{\fig{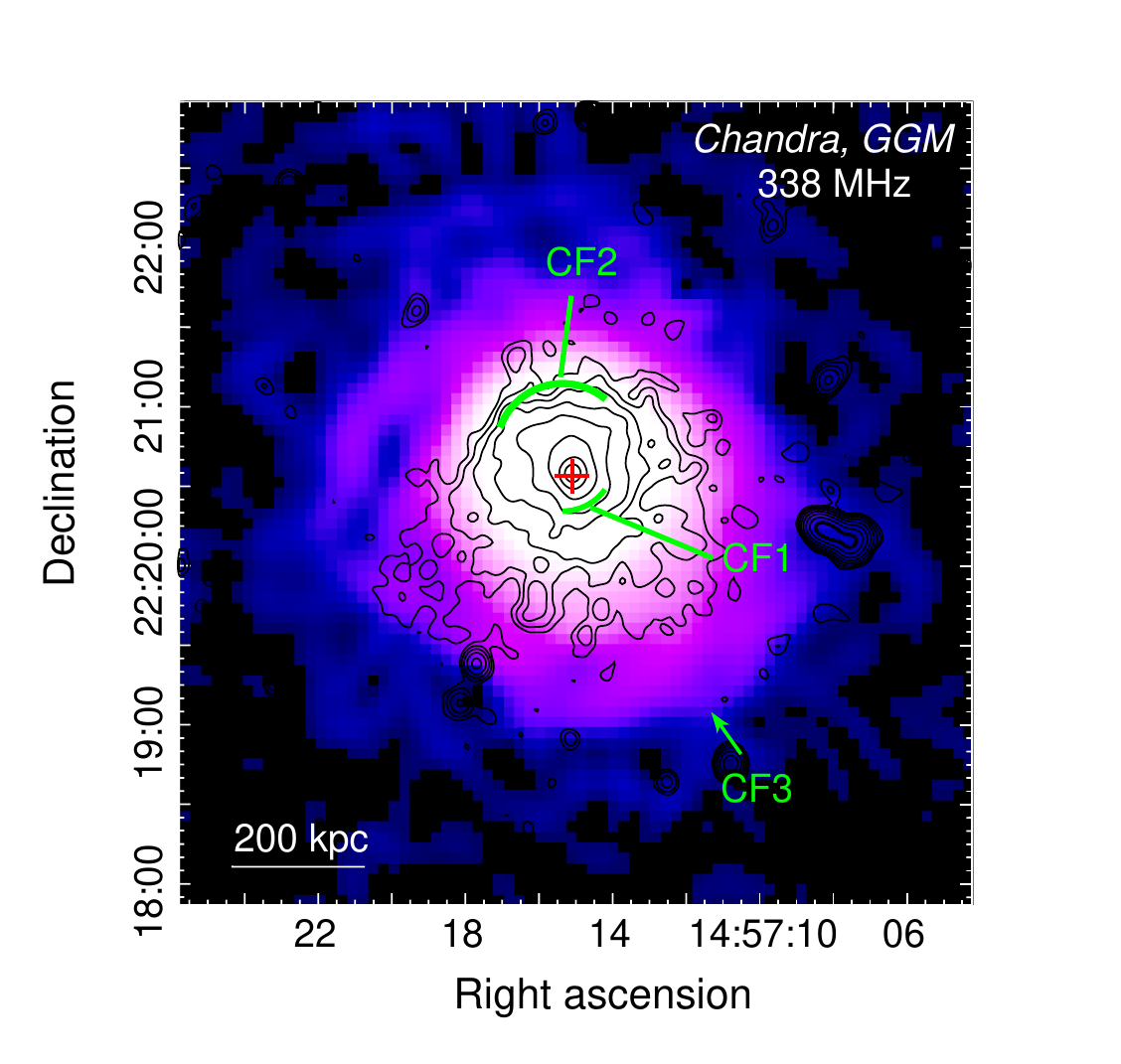}{0.5\textwidth}{(c)}
          \fig{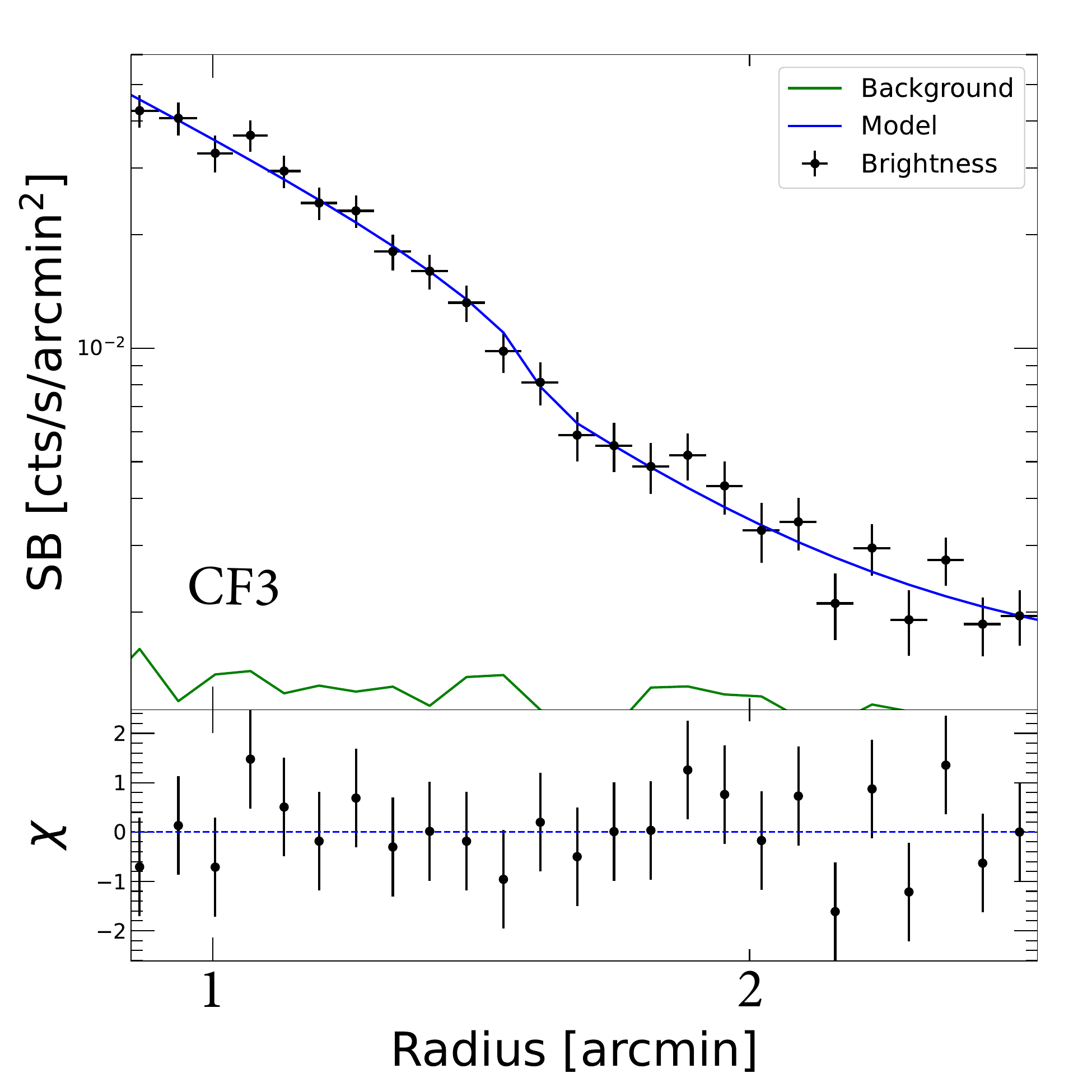}{0.45\textwidth}{(d)}}     \smallskip
\caption{MS\,1455.0+2232.  (a) {\em Chandra} X-ray image in the 0.5-2.5 keV band,
background subtracted, divided by the exposure map and binned by a factor of 4  (1 pixel is $2^{\prime\prime}$). The red cross is the X-ray peak. The known cold fronts are marked as CF1 and CF2. (b) X-ray surface brightness
profile extracted using the sector marked by white lines in (a). Vertical, black-dotted lines mark the position of CF1 and of an additional edge CF3. 
(c) GGM-filtered {\em Chandra} image with $\sigma=2.5$ pixel ($5^{\prime\prime}$). 
The red cross is the X-ray peak. Arcs show CF1 and CF2. The outermost cold front is indicated as CF3. uGMRT 383 MHz contours at $7^{\prime\prime}$ resolution are overlaid in black, spaced by a factor of 2 from 0.05 mJy beam$^{-1}$.
(d) X-ray surface brightness profiles across CF3 extracted using the sector in (a). The blue lines are the best-fit broken power-law model with residuals ($\chi^{2}$/d.o.f.=8/12).} 
\label{fig:ms1455}         
\end{figure*}

\subsection{Large-scale sloshing in MS\,1455.0+2232}

Our findings for A\,3444 raise the question whether there are outer cold fronts also in MS\,1455.0+2232 that may contain 
the radio emission seen outside of the sloshing cool core. To check whether there are indeed signatures of large-scale sloshing,
we used an archival {\em Chandra} X-ray observation (OBS ID 4192) and 
produced images of MS\,1455.0+2232 following the procedure described in  \cite{2017ApJ...841...71G}. In the following, all radial distances $r$ are computed from the position of the X-ray peak (14h57m15.1s,+22$^{\circ}$20$^{\prime}$33.9$^{\prime\prime}$).

In Fig.~\ref{fig:ms1455}(a), we show a {\em Chandra} image 
in the 0.5--2.5 keV band with the well-known sloshing cold fronts 
marked as CF1 and CF2 \citep{2001astro.ph..8476M,2008ApJ...675L...9M}. 
We extracted an X-ray surface brightness profile in a sector that contains 
both CF1 and the radio extension seen at larger radii. 
The sector is marked by white lines in Fig.~\ref{fig:ms1455}(a) and is 
centered on the center of curvature of CF1. 
The resulting profile is shown in (b), where the CF1 radius 
is indicated by the vertical, black-dotted line at $d\sim 0.3^{\prime}$ 
from the center of the sector ($r\sim 36$ kpc from the X-ray peak).
The profile reveals an additional prominent edge (CF3) at $d\sim 1.9^{\prime}$ 
($r\sim 425$ kpc), probably a larger and older cold front.
To quantify this surface brightness edge, we fit 
the brightness profile in the immediate vicinity of the edge
using a broken power-law model, with power-law slopes and the position 
and amplitude of the jump being free parameters. The best-fit model 
(blue line in Fig.~\ref{fig:ms1455}(d)) gives a jump 
factor of $C_{\rm CF3}=1.4\pm0.2$.

To highlight the position of this newly-detected front, we applied 
a Gaussian Gradient Magnitude \citep[GGM;][]{2016MNRAS.460.1898S} filter 
to a point-source subtracted {\em Chandra} image assuming Gaussian derivatives 
with a width $\sigma=2.5$ pixel ($5^{\prime\prime}$). In black, we overlay the
uGMRT radio contours at 338 MHz, showing that the diffuse radio emission is
well contained within CF3. 

As in A\,3444, the detection of an outer X-ray front points to gas sloshing 
motions on cluster scale and suggests that the whole diffuse radio emission 
seen in MS\,1455.0+2232 is sustained, through turbulent particle reacceleration, 
by large-scale sloshing motions of the hot ICM.

\subsection{Origin of minihalos}

Key information that can discriminate between the possible physical mechanisms 
for the origin of the relativistic electrons in minihalos is the slope and 
shape of the integrated radio spectrum and spectral index distribution across
the region occupied by the diffuse emission \citep[e.g.,][]{2014IJMPD..2330007B}.
If minihalos are sustained by turbulent re-acceleration of relativistic electrons 
(primaries or secondaries), depending on the spatial distribution of turbulent dissipation processes and magnetic field in the emitting region, the radio spectral index can range within a large interval of values, including very steep values   
\citep[$\alpha \gax 1.5$; e.g.,][]{2013ApJ...762...78Z, 2014IJMPD..2330007B}. 
Significant spatial variations of the spectral index are also expected 
across the emitting region. Furthermore, in the case of homogeneous 
magnetic field distribution and turbulence in the emitting region, the emitted 
synchrotron spectrum may show a break resulting from a cutoff in the electron 
energy distribution, at a frequency that depends on the acceleration 
efficiency \citep[e.g.,][]{2013ApJ...762...78Z}. 

The minihalo in A\,3444 has a total spectral index 
of $\alpha_{\rm fit,\, total}=1.0\pm0.1$ in the 333--1435 MHz range. 
Its innermost region ($12 \le r \le 50$ kpc) has a slighly 
flatter slope ($\alpha_{\rm fit,\, inner}=0.8\pm0.1$), whereas the spectrum of its 
outer region (50 kpc $< r < 160$ kpc) steepens to $\alpha_{\rm fit,\, outer}=1.3\pm0.1$. 
All of these spectra are consistent with an unbroken power law in the 
interval of frequency covered by our data, which is insufficient 
to discriminate between a re-acceleration or pure hadronic origin for the minihalo 
based on its integrated spectrum. A spectral break may still be present above 1435 MHz, 
as possibly indicated by a steeper spectral index measured for the minihalo emission 
by MeerKAT in the 0.8-1.7 GHz band, even though with large uncertainties \citep{2023MNRAS.520.4410T}. 

Similar to A\,3444, the integrated spectrum of the minihalo in MS\,1455.0+2232 
is a single power-law with $\alpha=0.97\pm0.05$ in 
145-1283 MHz range \citep{2022MNRAS.512.4210R}.

For both minihalos, sensitive observations above $2$ GHz would be therefore needed to search for a possible spectral break at high frequency. 
We note that power-law spectra extending to frequencies as high as 10-20 GHz have been reported for a couple of minihalos \citep{2021A&A...646A..38T,2021MNRAS.508.2862P}, 
indicating the absence of a high--frequency break. The lack of a spectral break in the 
integrated spectrum  does not necessary rule out re-acceleration. The radio spectrum may be, 
 in fact, stretched into a power--law shape if it results from the combination of different 
 spectral components tracing local fluctuations in the physical properties 
 (magnetic field intensity, electron energy distribution, turbulence strength)  within the emitting region \citep[e.g.,][]{2013MNRAS.429.3564D}. 

In Sec.~\ref{sec:sp_index_map}, we studied the spatial distribution 
of the radio spectral index over the minihalo in A\,3444, 
which can provide important constraints on the origin of the diffuse emission. Our 8$^{\prime\prime}$-resolution spectral index image 
shows that the spectrum is relatively uniform at $\alpha \sim 0.8-1.2$ 
(with $<0.1$ uncertainty) within the central $50$ kpc. 
Spatial fluctuations from 1.2 to 1.5 (with uncertainties of 
$\sim 0.1-0.2$) are seen outside of this region, with a hint 
of a steepening toward the outskirts. A radial profile confirms 
this trend, showing a gradual steepening from $\alpha=0.8\pm0.1$ to 
$\alpha=1.3\pm0.3$ from $r\sim 20$ kpc to $r\sim 80$ kpc. 
This behavior is expected if the minihalo originates from 
turbulence re-acceleration of relativistic particles, in 
particular if the magnetic field (or re-acceleration rate) 
is declining radially, as discussed for giant radio halos 
\citep[e.g.,][]{2001MNRAS.320..365B,2022ApJ...933..218B}.
As a note of caution, it is possible that the innermost and 
flatter ($\alpha \sim 0.8$) region of the minihalo in A\,3444 
is contaminated by the activity of the BCG. No large-scale jets 
and lobes are seen in our high-resolution images, where the BCG 
is detected as a compact source ($<8$ kpc). Furthermore, no hints 
of AGN-driven X-ray cavities are seen in the {\em Chandra} image, 
further suggesting the absence of AGN extended emission on large scale. However, we cannot completely rule out the possibility that the 
AGN has faint extended structure projected over the innermost region 
of the minihalo, thus affecting our spectral index measurement 
within this region.  
 
Spectral index images of the minihalo in MS\,1455.0+2232 were presented by \cite{2022MNRAS.512.4210R} at $15^{\prime\prime}$ and 8$^{\prime\prime}$ resolution. A relatively uniform spectral index was found over 
the minihalo extent, with possible fluctuations seen at large radii 
at the highest resolution. However, no clear radial spectral steepening was observed, with the exception of a possible steeper spectral index 
in the region of the SW radio tail.
The lack of large variations in spectral index may result from the integration of different turbulent substructures along the line of 
sight, and therefore it may still be consistent with a turbulent reacceleration origin for the minihalo \citep{2022MNRAS.512.4210R}.

\section{Summary and conclusions}

We presented radio and X-ray studies of A\,3444 and
MS\,1455.0+2232, two cool-core galaxy clusters that host diffuse radio minihalos 
in their centers.

In our GMRT (333, 607 and 1300 MHz) and VLA (1435 MHz) images of A\,3444,
the minihalo fills the inner $r\sim 120$ kpc region of the cluster 
and encloses the central radio galaxy, that remains unresolved in our highest-resolution image ($2^{\prime\prime}$). 
South-West of the cluster center, a much fainter radio extension is detected at 607 MHz, out to a distance 
of $\sim 380$ kpc. This structure is also seen in MeerKAT images at 1283 MHz \citep{2023MNRAS.520.4410T}.
We derived the integrated radio spectrum of the A\,3444 minihalo, 
which is consistent with a power law with $\alpha=1.0\pm0.1$, 
and studied the distribution of the spectral index between 607 
MHz and 1435 MHz. We found a systematic steepening of the spectrum 
with increasing distance from the center. This behavior is expected 
if the minihalo originates from turbulence re-acceleration of 
relativistic particles.

We analyzed an archival uGMRT observation at 383 MHz of MS\,1455.0+2232. 
Similarly to A\,3444, a large-scale faint radio extension beyond 
the previously-known minihalo is detected in the 383 MHz images. 
This radio extension was previously reported by \cite{2022MNRAS.512.4210R} 
using LOFAR and MeerKAT observations.

We found three sloshing cold fronts in the {\em Chandra} X-ray images 
of the cool core of A\,3444, at $r=60$, $120$ and $230$ kpc. A fourth,
larger and older cold front is located well outside the core  
at $r=400$ kpc --- in the region of the SW radio extension --- 
suggesting that the ICM is sloshing on a cluster-wide scale. 
We also found a prominent cold front at $r=425$ kpc in MS\,1455.0+2232, 
well beyond its cool core. In both clusters, the diffuse radio emission 
is contained within these outer fronts. This strongly suggests that the 
whole radio emission detected in these clusters arises from electrons 
confined and re-accelerated in turbulent fields in large-scale 
sloshing motions 
of the hot ICM. 
\\
\\
  {\it Acknowledgements.}
We thank the anonymous referee for the timely report and comments.
Basic research in radio astronomy at the Naval Research Laboratory is 
supported by 6.1 Base funding.
We thank the staff of the GMRT that made the observations possible. 
GMRT is run by the National Centre for Astrophysics of the Tata Institute 
of Fundamental Research. RK acknowledges the support of the Department of Atomic Energy, Government of India, under project no. 12-R\&D-TFR-5.02-0700. The National Radio Astronomy Observatory is a 
facility of the National Science Foundation operated under cooperative 
agreement by Associated Universities, Inc. This scientific work makes use of the data from the Murchison Radio-astronomy Observatory and Australian SKA Pathfinder, managed by CSIRO. Support for the operation of the MWA and ASKAP is provided by the Australian Government (NCRIS). ASKAP and MWA use the resources of the Pawsey Supercomputing Centre. Establishment of ASKAP, MWA and the Pawsey Supercomputing Centre are initiatives of the Australian Government, with support from the Government of Western Australia and the Science and Industry Endowment Fund. We acknowledge the Wajarri Yamatji people as the traditional owners of the observatory sites. This paper employs a list of Chandra datasets, obtained by the Chandra X-ray Observatory, contained in~\dataset[Chandra Data Collection (CDC) 184]{https://doi.org/10.25574/cdc.184}. This research has made use of software provided by the Chandra X-ray Center (CXC) in the application packages CIAO. This research has made use of the CIRADA cutout service at URL cutouts.cirada.ca, operated by the Canadian Initiative for Radio Astronomy Data Analysis (CIRADA). CIRADA is funded by a grant from the Canada Foundation for Innovation 2017 Innovation Fund (Project 35999), as well as by the Provinces of Ontario, British Columbia, Alberta, Manitoba and Quebec, in collaboration with the National Research Council of Canada, the US National Radio Astronomy Observatory and Australia’s Commonwealth Scientific and Industrial Research Organisation.

{}


\begin{thebibliography}{}
\bibitem[Ascasibar 
\& Markevitch(2006)]{2006ApJ...650..102A} Ascasibar, Y., \& Markevitch, M.\ 2006, \apj, 650, 102 

\bibitem[B{\'e}gin et al.(2023)]{2023MNRAS.519..767B} B{\'e}gin, T., Hlavacek-Larrondo, J., Rhea, C.~L., et al.\ 2023, \mnras, 519, 767. doi:10.1093/mnras/stac3526

\bibitem[Bellomi et al.(2023)]{2023arXiv231009422B} Bellomi, E., ZuHone, J., Weinberger, R., et al.\ 2023, arXiv:2310.09422

\bibitem[Biava et al.(2021)]{2021MNRAS.508.3995B} Biava, N., de Gasperin, F., Bonafede, A., et al.\ 2021, \mnras, 508, 3995. doi:10.1093/mnras/stab2840

\bibitem[Bonafede et al.(2022)]{2022ApJ...933..218B} Bonafede, A., Brunetti, G., Rudnick, L., et al.\ 2022, \apj, 933, 218. doi:10.3847/1538-4357/ac721d

\bibitem[Bravi et al.(2016)]{2016MNRAS.455L..41B} Bravi, L., Gitti, M., \& Brunetti, G.\ 2016, \mnras, 455, L41 


\bibitem[Briggs(1995)]{1995PhDT.......238B} Briggs, D.~S.\ 1995, Ph.D. Thesis, New Mexico Institute of Mining and Technology


\bibitem[Brunetti et al.(2001)]{2001MNRAS.320..365B} Brunetti, G., Setti, G., Feretti, L., et al.\ 2001, \mnras, 320, 365. doi:10.1046/j.1365-8711.2001.03978.x

\bibitem[Brunetti \& Jones(2014)]{2014IJMPD..2330007B} Brunetti, G., \& Jones, T.~W.\ 2014, International Journal of Modern Physics D, 23, 1430007 


\bibitem[Bruno et al.(2023)]{2023arXiv230807603B} Bruno, L., Botteon, A., Shimwell, T., et al.\ 2023, arXiv:2308.07603. doi:10.48550/arXiv.2308.07603

\bibitem[CASA Team et al.(2022)]{2022PASP..134k4501C} CASA Team, Bean, B., Bhatnagar, S., et al.\ 2022, \pasp, 134, 114501. doi:10.1088/1538-3873/ac9642
  
\bibitem[Cassano et 
al.(2008)]{2008A&A...486L..31C} Cassano, R., Gitti, M., \& Brunetti, G.\ 2008, \aap, 486, L31 

\bibitem[Chambers et al.(2016)]{2016arXiv161205560C} Chambers, K.~C., Magnier, E.~A., Metcalfe, N., et al.\ 2016, arXiv:1612.05560

\bibitem[Chandra et al.(2004)]{2004ApJ...612..974C} Chandra, P., Ray, A., \& Bhatnagar, S.\ 2004, \apj, 612, 974. doi:10.1086/422675

\bibitem[Donnert et al.(2013)]{2013MNRAS.429.3564D} Donnert, J., Dolag, K., Brunetti, G., et al.\ 2013, \mnras, 429, 3564. doi:10.1093/mnras/sts628


\bibitem[Duchesne et al.(2021)]{2021PASA...38...53D} Duchesne, S.~W., Johnston-Hollitt, M., \& Bartalucci, I.\ 2021, \pasa, 38, e053. doi:10.1017/pasa.2021.45

\bibitem[Eckert et al.(2020)]{2020OJAp....3E..12E} Eckert, D., Finoguenov, A., Ghirardini, V., et al.\ 2020, The Open Journal of Astrophysics, 3, 12. doi:10.21105/astro.2009.13944

\bibitem[Fujita et al.(2004)]{2004ApJ...612L...9F} Fujita, Y., Matsumoto, T., \& Wada, K.\ 2004, \apjl, 612, L9. doi:10.1086/424483

\bibitem[Fujita et al.(2007)]{2007ApJ...663L..61F} Fujita, Y., Kohri, K., 
Yamazaki, R., \& Kino, M.\ 2007, \apjl, 663, L61 

\bibitem[Fujita \& Ohira(2013)]{2013MNRAS.428..599F} Fujita, Y., \& Ohira, Y.\ 2013, \mnras, 428, 599 

\bibitem[Gendron-Marsolais et al.(2017)]{2017MNRAS.469.3872G} Gendron-Marsolais, M., Hlavacek-Larrondo, J., van Weeren, R.~J., et al.\ 2017, \mnras, 469, 3872. doi:10.1093/mnras/stx1042


\bibitem[Giacintucci et al.(2014a)]{2014ApJ...781....9G} Giacintucci, S., 
Markevitch, M., Venturi, T., et al.\ 2014a, \apj, 781, 9 

\bibitem[Giacintucci et al.(2014b)]{2014ApJ...795...73G} Giacintucci, S., 
Markevitch, M., Brunetti, G., et al.\ 2014b, \apj, 795, 73 


\bibitem[Giacintucci et al.(2017)]{2017ApJ...841...71G} Giacintucci, S., Markevitch, M., Cassano, R., et al.\ 2017, \apj, 841, 71. doi:10.3847/1538-4357/aa7069 

\bibitem[Giacintucci et al.(2019)]{2019ApJ...880...70G} Giacintucci, S., Markevitch, M., Cassano, R., et al.\ 2019, \apj, 880, 70. doi:10.3847/1538-4357/ab29f1 

\bibitem[Giacintucci et al.(2020)]{2020ApJ...891....1G} Giacintucci, S., Markevitch, M., Johnston-Hollitt, M., et al.\ 2020, \apj, 891, 1. doi:10.3847/1538-4357/ab6a9d

\bibitem[Giacintucci et al.(2022)]{2022ApJ...934...49G} Giacintucci, S., Venturi, T., Markevitch, M., et al.\ 2022, \apj, 934, 49. doi:10.3847/1538-4357/ac7805

\bibitem[Giovannini et 
al.(2009)]{2009A&A...507.1257G} Giovannini, G., Bonafede, A., Feretti, L., et al.\ 2009, \aap, 507, 1257 (GBF09)

\bibitem[Giovannini et al.(2020)]{2020A&A...640A.108G} Giovannini, G., Cau, M., Bonafede, A., et al.\ 2020, \aap, 640, A108. doi:10.1051/0004-6361/202038263

\bibitem[Gitti et al.(2002)]{2002A&A...386..456G} Gitti, M., Brunetti, G., \& Setti, G.\ 2002, \aap, 386, 456 

\bibitem[Gitti et al.(2018)]{2018A&A...617A..11G} Gitti, M., Brunetti, G., Cassano, R., et al.\ 2018, \aap, 617, A11. doi:10.1051/0004-6361/201832749

\bibitem[Greisen(2003)]{2003ASSL..285..109G} Greisen, E.~W.\ 2003, Information Handling in Astronomy - Historical Vistas, 285, 109. doi:10.1007/0-306-48080-8\_7

  
\bibitem[Hale et al.(2021)]{2021PASA...38...58H} Hale, C.~L., McConnell, D., Thomson, A.~J.~M., et al.\ 2021, \pasa, 38, e058. doi:10.1017/pasa.2021.47


\bibitem[Hamer et al.(2012)]{2012MNRAS.421.3409H} Hamer, S.~L., Edge, A.~C., Swinbank, A.~M., et al.\ 2012, \mnras, 421, 3409. doi:10.1111/j.1365-2966.2012.20566.x

\bibitem[Hamer et al.(2016)]{2016MNRAS.460.1758H} Hamer, S.~L., Edge, A.~C., Swinbank, A.~M., et al.\ 2016, \mnras, 460, 1758. doi:10.1093/mnras/stw1054

\bibitem[Hlavacek-Larrondo et al.(2013)]{2013ApJ...777..163H} 
Hlavacek-Larrondo, J., Allen, S.~W., Taylor, G.~B., et al.\ 2013, \apj, 
777, 163 

\bibitem[Hurley-Walker et al.(2017)]{2017MNRAS.464.1146H} Hurley-Walker, N., Callingham, J.~R., Hancock, P.~J., et al.\ 2017, \mnras, 464, 1146. doi:10.1093/mnras/stw2337

\bibitem[Ichinohe et al.(2021)]{2021MNRAS.504.2800I} Ichinohe, Y., Simionescu, A., Werner, N., et al.\ 2021, \mnras, 504, 2800. doi:10.1093/mnras/stab1060

\bibitem[Ignesti et al.(2020)]{2020A&A...640A..37I} Ignesti, A., Brunetti, G., Gitti, M., et al.\ 2020, \aap, 640, A37. doi:10.1051/0004-6361/201937207

\bibitem[Intema et al.(2009)]{2009A&A...501.1185I} Intema, H.~T., van der Tol, S., Cotton, W.~D., et al.\ 2009, \aap, 501, 1185. doi:10.1051/0004-6361/200811094

\bibitem[Intema et al.(2017)]{2017A&A...598A..78I} Intema, H.~T., Jagannathan, P., Mooley, K.~P., et al.\ 2017, \aap, 598, A78. doi:10.1051/0004-6361/201628536

\bibitem[Jacob \& Pfrommer(2017)]{2017MNRAS.467.1478J} Jacob, S. \& Pfrommer, C.\ 2017, \mnras, 467, 1478. doi:10.1093/mnras/stx132

\bibitem[Johnson et al.(2010)]{2010ApJ...710.1776J} Johnson, R.~E., Markevitch, M., Wegner, G.~A., et al.\ 2010, \apj, 710, 1776. doi:10.1088/0004-637X/710/2/1776

\bibitem[Kale et al.(2015)]{2015A&A...579A..92K} Kale, R., Venturi, T., Giacintucci, S., et al.\ 2015, \aap, 579, A92 


\bibitem[Kale et al.(2022)]{2022MNRAS.514.5969K} Kale, R., Parekh, V., Rahaman, M., et al.\ 2022, \mnras, 514, 5969. doi:10.1093/mnras/stac1649

\bibitem[Keshet(2010)]{2010arXiv1011.0729K} Keshet, U.\ 2010, 
arXiv:1011.0729 

\bibitem[Keshet \& Loeb(2010)]{2010ApJ...722..737K} Keshet, U., \& Loeb, A.\, 2010, \apj, 722, 737 


\bibitem[Knowles et al.(2022)]{2022A&A...657A..56K} Knowles, K., Cotton, W.~D., Rudnick, L., et al.\ 2022, \aap, 657, A56. doi:10.1051/0004-6361/202141488

\bibitem[Lacy et al.(2020)]{2020PASP..132c5001L} Lacy, M., Baum, S.~A., Chandler, C.~J., et al.\ 2020, \pasp, 132, 035001. doi:10.1088/1538-3873/ab63eb

\bibitem[Lane et al.(2014)]{2014MNRAS.440..327L} Lane, W.~M., Cotton, W.~D., van Velzen, S., et al.\ 2014, \mnras, 440, 327. doi:10.1093/mnras/stu256


\bibitem[Lusetti et al.(2023)]{2023arXiv230801884L} Lusetti, G., Bonafede, A., Lovisari, L., et al.\ 2023, arXiv:2308.01884. doi:10.48550/arXiv.2308.01884

\bibitem[Markevitch et al.(2000)]{2000ApJ...541..542M} Markevitch, M., Ponman, T.~J., Nulsen, P.~E.~J., et al.\ 2000, \apj, 541, 542. doi:10.1086/309470

\bibitem[Markevitch 
\& Vikhlinin(2007)]{2007PhR...443....1M} Markevitch, M., \& Vikhlinin, A.\ 2007, \physrep, 443, 1 


\bibitem[Mazzotta et al.(2001)]{2001astro.ph..8476M} Mazzotta, P., Markevitch, M., Forman, W.~R., et al.\ 2001, astro-ph/0108476. doi:10.48550/arXiv.astro-ph/0108476

\bibitem[Mazzotta \& Giacintucci(2008)]{2008ApJ...675L...9M} Mazzotta, P., \& Giacintucci, S.\ 2008, \apjl, 675, L9 

\bibitem[McConnell et al.(2020)]{2020PASA...37...48M} McConnell, D., Hale, C.~L., Lenc, E., et al.\ 2020, \pasa, 37, e048. doi:10.1017/pasa.2020.41
  
\bibitem[McDonald et al.(2017)]{2017ApJ...843...28M} McDonald, M., Allen, S.~W., Bayliss, M., et al.\ 2017, \apj, 843, 28. doi:10.3847/1538-4357/aa7740

\bibitem[Offringa et al.(2014)]{2014MNRAS.444..606O} Offringa, A.~R., McKinley, B., Hurley-Walker, N., et al.\ 2014, \mnras, 444, 606. doi:10.1093/mnras/stu1368

\bibitem[Offringa \& Smirnov(2017)]{2017MNRAS.471..301O} Offringa, A.~R. \& Smirnov, O.\ 2017, \mnras, 471, 301. doi:10.1093/mnras/stx1547
  
\bibitem[Pasini et al.(2019)]{2019ApJ...885..111P} Pasini, T., Gitti, M., Brighenti, F., et al.\ 2019, \apj, 885, 111. doi:10.3847/1538-4357/ab4808

\bibitem[Pasini et al.(2021)]{2021ApJ...911...66P} Pasini, T., Gitti, M., Brighenti, F., et al.\ 2021, \apj, 911, 66. doi:10.3847/1538-4357/abe85f

\bibitem[Perley \& Butler(2017)]{2017ApJS..230....7P} Perley, R.~A. \& Butler, B.~J.\ 2017, \apjs, 230, 7. doi:10.3847/1538-4365/aa6df9

\bibitem[Pfrommer \& En{\ss}lin(2004)]{2004A&A...413...17P} Pfrommer, C., \& En{\ss}lin, T.~A.\ 2004, \aap, 413, 17 

\bibitem[Perrott et al.(2021)]{2021MNRAS.508.2862P} Perrott, Y.~C., Carvalho, P., Elwood, P.~J., et al.\ 2021, \mnras, 508, 2862. doi:10.1093/mnras/stab2706


\bibitem[Prasow-{\'E}mond et al.(2020)]{2020AJ....160..103P} Prasow-{\'E}mond, M., Hlavacek-Larrondo, J., Rhea, C.~L., et al.\ 2020, \aj, 160, 103. doi:10.3847/1538-3881/ab9ff3

\bibitem[Roediger \& Zuhone(2012)]{2012MNRAS.419.1338R} Roediger, E. \& Zuhone, J.~A.\ 2012, \mnras, 419, 1338. doi:10.1111/j.1365-2966.2011.19794.x


\bibitem[Riseley et al.(2022)]{2022MNRAS.512.4210R} Riseley, C.~J., Rajpurohit, K., Loi, F., et al.\ 2022, \mnras, 512, 4210. doi:10.1093/mnras/stac672


\bibitem[Riseley et al.(2023)]{2023MNRAS.524.6052R} Riseley, C.~J., Biava, N., Lusetti, G., et al.\ 2023, \mnras, 524, 6052. doi:10.1093/mnras/stad2218

\bibitem[Richard-Laferri{\`e}re et al.(2020)]{2020MNRAS.499.2934R} Richard-Laferri{\`e}re, A., Hlavacek-Larrondo, J., Nemmen, R.~S., et al.\ 2020, \mnras, 499,2934. doi:10.1093/mnras/staa2877

\bibitem[Rossetti et al.(2013)]{2013A&A...556A..44R} Rossetti, M., Eckert, D., De Grandi, S., et al.\ 2013, \aap, 556, A44. doi:10.1051/0004-6361/201321319

\bibitem[Sanders et al.(2016)]{2016MNRAS.460.1898S} Sanders, J.~S., Fabian, A.~C., Russell, H.~R., et al.\ 2016, \mnras, 460, 1898. doi:10.1093/mnras/stw1119


\bibitem[Savini et al.(2018)]{2018MNRAS.478.2234S} Savini, F., Bonafede, A., Br{\"u}ggen, M., et al.\ 2018, \mnras, 478, 2234. doi:10.1093/mnras/sty1125

\bibitem[Savini et al.(2019)]{2019A&A...622A..24S} Savini, F., Bonafede, A., Br{\"u}ggen, M., et al.\ 2019, \aap, 622, A24. doi:10.1051/0004-6361/201833882


\bibitem[Scaife 
\& Heald(2012)]{2012MNRAS.423L..30S} Scaife, A.~M.~M., \& Heald, G.~H.\ 2012, \mnras, 423, L30 

\bibitem[Simionescu et al.(2012)]{2012ApJ...757..182S} Simionescu, A., Werner, N., Urban, O., et al.\ 2012, \apj, 757, 182. doi:10.1088/0004-637X/757/2/182

\bibitem[Storm et al.(2015)]{2015MNRAS.448.2495S} Storm, E., Jeltema, T.~E., \& Rudnick, L.\ 2015, \mnras, 448, 2495. doi:10.1093/mnras/stv164


\bibitem[Timmerman et al.(2021)]{2021A&A...646A..38T} Timmerman, R., van Weeren, R.~J., McDonald, M., et al.\ 2021, \aap, 646, A38. doi:10.1051/0004-6361/202039075

\bibitem[Tittley \& Henriksen(2005)]{2005ApJ...618..227T} Tittley, E.~R. \& Henriksen, M.\ 2005, \apj, 618, 227. doi:10.1086/425952


\bibitem[Trehaeven et al.(2023)]{2023MNRAS.520.4410T} Trehaeven, K.~S., Parekh, V., Oozeer, N., et al.\ 2023, \mnras, 520, 4410. doi:10.1093/mnras/stad391


\bibitem[Ubertosi et al.(2021)]{2021MNRAS.503.4627U} Ubertosi, F., Gitti, M., Torresi, E., et al.\ 2021, \mnras, 503, 4627. doi:10.1093/mnras/stab819


\bibitem[van Weeren et al.(2019)]{2019SSRv..215...16V} van Weeren, R.~J., de Gasperin, F., Akamatsu, H., et al.\ 2019, \ssr, 215, 16. doi:10.1007/s11214-019-0584-z

\bibitem[Vazza et al.(2012)]{2012A&A...544A.103V} Vazza, F., Roediger, E., \& Br{\"u}ggen, M.\ 2012, \aap, 544, A103. doi:10.1051/0004-6361/201118688

\bibitem[Venturi et al.(2007)]{2007A&A...463..937V} Venturi, T., Giacintucci, S., Brunetti, G., et al.\ 2007, \aap, 463, 937 


\bibitem[Venturi et al.(2008)]{2008A&A...484..327V} Venturi, T., Giacintucci, S., Dallacasa, D., et al.\ 2008, \aap, 484, 327. doi:10.1051/0004-6361:200809622

\bibitem[Venturi et al.(2017)]{2017A&A...603A.125V} Venturi, T., Rossetti, M., Brunetti, G., et al.\ 2017, \aap, 603, A125. doi:10.1051/0004-6361/201630014


\bibitem[Walker et al.(2014)]{2014MNRAS.441L..31W} Walker, S.~A., Fabian, A.~C., \& Sanders, J.~S.\ 2014, \mnras, 441, L31. doi:10.1093/mnrasl/slu040


\bibitem[Walker et al.(2018)]{2018NatAs...2..292W} Walker, S.~A., ZuHone, J., Fabian, A., et al.\ 2018, Nature Astronomy, 2, 292. doi:10.1038/s41550-018-0401-8


\bibitem[Werner et al.(2016)]{2016MNRAS.460.2752W} Werner, N., Zhuravleva, I., Canning, R.~E.~A., et al.\ 2016, \mnras, 460, 2752. doi:10.1093/mnras/stw1171


\bibitem[Zandanel et al.(2014)]{2014MNRAS.438..124Z} Zandanel, F., Pfrommer, C., \& Prada, F.\ 2014, \mnras, 438, 124. doi:10.1093/mnras/stt2250



\bibitem[ZuHone et al.(2010)]{2010ApJ...717..908Z} ZuHone, J.~A., Markevitch, M., \& Johnson, R.~E.\ 2010, \apj, 717, 908. doi:10.1088/0004-637X/717/2/908


\bibitem[ZuHone et al.(2013)]{2013ApJ...762...78Z} ZuHone, J.~A., 
Markevitch, M., Brunetti, G., \& Giacintucci, S.\ 2013, \apj, 762, 78 

\bibitem[ZuHone et al.(2015)]{2015ApJ...801..146Z} ZuHone, J.~A., Brunetti, G., Giacintucci, S., et al.\ 2015, \apj, 801, 146. doi:10.1088/0004-637X/801/2/146


\bibitem[ZuHone et al.(2021)]{2021ApJ...914...73Z} ZuHone, J.~A., Markevitch, M., Weinberger, R., et al.\ 2021, \apj, 914, 73. doi:10.3847/1538-4357/abf7bc

\bibitem[ZuHone \& Su(2022)]{2022hxga.book...93Z} ZuHone, J. \& Su, Y.\ 2022, Handbook of X-ray and Gamma-ray Astrophysics, 93. doi:10.1007/978-981-16-4544-0\_124-1

\end{thebibliography}
\end{document}